\documentclass[%
  onecolumn
]{mpi2015-cscpreprint}

\usepackage[american]{babel}

\usepackage{graphicx}

\usepackage[utf8]{inputenc}
\usepackage{epsfig}
\usepackage{xcolor}
\usepackage{amsmath,amssymb,rotating}
\usepackage{comment,longtable}
\usepackage{mathdots,pstricks,fancybox,tcolorbox}
\usepackage{todonotes}
\usepackage{newtxtext}
\usepackage{newtxmath}
\usepackage{lipsum}
\usepackage{amsfonts}
\usepackage{algorithmic}
\usepackage{dsfont,url}
\usepackage{here}
\usepackage{float}

\usepackage{graphbox}
\usepackage{tikz}
\usetikzlibrary{arrows,decorations.pathmorphing,backgrounds,fit,positioning,shapes.symbols,chains,mindmap,trees,snakes,calc}

\definecolor{bleuONERA}{RGB}{16,97,169}
\definecolor{grisONERA}{RGB}{64,64,66}

\providecommand{\eg}[0]{\textit{e.g.}~} 
\providecommand{\ie}[0]{\textit{i.e.}~}

\newcommand{\bA}{{\mathbf A}}
\newcommand{\bB}{{\mathbf B}}
\newcommand{\bC}{{\mathbf C}}
\newcommand{\bD}{{\mathbf D}}
\newcommand{\bE}{{\mathbf E}}
\newcommand{\bF}{{\mathbf F}}
\newcommand{\bG}{{\mathbf G}}
\newcommand{\bH}{{\mathbf H}}
\newcommand{\bI}{{\mathbf I}}
\newcommand{\bL}{{\mathbf L}}
\newcommand{\bM}{{\mathbf M}}
\newcommand{\bN}{{\mathbf N}}
\newcommand{\bQ}{{\mathbf Q}}
\newcommand{\bR}{{\mathbf R}}
\newcommand{\bS}{{\mathbf S}}

\newcommand{\bU}{{\mathbf U}}
\newcommand{\bV}{{\mathbf V}}
\newcommand{\bW}{{\mathbf W}}
\newcommand{\bX}{{\mathbf X}}
\newcommand{\bY}{{\mathbf Y}}

\newcommand{\bfe}{{\mathbf e}}
\newcommand{\bff}{{\mathbf f}}
\newcommand{\bg}{{\mathbf g}}
\newcommand{\bh}{{\mathbf h}}
\newcommand{\bq}{{\mathbf q}}
\newcommand{\br}{{\mathbf r}}
\newcommand{\bu}{{\mathbf u}}
\newcommand{\bv}{{\mathbf v}}
\newcommand{\bw}{{\mathbf w}}
\newcommand{\bx}{{\mathbf x}}

\newcommand{\by}{{\mathbf y}}
\newcommand{\bl}{{\mathbf l}}
\renewcommand{\br}{{\mathbf r}}

\newcommand{\cS}{{\mathcal S}}

\newcommand{\bPhi}{ \boldsymbol{\Phi} }
\def\IR{{\mathbb R}}
\def\IC{{\mathbb C}}
\def\IN{{\mathbb N}}
\newcommand{\cO}{ {\cal O} }
\newcommand{\cR}{ {\cal R} }
\def\IL{{\mathbb L}}
\def\IT{{\mathbb T}}
\def\IV{{\mathbb V}}
\def\IW{{\mathbb W}}
\newcommand{\sIL}{{{{\mathbb M}}}}

\newcommand{\bLambda}{{\boldsymbol{\Lambda}}}
\newcommand{\bPi}{{\boldsymbol{\Pi}}}

\newcommand{\bfz}{{\mathbf 0}}
\newcommand{\Si}{{\boldsymbol{\Sigma}}}
\providecommand{\rank}[0]{\textbf{rank}} 
\providecommand{\kernel}[0]{\textbf{ker}} 
\newcommand{\blambda}{\boldsymbol{\lambda}}
\newcommand{\bmu}{\boldsymbol{\mu}}
\providecommand{\x}[0]{\mathbf{x}} %
\providecommand{\dx}[0]{\mathbf{\dot{x}}} %
\renewcommand{\u}{\mathbf{u}} %
\providecommand{\y}[0]{\mathbf{y}} %
\providecommand{\bv}[0]{\mathbf{b}} %
\providecommand{\Hreal}[0]{\mathcal{S}} %
\providecommand{\Htran}[0]{\mathbf{H}} %

\providecommand{\norm}[1]{|| #1 ||} %
\providecommand{\Hinf}[0]{\mathcal H_\infty} %
\providecommand{\Htwo}[0]{\mathcal H_2} %

\providecommand{\Cplx}[0]{\mathbb{C}} %
\providecommand{\Real}[0]{\mathbb{R}} %
\providecommand{\lti}[0]{\textbf{LTI}~} %

\newcommand{\udisk}{\mathcal{D}}
\newcommand{\ucir}{\partial \mathcal{D}}
\newcommand{\udiskcomp}{\overline{\mathcal{D}}}
\newcommand{\cp}{\mathbb{C}_+}

\newtheorem{theorem}{Theorem}

\newtheorem{remark}{Remark}

\newtheorem{example}{Example}

\providecommand{\red}[1]{\textcolor[rgb]{0.98,0.00,0.00}{#1}}

\providecommand{\matrixtwo}[4]{ \left[\begin{array}{cc} #1 & #2 \\ #3 & #4 \end{array}\right] } %
\providecommand{\matrixthree}[9]{ \left[\begin{array}{ccc} #1 & #2 & #3 \\ #4 & #5 & #6 \\ #7 & #8 & #9\end{array}\right] } %
\providecommand{\vectorthree}[3]{ \left[\begin{array}{c} #1 \\ #2 \\ #3 \end{array}\right] } %
\begin{document}
  
%%%%%%%%%%%%%%%%%%%%%%%%%%%%%%%%%%%%%%%%%%%%%%%%%%%%%%%%%%%%%%%%%%%%%%%%%%%%%%%%
% PAPER INFORMATION.                                                           %
%%%%%%%%%%%%%%%%%%%%%%%%%%%%%%%%%%%%%%%%%%%%%%%%%%%%%%%%%%%%%%%%%%%%%%%%%%%%%%%%

\title{Data-driven modeling and control of large-scale dynamical systems in the Loewner framework}
\subtitle{Methodology and Applications}
  
\author[$\ast$]{Ion Victor Gosea}
\affil[$\ast$]{Max Planck Institute for Dynamics of Complex Technical Systems,
	Sandtorstr. 1, 39106 Magdeburg, Germany.\authorcr
	\email{gosea@mpi-magdeburg.mpg.de}, \orcid{0000-0003-3580-4116}}  
  
\author[$\dagger$]{Charles Poussot-Vassal}
\affil[$\dagger$]{Information Processing and Systems Department, ONERA, 2 Av. Edouard Belin, 31000, Toulouse, France.\authorcr
  \email{charles.poussot-vassal@onera.fr}, \orcid{0000-0001-9106-1893}}

\author[$\ddagger$]{Athanasios C. Antoulas}
\affil[$\ddagger$]{Department of Electrical and Computer Engineering, Rice University, Houston, Texas 77005, and
	Max-Planck Institute for the Dynamics of Complex Technical Systems,
	D-39106 Magdeburg.\authorcr
	\email{aca@rice.edu}}  
  
\shorttitle{Data-driven Loewner modeling and control}
\shortauthor{I. V. Gosea, C. Poussot-Vassal, A. C. Antoulas}
\shortdate{}


%\msc{MSC1, MSC2, MSC3}

\abstract{%
   In this contribution, we discuss the modeling and model reduction framework known as the Loewner framework. This is a data-driven approach, applicable to large-scale systems, which was originally developed for applications to linear time-invariant systems. In recent years, this method has been extended to a number of additional more complex scenarios, including linear parametric or nonlinear dynamical systems. We will provide here an overview of the latter two, together with time-domain extensions. Additionally, the application of the Loewner framework is illustrated by a collection of practical test cases. Firstly, for data-driven complexity reduction of the underlying model, and secondly, for dealing with control applications of complex systems (in particular, with feedback controller design).
}

\maketitle

%%%%%%%%%%%%%%%%%%%%%%%%%%%%%%%%%%%%%%%%%%%%%%%%%%%%%%%%%%%%%%%%%%%%%%%%%%%%%%%%
% PAPER CONTENT.                                                               %
%%%%%%%%%%%%%%%%%%%%%%%%%%%%%%%%%%%%%%%%%%%%%%%%%%%%%%%%%%%%%%%%%%%%%%%%%%%%%%%%

\section{Introduction: Data-driven modeling and control}
\label{sec:intro}
The physical complexity of dynamical systems describing time-dependent processes stems from underlying non-linearities, the coupling dynamics, or the large amount of degrees of freedom (variables or parameters). The latter aspect is also related to enforcing specific accuracy requirements, that yield models of large dimension which are hence challenging to use for control purposes or for numerical simulations.

Simulating such complex dynamical systems is currently a common feature of many numerical software toolboxes, and is widely used both in industry and in academia. As numerical simulations become more involved, processing of increased amounts of data is required. Consequently, the number of variables under analysis is limited to the physical ones (even in the era of machine learning), while the rest are discarded. Computing simplified, easy to use dynamical models is one purpose of the model approximation and reduction discipline. Such models may then be used in a many query optimisation and simulation processes.
That is why it is of critical importance to construct reliable surrogate models. Model reduction typically refers to a class of methodologies used for reducing the computational complexity of large-scale models of dynamical systems. The goal generally is to approximate the original system with a smaller and simpler system with the same structure
and similar response characteristics as the original. For an overview of model reduction methods, we refer the reader to \cite{ACA05,benner15,BOCW17,ABG20}, and to the references therein. In many practical scenarios, a complete mathematical description of the dynamical system under study is not always available or not fully known. Instead of depending only on physical laws (describing partial or ordinary differential equations), one can infer important properties of the system directly from measured or computed data.

With the increasing emergence of data-driven applications in many fields of the applied sciences, the need for incorporating measurements in the modeling and controlling stage has steadily grown over the last decades. The main challenge consists in using the available data in order to effectively construct surrogate models or controllers. In this latter case, the controller has to be designed based on experimental measurements, instead of a model description.  Model-based methods can hence be replaced by data-driven strategies that construct the controller, directly from experimental data. Such techniques are also known as direct methods and can be useful when control-oriented models are either too complex or too costly to obtain. 

The Loewner Framework (LF) is a data-driven model identification and reduction technique that was originally introduced in \cite{MA07}. It is based on the Loewner-matrix interpolation method elaborated by the third author of the current paper, more than 20 years earlier, in the seminal contribution \cite{AA86}. Using only measured data, the LF constructs surrogate models directly and with low computational effort. For recent tutorial papers on LF applied to linear dynamical systems, we refer the reader to \cite{ALI17,Karachalios:2020}. Extensions that use time-domain data were provided in \cite{ia12,AntoulasPLOSOne:2018} (for a Hankel matrix approach) as well as in \cite{PGW17} (for inferring transfer function measurements from time series). The Loewner framework has been recently extended to certain classes of nonlinear systems, such as bilinear systems in \cite{AGI16}, and quadratic-bilinear systems in \cite{morGosA15,GA18}. An adaptive extension of the original Loewner-based method in \cite{AA86}, named the AAA (Adaptive-Antoulas-Anderson) algorithm, was recently proposed in \cite{NST18}; it is a data-driven rational approximation method that combines interpolation and least-squares (LS) fitting. 

In the first part of this contribution, the Loewner framework is mainly used as a model identification and reduction tool. In the second part, the same framework is used for feedback controller design. In the proposed setup, the reference controller is not computed by means of a given model, but using input-output data of the system. Consequently,  the Loewner framework is used for synthesizing a controller directly from measured data, which shows that it is also a data-driven control tool. Data-driven control strategies consist in recasting the control design problem as an identification one. By doing so, the model simplification process is shifted directly to the controller design step. The Loewner-based data-driven control methodology was extensively studied in recent years, starting with the original contribution \cite{KergusWC:2017} and subsequently with \cite{VuilleminWC:2020,GoseaECC:2021,PoussotCST:2021,PoussotGTA:2021}.\\
\noindent
The main philosophy of the Loewner framework is as follows: starting with frequency response measurements (or, alternatively, with time-domain sequences of measured inputs and outputs), the data is arrange in a specific matrix format. Then, the dominant characteristics of the model are extracted by means of an appropriate projection (the \textbf{SVD} is the relevant tool here). Thus, simplified/reduced surrogate models can be computed without access to the specific system's description.

%\charles{Balanced truncation methods develop a projecting function ranking the dynamically reachable and observable variables from the largest to the smallest ones. Then discarding the less important variables, while retaining the important ones leads to the reduced model. Rational interpolation constructs a phenomenological model, matching some input-output transfer at well chosen phase and gains (support complex points). As such, rational interpolation provides a generalisation of the Pad\'e method to an arbitrary (set of) point(s). Rational interpolation benefit also of the fact that they only require the transfer function evaluation whereas projection methods require the internal model. The latter are thus referred to as intrusive methods, while the former are non-intrusive ones.}

\subsection{Notations}

We denote by $\Real$ the set of real numbers, $\Cplx$ the set of complex numbers, $\cp$ ($\Cplx_-$) the open right (left) half plane, $\udisk$ the open unit disk, $\ucir$ its boundary and $\udiskcomp$ the complementary of the closed unit disk, respectively. The complex variable is denoted by $\imath=\sqrt{-1}$. $\mathcal{L}_2(\mathcal{I})$ ($\mathcal{I} = \imath\mathbb{R}$ or $\ucir$) denotes the set of functions that are square integrable on $\mathcal{I}$, while $\mathcal{H}_2(\mathcal{\udisk})$ (resp. $\mathcal{H}_2(\mathcal{\udiskcomp})$) is the subset of $\mathcal{L}_2(\ucir)$ containing the functions analytic in $\udisk$ (resp. $\udiskcomp$). Let $\Htwo(\cp)$, shortly $\Htwo$,  (resp. $\mathcal{H}_2(\Cplx_-)$) be the subset of $\mathcal{L}_2(\imath\Real)$ containing the functions analytic in $\cp$ (resp. $\Cplx_-$). Similarly, $\mathcal{L}_\infty (\mathcal{I})$ ($\mathcal{I} = \imath\mathbb{R}$ or $\ucir$) denotes the set of functions that are bounded on $\mathcal{I}$. $\mathcal{H}_\infty(\mathcal{\udisk})$ (resp. $\mathcal{H}_\infty(\mathcal{\udiskcomp})$) denotes the subset of $\mathcal{L}_\infty(\ucir)$ containing the functions analytic in $\udisk$ (resp. $\udiskcomp$) and $\mathcal{H}_\infty(\cp)$, shortly $\Hinf$, the subset of $\mathcal{L}_\infty(\imath\Real)$ of functions analytic in $\cp$. %Let $\mathcal{L}_\infty$ be the set of functions bounded on the imaginary axis and $\chinf \subseteq \mathcal{L}_\infty$ its that are analytic on $\cp$. 
The Fourier transform of a time-domain signal $v \in \mathcal{L}_2(\mathbb{R})$ is denoted by $\overline{v} = \mathcal{F}(v)$.

\subsection{Organizational plan}

The paper is organized as follows: after the introduction on data-driven modeling and control in Section \ref{sec:intro}, we continue with an overview on the Loewner framework for data-driven modeling in Section \ref{sec:loewner} with various subsections that cover specific extensions of the framework. Section \ref{sec:examples} contains three model reduction examples in the Loewner framework, while Section \ref{sec:control} deals with the Loewner data-driven control rationale. This illustrates how the Loewner tool can be effective for both model-based or data-driven control approaches. Finally, \ref{sec:conc} contains the concluding remarks together with a short summary of the paper.

\section{The Loewner framework for data-driven modeling: an overview}
\label{sec:loewner}

\subsection{Generalities on the Loewner framework and model structures}
\label{ssec:loe_gene}

The Loewner framework is a data-driven method aimed at building a time invariant differential algebraic equation model / realization, with associated transfer function $\Htran_{\mathcal I}$ or $\Htran^{(\mathcal J)}$  (defined later). This model interpolates data obtained from experimental measurements or the evaluation of a (collection of) transfer function(s). As made clearer later in this section, according to the mathematical structure and nature of the underlying system, $\Htran_{\mathcal I}$ has some specific properties. 

%$\Htran_M$ of a given dimension $M\in\IN$ which interpolates given complex data. These data may come from any dynamical system. 
%dynamical model given in a

In its original form presented in \cite{MA07}, $\Htran^{(N)}$ (${\mathcal J} = \{N\}$) is a descriptor linear time invariant (\textbf{LTI}) dynamical model with transfer function $\Htran^{(N)}: \IC \rightarrow \IC^{p \times m}$, where $N\in\IN$ denotes the number of collected data. We also denote with $\Htran_n$ the transfer function with Mc Millan degree $n$ (${\mathcal I} = \{n\}$). A complete description of this case is available in the recent surveys \cite{AntoulasSurvey:2016,Karachalios:2020}. Extension to parametric \lti  (\textbf{pLTI}) model structure also exist \cite{IA14}. In this case, one obtains a multi-valued rational transfer function $\Htran^{(N,M)}: (\IC\times\IR) \rightarrow \IC^{p \times m}$ where ${\mathcal J} = \{N,M\}$ data are used (or $\Htran_{r,q}$, where ${\mathcal I} = \{r,q\}$), where $M\in\IN$ is the number of data along the parameter variable (and $q\in\IN$ is the rational order along the parameter). The resulting rational function both interpolates the complex and real parametric variables. From a different perspective, extensions to nonlinear model structures have also been investigated. Among them, one can mention the bilinear and/or quadratic forms, explored in a series of papers \cite{morGosA15,AGI16,GA18,morAntGH19,morGosKA20a,morAntGH20a}. In these cases, the associated  transfer function is a collection of multivariate coupled infinite cascade of linear systems reading as $\Htran^{(N_1)}: \IC \rightarrow \IC^{p \times m}$, $\Htran^{(N_1,N_2)}: (\IC\times\IC) \rightarrow \IC^{p \times m}$ and $\Htran^{(N_1,N_2,\dots)}: (\IC\times\IC\times\dots) \rightarrow \IC^{p \times m}$ (${\mathcal J} = \{N_1,N_2,\dots\}$). The Loewner interpolation framework seek for function that interpolates the $N_1,N_2,\dots$ data along each related multi-valued transfer functions $\Htran^{(N_1)}$, $\Htran^{(N_1,N_2,\dots)}$ ($N_1,N_2,\dots\in\IN$). Similarly, we denote with $\Htran_{r_1,r_2,\dots}$ the associated transfer function of order $r_1,r_2,\dots\in\IN$.

%Whatever the chosen model structure is, the Loewner framework constructs a differential algebraic equation set that interpolates some data obtained from measurement or evaluation of the (collection of) transfer functions. 

In all the cases mentioned here, the transfer function (or the set of transfer functions) is rational, and it interpolates the data. In comparison to realization-driven model reduction,  data-driven methods based on rational interpolation construct models that match the original transfer function(s) at well chosen points in the complex plane (also denoted as support points for barycentric representations \cite{NST18}). As such, it provides a generalization of the Pad\'e method to an arbitrary (set of) point(s). Data-driven methods based on rational interpolation benefits also from the fact that it only requires the transfer function evaluation, whereas projection methods require the internal model (system matrices or operators). The latter are thus referred to as \emph{intrusive} methods, while the former are \emph{non-intrusive} or \emph{data-driven} ones.

In this section, an extensive review of the Loewner framework is provided, together with some of its extensions. More specifically, section \ref{ssec:loe_lti} presents the Loewner framework in its original form, leading to a linear time invariant model. Extensions to linear parametric and to bilinear systems are sequentially illustrated in sections \ref{ssec:loe_plti} and \ref{ssec:loe_blti}. As a direct extension, the time-domain Loewner, dealing with sampled time-domain data instead of frequency domain data, is covered in section \ref{ssec:loe_time}.

\subsection{Loewner framework in the rational LTI case}
\label{ssec:loe_lti}

The main ingredient of the Loewner framework are summarized next in the multi-input multi-output (\textbf{MIMO})  rational \lti case. Let us consider that such system is a $m$ inputs $p$ outputs dynamical one described by a $n$-th order differential algebraic equation (\textbf{DAE}) model $\Hreal:(\bE,\bA,\bB,\bC,\bfz)$ which explicitly reads as
\begin{equation}
\Hreal:\left \lbrace
\begin{array}{l}
\bE   \dot{\bx}(t) = \bA \bx(t) + \bB \bu(t)\text{ , } \by(t) =\bC \bx(t) \text{ where }\\
\bE,\bA \in \IR^{n\times n}, \bB \in \IR^{n\times m}, \bC \in \IR^{p\times n}.
\end{array}
\right. 
\label{eq:ode_real}
\end{equation}
Its associated transfer function $\Htran: \IC \rightarrow \IC^{p \times m}$ is
\begin{equation}
\Htran(\xi) = \bC \bPhi(\xi) \bB \text{ where } \bPhi(\xi) = (\xi\bE-\bA)^{-1} \in \IC^{n\times n}.
\label{eq:ode_tran}
\end{equation}
Importantly, as any rational function, relation \eqref{eq:ode_tran} can be characterized in its Lagrangian basis with distinct Lagrange nodes (or support points) $\lambda_i\in\IC$. Then one can  rewrite it in its rational barycentric formula as follows (for $\alpha_{i} \neq 0$),
\begin{equation}
\Htran(\xi) = \dfrac{\sum_{i=1}^{n+1}\beta_{i}\mathbf q_{i}(\xi)}{\sum_{i=1}^{n+1}\alpha_{i}\mathbf q_{i}(\xi)}
\text{ where } \mathbf q_{i}(\xi) = \dfrac{1}{\xi-\lambda_i}.
\label{eq:ode_rat}
\end{equation}
Let this system generate the right (or column) data together with the left (or row)  data, as follows
\begin{equation}
\left.
\begin{array}{c}
(\lambda_i,\br_i,\bw_i) \\
\text{for $i=1,\dots ,\overline n$}
\end{array}
\right\}
\text{ and }
\left\{
\begin{array}{c}
(\mu_j,\bl_j^T,\bv_j^T) \\
\text{for $j=1,\dots,\underline n$}
\end{array}
\right. ,
\label{eq:loewnerInput}
\end{equation}
where $\bw_i=\Htran(\lambda_i)\br_i$ and $\bv_j^T=\bl_j^T\Htran(\mu_j)$, with $\br_i\in\IC^{m \times 1}$, $\bl_j\in\IC^{p \times 1}$,  $\bw_i\in\IC^{p \times 1}$ and $\bv_j\in\IC^{m \times 1}$ ($m,p\geq 1$). In addition, we define the set of distinct interpolation points $\{z_k\}_{k=1}^{N} \subset \IC$, leading to responses $\{\Phi_{k}\}_{k=1}^N\in\IC^{p \times m}$, rearranged as follows ($N=\overline n+\underline n$),
\begin{equation}\label{eq:shift}
\{z_k\}_{k=1}^{N} = \{\lambda_i\}_{i=1}^{\overline n} \cup \{\mu_j\}_{j=1}^{\underline n} \text{ and }
\{\Phi_k\}_{k=1}^{N} = \{\bw_i\}_{i=1}^{\overline n} \cup \{\bv_j\}_{j=1}^{\underline n} .
\end{equation}
% that can be split up into two subsets ($\lambda_i, \ \mu_j\in\IC$),
The method then consists in building the \emph{Loewner} matrix $\IL \in \IC^{\underline n\times \overline n}$ and \emph{shifted Loewner} matrix $\sIL \in \IC^{\underline n\times \overline n}$ defined as follows, for $i=1,\dots,\overline n$ and $j=1,\dots,\underline n$:
\begin{equation}
\label{eq:loewnerMatrices}
\begin{array}{rcccl}
%[\IL]_{j,i} 
\IL_{(j,i)} &=& \dfrac{\bv_j^T\br_i - \bl_j^T\bw_i}{\mu_j - \lambda_i} 
&=& \dfrac{\bl_j^T\big( \Htran(\mu_j) - \Htran(\lambda_i) \big) \br_i}{\mu_j - \lambda_i}, \\
%\,[\sIL]_{j,i} 
\sIL_{(j,i)}&=& \dfrac{\mu_j\bv_j^T\br_i - \lambda_i\bl_j^T\bw_i}{\mu_j - \lambda_i}
&=& \dfrac{ \bl_j^T\big( \mu_j\Htran(\mu_j) - \lambda_i\Htran(\lambda_i) \big) \br_i}{\mu_j - \lambda_i}.
\end{array}
\end{equation}
Additionally, let $\IW = [\bw_1,\cdots,\bw_{\overline n}]$ and $\IV = [\bv_1,\cdots,\bv_{\underline n}]^T$. Finally, let $\bLambda = \text{diag}\left(\lambda_1,\cdots,\lambda_{\overline n} \right)$, $\bM = \text{diag}\left(\mu_1,\cdots,\mu_{\underline n} \right)$, $\bR = [\br_1, \cdots , \br_{\overline n}]$ and $\bL = [ \bl_1, \cdots,\bl_{\underline n}]$. The following Sylvester equations are hence satisfied by the Loewner $\IL$ and shifted Loewner $\sIL$ matrices:  
\begin{equation}
    %\begin{array}{rcl}
    \bM \IL - \IL \bLambda = \IV \bR - \bL \IW\text{ and }
    \bM \sIL - \sIL \bLambda =  \bM \IV \bR -  \bL  \IW \bLambda.
%\end{array} .
\label{eq:sylv_loe}
\end{equation}
Then, the descriptor realization\footnote{Note here that the capital subscript $N$ denotes the number of considered data.},
\begin{equation}
\Hreal^{(N)}:\left \lbrace
\begin{array}{l}
\bE^{(N)}   \dot{\bx}(t) = \bA^{(N)} \bx(t) + \bB^{(N)} \bu(t)\text{ , } \by(t) =\bC^{(N)} \bx(t) \text{ where , }\\
\bE^{(N)},\bA^{(N)} \in \IC^{\underline n\times \overline n}, \bB^{(N)} \in \IC^{\underline n \times m}, \bC^{(N)} \in \IC^{p\times \overline n}.
\end{array}
\right. 
\label{eq:ode_loe1}
\end{equation}
where $\bE^{(N)} = -\IL$, $\bA^{(N)} = -\sIL$, $\bB^{(N)} = \IV$ and $\bC^{(N)}= \IW$ and which associated transfer function $\Htran^{(N)}:\IC\rightarrow\IC^{p\times m}$
\begin{equation}
\Htran^{(N)}(\xi) = \bC^{(N)} \bPhi^{(N)}(\xi) \bB^{(N)}\text{ where } \bPhi^{(N)}(\xi) = (\xi \bE^{(N)}-\bA^{(N)})^{-1} \in \IC^{\underline n\times \overline n}
\label{eq:rat1}
\end{equation}
tangentially interpolates $\Htran$ at the given support points and directions defined in \eqref{eq:loewnerInput}, \ie satisfies the conditions
\begin{equation}
    \Htran^{(N)}(\lambda_i)\br_i = \Htran(\lambda_i) \br_i
    \text{ and }
    \bl_j^T\Htran^{(N)}(\mu_j) = \bl_j^T\Htran(\mu_j).
\label{eq:loewnerIntep}
\end{equation}
Note that $\Htran^{(N)}$ or $\Hreal^{(N)}$ is an interpolant of the data without any reduction. It refers to the realization constructed using the $N$ available data.

From now on, let us assume that $\underline n=\overline n$, also referred to as the square case\footnote{The term square refers to the square shape of the dynamic matrices $\bA$ and $\bE$. More details can be found in \cite{Antoulas:2016}.}. Moreover, assuming that the number $N=\underline n+\overline n$ of available data is large enough, then it was shown in \cite{MA07} that a minimal model $\Htran_r$ of dimension $r < \overline n=\underline n$ still satisfying the interpolatory conditions \eqref{eq:loewnerIntep} can be computed by projecting the realization \eqref{eq:ode_loe1},  provided that the following holds (for $k=1,\ldots,N$)\footnote{Note here that the lower subscript letter $r$ denotes the dimension of the realization instead of the number of data (in the capital case).} 
\begin{equation}
    \rank (z_k \IL - \sIL) = 
    \rank ([\IL,\sIL]) = 
    \rank ([\IL^H,\sIL^H]^H) = r,
    \label{eq:rankCond}
\end{equation}
where $z_k$ are as in \eqref{eq:shift}. Let $\bY \in \IC^{\underline n \times  r}$ (resp. $\bX \in \IC^{\overline n \times  r}$) be the matrix containing the first $r$ left (resp. right) singular vectors of $[\IL,\sIL]$  (resp. $[\IL^H,\sIL^H]^H$). Then,  $\Hreal_r:(\bE_r,\bA_r,\bB_r,\bC_r,\bfz)$ where
\begin{equation}
    \bE_r = \bY^H \bE^{(N)} \bX \text{ , }
    \bA_r = \bY^H \bA^{(N)} \bX \text{ , }
    \bB_r = \bY^H \bB^{(N)} \text{ and }
    \bC_r = \bC^{(N)} \bX,
\label{eq:proj1}
\end{equation}
is a descriptor realization of $\Htran_r$, given as
\begin{equation}
\Htran_r(\xi) = \bC_r \bPhi_r(\xi) \bB_r \text{ where } \bPhi_r(\xi) = (\xi \bE_r-\bA_r)^{-1} \in \IC^{r\times r},
%\label{eq:loewnerDescrCn}
\end{equation}
encoding a \emph{minimal Mc Millan degree} equal to $\nu=\rank(\IL)$. Note that if $r$ in \eqref{eq:rankCond} is greater than $\rank(\IL)$, then $\Htran_r$ may either have a direct feed-through term or a polynomial part. Finally, the number $r$ of singular vectors composing $\bY$ and $\bX$ used to project the system $\Htran_r$ in \eqref{eq:proj1} may be decreased to at the cost of imposing an approximate interpolation of data, leading to the reduced order rational model. This allows a trade-off between complexity of the resulting model and accuracy of the interpolation. The Loewner framework thus is a landmark appropriate for identification, approximation and order reduction. %More detailed insight can be fund in \cite{AntoulasSurvey:2016,Karachalios:2020}.

%\begin{remark}[About $\delta\left\{\cdot\right\}$ and $\xi$ notations]
%In \eqref{eq:loewnerDescrR}, ``$(\cdot)$'' denotes the considered time-domain variable; this latter can either be ``$(t)$'' for continuous-time models ($t\in\Real_+$) or ``$[q]$'' for sampled-time models ($q\in\mathbb Z$). Similarly, in \eqref{eq:loewnerDescrR} ``$\delta\left\{\cdot\right\}$'' stands as the shift operator being either $\delta\{\x(t)\}=\dx(t)$ in the continuous-time case and $\delta\{\x(q)\}=\x[q+1]$ in the sampled one. With reference to \eqref{eq:loewnerDescrC} $\mathbf \xi$ stands as the associated complex version being the Laplace variable $\mathbf \xi=s$ in the continuous-time case and the forward shift $\mathbf \xi=z$ in the sampled-time one.
%\end{remark}

Let us close this first part with two linear differential algebraic  equations examples where the Loewner framework is applied. Both continuous and sampled-time cases are considered, highlighting how versatile this landmark is. More detailed and didactic examples may be found in the surveys \cite{AntoulasSurvey:2016,Karachalios:2020}.

\begin{example}[Continuous-time rational and polynomial model interpolation]
{\textrm 
Let us consider the following rational and polynomial (improper) model,
$\Htran(s)= s+1/(s+1) = (s^2+s+1)/(s+1)$ which a realization $\Hreal:(\bE,\bA,\bB,\bC,\bfz)$ can be described as follows:
\begin{equation}
\bE=\matrixthree{0}{1}{0}{0}{0}{1}{0}{0}{1} \text{ , }
\bA=\matrixthree{1}{0}{0}{0}{1}{0}{0}{0}{-1} \text{ , }
\bB=\vectorthree{0}{0}{1} \text{ and }
\bC^T=\vectorthree{1}{1}{1}.
\end{equation}
%\begin{equation}
%\Hreal:\left\{
%\begin{array}{rcl}
%\dot x_2 &=& x_1\\
%\dot x_3 &=& x_2 \\
%x_2 &=& -x_3 + u\\
%y &=& x_1+x_2+x_3
%\end{array}
%\right.
%\end{equation}
By sampling $\Htran$ with the following support points $\lambda_i=\{1, 3, 5, 7\}$ and $\mu_j=\{2, 4, 6, 8\}$ and tangential directions $\br_i=\bl_j=1$ for $i,j=1,\dots,4=\overline n=\underline n$ ($N=8$), leads to the measurements $\bw_i=\{3/2,13/4,31/6,57/8\}$ and $\bv_j=\{7/3,21/5,43/7,73/9\}$. Constructing the Loewner matrices as in \eqref{eq:loewnerMatrices}, one obtains a $4$-th order realization $\Hreal_N:(-\IL,-\sIL,\IV,\IW)$. Following \eqref{eq:rankCond}, the rank of the $[\IL,\sIL]$ matrix is equal to $r=3$. Practically, by computing the \textbf{SVD} of the $[\IL,\sIL]$ matrix leads to the following normalized singular values $\sigma=\{1,5.59\cdot 10^{-2},6.8804\cdot 10^{-4},5.8311\cdot 10^{-17}\}$ and thus suggests to preserve the $r=3$ first columns of $\bY$ and $\bX$, as in \eqref{eq:proj1}. After projection, this leads to a minimal order realization which related transfer function exactly recovers the original $\Htran$ one, with Mc Millan degree of $\nu=2$ and associated realization $r=3$. In addition, computing the singularities of the associated pencil $(\sIL,\IL)$ gives $\{-1,\infty,\infty\}$, being exactly the one of the original model $\Htran$. The singularity in $-1$ is related to the rational part of $\Htran$, $1/(s+1)$, being the finite dynamic mode. Then, the two singularities in $\infty$ are related to the impulsive (irrational part) and non-dynamic (direct feed-through term) modes.}
\end{example}

\begin{example}[Interpolation in the sampled-time]
{\textrm Let us now consider the discrete-time model $\Htran(z)=z / (z-1/2)$, sampled with a constant period $h=1$ second. One may evaluate the function on the unit circle centered in zero, being the projection of the imaginary axis classically considered in continuous-time. Then, by choosing $\lambda_i=\{e^{-\imath 0.1h},e^{\imath 0.1h},e^{-\imath 2h},e^{\imath 2h}\}$, $\mu_j=\{e^{-\imath h},e^{\imath h},e^{-\imath 3h},e^{\imath 3h}\}$ and tangential directions $\br_i=\bl_j=1$ for $i,j=1,\dots,4=\overline n=\underline n$ ($N=8$), one respectively obtains $\bw_i$ and $\bv_j$ (notice that the Nyquist pulsation, being the maximal pulsation prior periodic frequency response is at $\omega_N=\pi/h$ rad/s). By construction, the Loewner matrices are complex of dimension $4\times 4$. As data are provided in complex conjugate form, one may work with real arithmetic instead of complex ones by projecting the data (see \S 2.5.4 \cite{Karachalios:2020} for details). Then, one obtains
\begin{equation}
\begin{array}{rcl}
    \bLambda &=& \textbf{blkdiag}\bigg(\matrixtwo{0.9950}{-0.0998}{0.0998}{0.9950},\matrixtwo{-0.4161}{-0.9093}{0.9093}{-0.4161}\bigg)\text{ , }\\
    \bR &=& [\sqrt{2},0,\sqrt{2},0] \text{ , }\\
    \bW &=& [2.7869,0.2768,1.0254,0.3859] \\
    \bM &=& \textbf{blkdiag}\bigg(\matrixtwo{0.5403}{-0.8415}{0.8415}{0.5403},\matrixtwo{-0.9900}{-0.1411}{0.1411}{-0.9900}\bigg) \text{ , } \\
    \bL &=& [\sqrt{2},0,\sqrt{2},0]^T \text{ and } \\
    \bV &=& [1.4544,-0.8384,0.9439,-0.0445]^T\text{ , } 
\end{array}
\end{equation}
By then solving \eqref{eq:sylv_loe}, one readily obtains $\IL$ and $\sIL$ and the associated 4-th order realization $\Hreal_N:(-\IL,-\sIL,\bV,\bW,\bfz)$. Applying the rank revealing factorisation \eqref{eq:rankCond} and \eqref{eq:proj1}, one obtains the Mc Millan degree $\nu=\rank(\IL)=1$ and $r=2$. This indicates a constant term. By applying the procedure detailed in \cite{GoseaDAE:2020}, one may reconstruct the direct term by the infinite eigenvalue computation of the $(\sIL,\IL)$ pencil (or zero eigenvalue of $(\IL,\sIL)$). In this case one finds $D=1$. By removing it from the raw data and re-compute the Loewner procedure one gets $\nu=r=1$ and the sampled realization $(\bE_1,\bA_1,\bB_1,\bC_1,\bD_1)=(2.897,1.448,-0.9632,-1.504,1)$, which transfer function $\Htran_1=(z-1.665\times 10^{-16})/ (z - 0.5)$, recovering almost perfectly the original model $\Htran$. Note that in this case, the realization is a sampled one and time-domain dynamical equation reads $\bE_1\bx(t_{k+1})=\bA_1\bx(t_k)+\bB_1\bu(t_k)$ and $\by(t_k)=\bC_1\bx(t_k)+\bD_1\bu(t_k)$, where $t_{k+1}=t+kh$.}
\end{example}

\subsection{Generalizations to parametric linear systems}
\label{ssec:loe_plti}

The Loewner framework has been extended to parametric \lti (\textbf{pLTI}) systems, first in \cite{AntoulasLAA:2012} and in a more detailed manner in \cite{IA14}\footnote{The approach developed in \cite{IA14} interpolates more combinations of frequencies and parameter than the one in  \cite{AntoulasLAA:2012}, which interpolates an extended Loewner matrix, leading to the coefficients of rational function given in Barycentric form.}. In parametric model approximation and reduction, the aim is to construct reduced-order models that match the response of the original model / data,  along the dynamical parameter $\xi$ (usually complex) and along the parameters $\rho$ (traditionally real). In what follows we will only show how the two variable case works, \ie with one single parameter $\rho\in\IR$ (for further extensions, see \cite{IA14}). We construct models which are reduced both with respect to the complex variable (or frequency) and to the real one (parameter).  In this configuration let us consider such a $m$ input $p$ output $\rho$-parametrized dynamical system described by a $n$-th order differential algebraic equation (\textbf{DAE}) model denoted $\Hreal(\rho):(\bE(\rho),\bA(\rho),\bB(\rho),\bC(\rho),\bfz)$  given as 
\begin{equation}
\Hreal(\rho):\left \lbrace
\begin{array}{l}
\bE(\rho)   \dot{\bx}(t) = \bA(\rho) \bx(t) + \bB(\rho) \bu(t)\text{ , } \by(t) =\bC(\rho) \bx(t) \text{ where }\\
\bE(\rho),\bA(\rho) \in \IR^{n\times n}, \bB(\rho) \in \IR^{n\times m}, \bC(\rho) \in \IR^{p\times n}, \rho \in \IR.
\end{array}
\right. 
\label{eq:pode_real}
\end{equation}
with associated transfer function $\Htran: (\IC\times\IR) \rightarrow \IC^{p \times m}$
\begin{equation}
\Htran(\xi,\rho) = \bC(\rho) \bPhi(\xi,\rho) \bB(\rho) \text{ where } \bPhi(\xi,\rho) = (\xi\bE(\rho)-\bA(\rho))^{-1} \in \IC^{n\times n}.
\label{eq:pode_tran}
\end{equation}
As for the Loewner case, let us assume that function \eqref{eq:pode_tran} can be expressed in the Lagrange, using the distinct Lagrange support points $\lambda_i$ and $\pi_j$, as (for $\alpha_{ij} \neq 0$)
\begin{equation}
\Htran(\xi,\rho) = \dfrac{\sum_{i=1}^{n+1}\sum_j^{m+1}\beta_{ij}\mathbf q_{ij}(\xi,p)}{\sum_{i=1}^{n+1}\sum_{j=1}^{m+1} \alpha_{ij}\mathbf q_{ij}(\xi,p)}
\text{ where } \mathbf q_{ij}(\xi,p) = \dfrac{1}{(\xi-\lambda_i)(\rho-\pi_j)}.
\label{eq:pode_rat}
\end{equation}
Computation of the approximant is done in a similar way as for the non-parametric rational case: one seeks the $\beta_{ij}$ and $\alpha_{ij}$ of the rational barycentric formula \eqref{eq:pode_rat}. Let us assume that the system $\Htran(\xi,\rho)$ is sampled along the dynamical parameter $\xi$ and the parametric one $\rho$ as follows
\begin{equation} 
\{z_k\}_{k=1}^{N} =\{\lambda_i\}_{i=1}^{\overline{n}}  \cup \{\mu_j\}_{j=1}^{\underline{n}}\text{ and } 
\{p_l\}_{l=1}^{M} = \{\pi_i\}_{i=1}^{\overline{m}} \cup \{\nu_j\}_{j=1}^{\underline{m}}.
\label{eq:supportParam}
\end{equation}
Each $k=1,\dots,N$ and $l=1,\dots,M$ provides $\Htran(z_k,p_l)=\Phi_{k,l}$. Thus the measurement matrix reads
\begin{equation}
    \Phi = \matrixtwo{\Phi_{(11)}}{\Phi_{(12)}}{\Phi_{(21)}}{\Phi_{(22)}} \in\IC^{N\times M},
    \label{eq:dataParam}
\end{equation}
where $\Phi_{(11)}=\Phi_{1,\dots,\overline n / 1,\dots,\overline m}\in\IC^{\overline n \times \overline m}$, $\Phi_{(12)}=\Phi_{1,\dots,\overline n / 1,\dots,\underline m}\in\IC^{\overline n \times \underline m}$, $\Phi_{(21)}=\Phi_{1,\dots,\underline n / 1,\dots,\overline m}\in\IC^{\underline n \times \overline m}$ and $\Phi_{(22)}=\Phi_{1,\dots,\underline n / 1,\dots,\underline m}\in\IC^{\underline n \times \underline m}$. The rows correspond to frozen values of $z_k$ related to the dynamical (complex) $\xi$ parameter. The columns correspond to frozen $p_l$ values related to the (real) $\rho$ parameter. Similarly to the non-parametric case mentioned in section \ref{ssec:loe_lti}, one may construct the following one variable Loewner matrices
\begin{equation}
    \begin{array}{ll}
    \IL_2 \in \IC^{\underline n \underline m \times \overline n \overline m} & 
    \text{ associated to $\Phi_{(11)}$ along $\lambda_i \bigcup \pi_j$ }\\
    \IL_{\lambda_i} \in \IC^{\underline m \times \overline m}& 
    \text{ associated to the $i$-th row of $[\Phi_{(11)},\Phi_{(12)}]$ along $p_l$}\\
    \IL_{\pi_j} \in \IC^{\underline n \times \overline n}& 
    \text{ associated to the $j$-th column of $[\Phi_{(11)}^H,\Phi_{(21)}^H]^H$ along $z_k$}\\
    \end{array}
\end{equation}
and the global two dimensional Loewner matrix $\widehat{{\IL_2}}\in\IC^{(\overline n\underline m + \underline n\overline m+\underline n\underline m)\times(\overline n\overline m)}$
\begin{equation}
    \widehat{{\IL_2}} = \vectorthree{\IL_\lambda}{\IL_\pi}{\IL_2}, 
    \text{ where }
    \IL_\lambda = \vectorthree{\bfe_1^T\otimes \IL_{\lambda_1}}{\vdots}{\bfe_{\overline n}^T\otimes \IL_{\lambda_{\overline n}}}
    \text{ and }
    \IL_\pi = \vectorthree{\IL_{\pi_1}\otimes\bfe_1^T}{\vdots}{\IL_{\pi_{\overline n}}\otimes\bfe_{\overline m}^T}. 
    \label{eq:paramLoewner}
\end{equation}

As in the non-parametric case, one important step is the determination of the minimal rational orders $n$ and $m$ in \eqref{eq:pode_rat} hidden in the data collection. Here again, this is computed by  a rank revealing operation, namely evaluating the null-space of the single variable Loewner matrices combinations
\begin{equation}
r=\max_l\rank\,\IL_{p_l} \text{ and } q=\max_k\rank\,\IL_{z_k}, 
\label{eq:loewnerParam_rank}
\end{equation}
where $\IL_{p_l}$ and $\IL_{z_k}$ are the one dimensional Loewner matrices associated to the $k$-th row and $l$-th column of $\Phi$, respectively. Then, one can simply set 
\begin{equation}
    (\overline n,\overline m) = (r+1,q+1),
\end{equation}
and partition the data \eqref{eq:supportParam}-\eqref{eq:dataParam}, and reconstruct \eqref{eq:paramLoewner}. The two dimensional Lowner matrices ensure $\rank\,\widehat{\IL_2}=\rank\,\IL_2=\overline n\overline m-(\overline n - r)(\overline m - q)=\overline n\overline m-1$. The coefficients $\alpha_{ij}$ and $\beta_{ij}$ of the rational two variables barycentric function interpolating the data, are obtained by computing the null-space of  $\widehat{\IL_2}$ as 
\begin{equation}
\mathbf c=\kernel\,\widehat{\IL_2} \text{ where } 
\mathbf c \in \IC^{(r+1) \times (q+1)}
\end{equation}
Note that it is usually preferred to work with real arithmetic, \eg for model time domain simulation or control design and analysis. In that case $z_k$ are compiled in a closed conjugate form and support points are doubled (refer to \S A.2 of \cite{IA14} for detailed exposition). Note also that a trade-off between accuracy and complexity with both the frequency and the parameter variables can be obtained by decreasing the order $r$ and $q$ below the one given by \eqref{eq:loewnerParam_rank}.

Following the barycentric formulae, as exposed in \cite{IA14} and \cite{AntoulasLAA:2012}, one may reconstruct the associated multi-valued transfer function $\Htran_{r,q}:(\IC\times\IR)\rightarrow\IC^{p\times m}$ as follows (where $N_{r,q}=r+2q+2$ and $\bPhi(\xi,\rho)\in \IC^{N_{r,q} \times N_{r,q}}$)
\small
\begin{equation}
\Htran_{r,q}(\xi,\rho) = \bC\bPhi^{-1}(\xi,\rho) \bB \text{ where } \bPhi(\xi,\rho) = \matrixthree{\mathbf J_{\lambda,r}(\xi)}{\bfz}{\bfz}{\mathbb A}{\mathbf J_{\pi,q}^T(\rho)}{\bfz}{\mathbb B}{\bfz}{[\mathbf J_{\pi,q}(\rho),\boldsymbol {\tau}]},
\label{eq:pode_tranRed}
\end{equation}
\normalsize
with $\bB=[\bfz,\boldsymbol \tau,\bfz]^T\in\IR^{N_{r,q}}$ and $\bC=[0,\dots,0,-1]\in\IR^{N_{r,q}}$. Moreover, the following holds (for $k=1,\dots,r$, $l=1,\dots,q$ and $\mathbf w=\text{vect}(\Phi_{(11)})$)
\begin{equation}
    \mathbb A_{:,k} = \vectorthree{\mathbf c_{k,1}}{\vdots}{\mathbf c_{k,q+1}}
    \text{ , }
    \mathbb B_{:,k} = \vectorthree{\mathbf c_{k,1}\mathbf w_{k,1}}{\vdots}{\mathbf c_{k,q+1}\mathbf w_{k,q+1}}
    \text{ and }
    \boldsymbol \tau_k = \bigg(\prod_{l=1,l\neq k}^{q+1} \pi_k-\pi_l\bigg)^{-1}
\end{equation}
and with 
\begin{equation}
    \mathbf J_{\eta,t}(x) =
    \left[\begin{array}{cccc} x-\eta_1 & \eta_2-x & &  \\ \vdots &  & \ddots & \\x-\eta_1 & & & \eta_{t+1}-x \end{array}\right] \in \IC^{t\times (t+1)}.
\end{equation}
Notice that \eqref{eq:pode_tranRed} only depends on the extended Loewner matrix null-space $\mathbf c$, the considered support points $\{\lambda_i\}_{i=1}^{r+1}$ and $\{\pi_j\}_{j=1}^{q+1}$ and the response data matrix $\{\Phi_{(11)}\}_{i,j=1}^{r+1,q+1}$. To stick with traditional tools deployed in simulation and control theory, one may also recover a descriptor realization $\Hreal_{r,q}$ where all the parametric dependency is contained in the $\bA_{r,q}(\rho)$ operator as \cite{AntoulasLAA:2012}
\begin{equation}
\Hreal_{r,q}:\left \lbrace
\begin{array}{l}
\bE_{r,q}   \dot{\bx}(t) = \bA_{r,q}(\rho) \bx(t) + \bB_{r,q} \bu(t)\text{ , } \by(t) =\bC_{r,q} \bx(t) \text{ where , }\\
\bE_{r,q},\bA_{r,q}(\rho) \in \IC^{N_{r,q}\times N_{r,q}}, \bB_{r,q} \in \IC^{N_{r,q}\times m}, \bC_{r,q} \in \IC^{p\times N_{r,q}}.
\end{array}
\right. 
\label{eq:pode_loep1}
\end{equation}

\begin{remark}[Minimal realization in the multi-parametric case]
By inspecting \eqref{eq:pode_tranRed}, the realization is no longer identical to the one in the single variable case. Indeed in \eqref{eq:pode_tranRed}, the resolvant $\bPhi(\xi,\rho)$ includes both the dynamic and parametric variables, leading to a realization of order $N_{r,q}$ instead of $r$. Finding a minimal order realization is actually an unsolved problem so far. It has been investigated and is an important research field that can also be connected to the linear fractional transformation research one, largely used in the control community (see \eg realization and control works \cite{MagniLFR:2006,PoussotECC:2018}). 
\end{remark}

\begin{remark}[\textbf{SIMO} and \textbf{MISO} cases]
The \textbf{SIMO} and \textbf{MISO} cases can also be addressed following the very same framework, by tangentially interpolating the data instead of the element-wisely (see \S A.1 of \cite{IA14} for details).
\end{remark}

\begin{remark}[About the \textbf{MIMO} case]
The parametric extension to the \textbf{MIMO} case is not solved yet. Indeed, the tangential approach used in the non-parametric case and in most of multi-port interpolation frameworks \cite{VanDooren:2008,Gallivan:2004} is not applicable as is. Indeed, the realization construction is no longer applicable. An alternative approach is presented in \cite{Lefteriu:2018} but which "only" interpolates a part of the data, namely $\Phi_{(11)}$, forgetting  $\Phi_{(12)}$ and  $\Phi_{(21)}$. This latter work also considers the same number of inputs and outputs.%, although the rectangular multi port can easily be handled.
\end{remark}

\begin{example}[Reynolds parameter dependent linearized Navier-Stokes model]
Let us consider a fluid-flow configuration. It consists of a two-dimensional open square cavity flow problem where air flows from left to right for three different Reynolds numbers. Such a configuration, illustrated on Figure \ref{fig:ns} (top right), is described in detail in the original work of \cite{Barbagallo:2008} and in \cite{PoussotLPVS:2015}. For simulation, Navier-Stokes equations are used along a mesh composed of $193,874$ triangles, corresponding to $n=680,974$ degrees of freedom for the velocity variables along the $x$ and $y$ axis. After linearization around three fixed points for varying Reynolds numbers $Re=\{4000,5250,6000\}$ and discretization along the flow axis, three dynamical models $\{\Htran_l\}_{l=1}^3$ can be described as a \textbf{DAE} realization of order $n=680,974$ where the input $\u(t)$ is the vertical pressure actuator located upstream of the cavity and the output $\y(t)$ is a shear stress sensor, located downstream of the cavity. Such a continuous-time $n$-th order realization  for $l=\{1,2,3\}$ $\Hreal_l:(\bE,\bA_l,\bB,\bC,\bfz)$ where the parameter is the Reynolds number $Re$. In \cite{PoussotLPVS:2015}, the \textbf{IRKA} approach \cite{GugercinSIAM:2008} (being a realization based $\mathcal H_2$-oriented reduction method) is used to sequentially approximate each realization with a low dimensional one. Then, the interpolation along the parameter is done in a second step by interpolating each coefficients in the canonical basis of the obtained realization. 

Here instead, the parametric Loewner framework is applied. The frequency response of each configuration along $\{z_k\}_{k=1}^N=z_0 \bigcup \{\imath \omega_{ k},-\imath \omega_{ k}\}_{k=1}^{100}$, where $z_0\in\Real_+$ and $\omega_k$ logarithmically-spaced frequencies. Then, twenty intermediate configurations between each Reynolds numbers $Re=\{4000,5250,6000\}$ are constructed by linear interpolation. We obtain $\{z_k\}_{k=1}^{N=201}$, $\{p_l\}_{l=1}^{N=41}$ and thus $\Phi \in \IC^{201 \times 41}$. Our objective is to come up with a parametrized linear model that is able to faithfully reproduce the original transfer function data on a particular range of frequencies as well as on a target parameter range\footnote{Additional details and the data are available at \url{https://morwiki.mpi-magdeburg.mpg.de/morwiki/index.php/Fluid_Flow_Linearized_Open_Cavity_Model}}.

\begin{figure}[H] 
\centering
\includegraphics[width=.6\columnwidth,align=c]{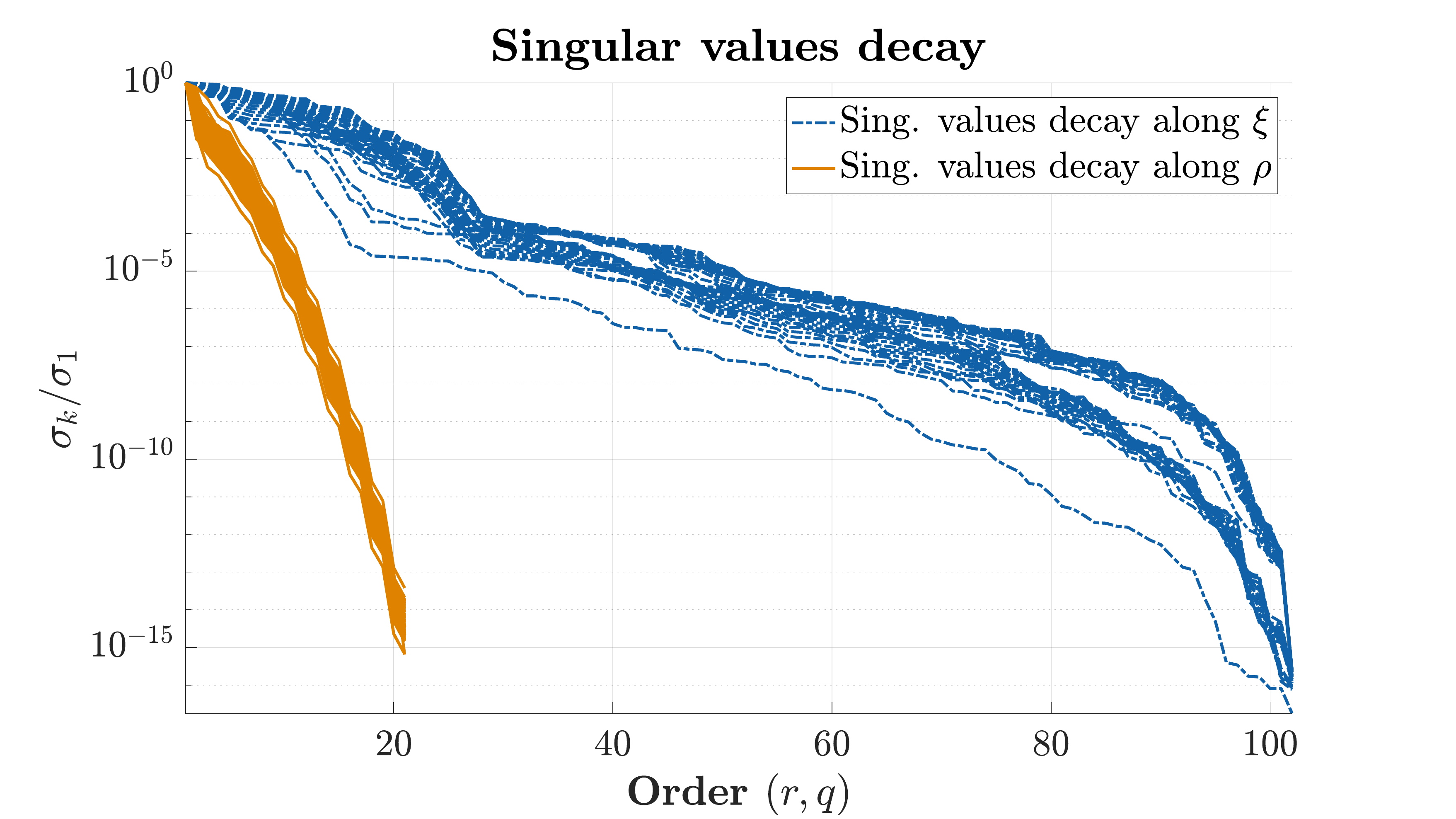}
\includegraphics[width=.35\columnwidth,align=c]{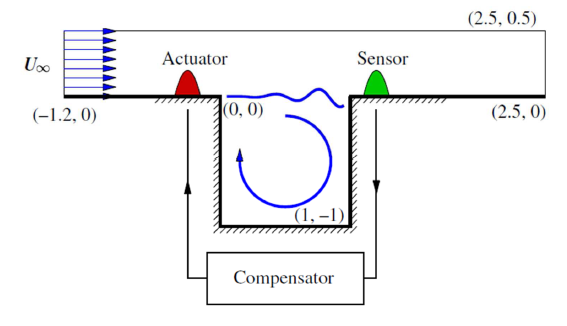}\\
\includegraphics[width=.8\columnwidth,align=c]{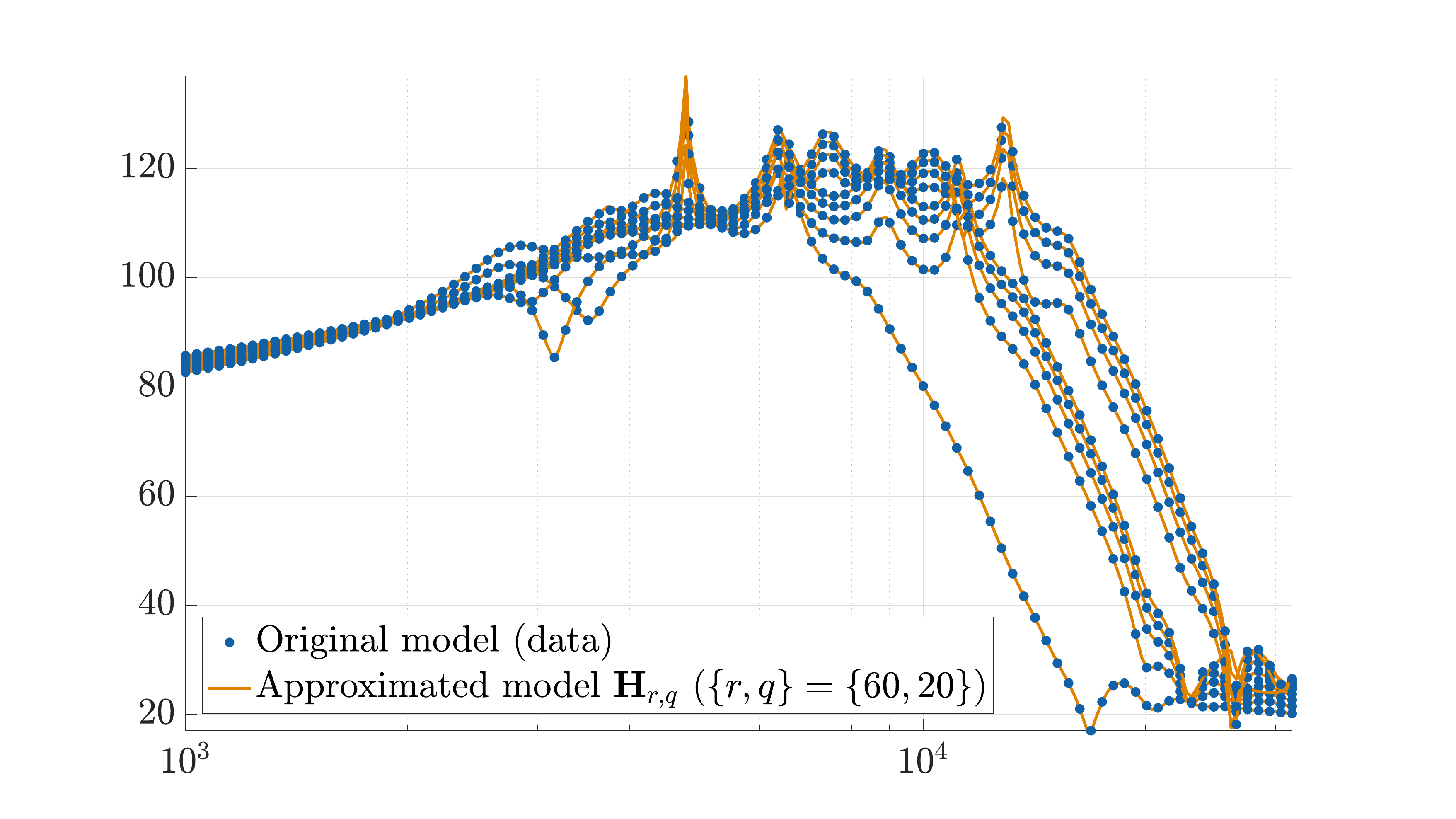} 
\includegraphics[width=.8\columnwidth,align=c]{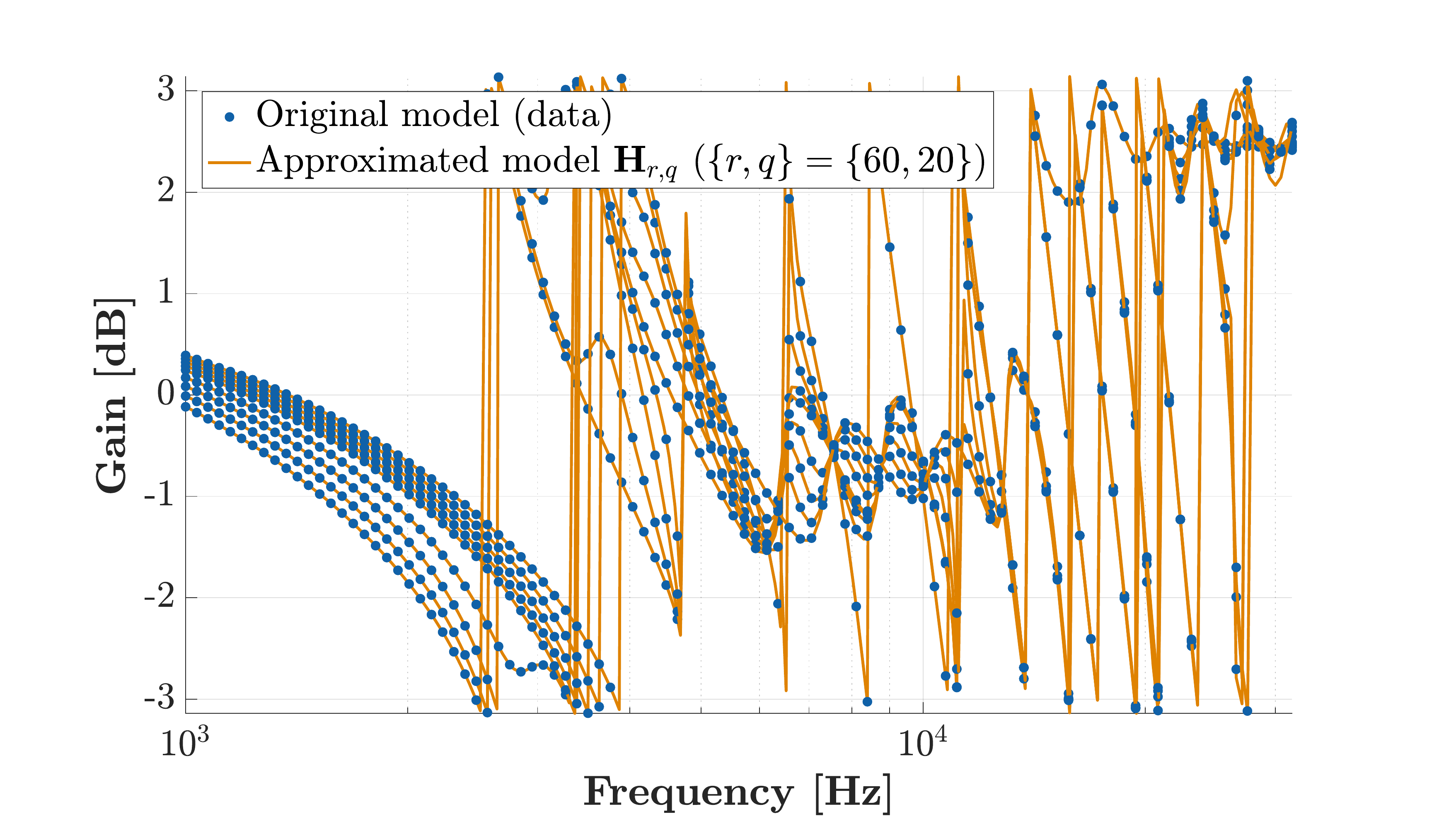}
\caption{Top right: singular values drop of the one variable Loewner matrices \eqref{eq:loewnerParam_rank}. Top right: schematic view of the geometry (with illustration of the control structure used in \cite{PoussotECC:2018}). Middle and bottom frames: frequency response gain and phase of the original sampled data (blue dots) and resulting parametric model $\Htran_{60,20}$ for some parametric values (solid orange lines).}
\label{fig:ns}
\end{figure}

On one hand, we form Loewner matrices by using measurements for varying frequency and constant parameter, while on the other hand we use varying parameter and constant frequency. Figure \ref{fig:ns} (top left) depicts the two types of singular values. By investigating the drop in the singular values plot, we decide to use reduction orders $r = 30$ and $q = 20$ for building the two dimensional Loewner matrix. As we want to find a real valued rational function rather than complex, the twice more support points are considered and realization size is increased. The reduced linear parametric model which is sampled over the same frequency and parameter range as before. When comparing to the original samples on Figure \ref{fig:ns} (middle-bottom), the overall result is satisfactory, with a model of complexity $r=60$ (instead of $n=680,974$) an $q=20$ (instead of a collection), enforcing a drastic memory saving, and hence, being a game changer for simulation and control design. 
\end{example}

\subsection{Generalization to discrete-time models from time-domain data}
\label{ssec:loe_time}

\noindent
For an LTI \textbf{SISO} system, let the impulse response be denoted with: 
$
\bh=\left\{~\cdots~ h_{-2}, h_{-1}, h_{0}, h_1, h_2,~\cdots~\right\}
$. The associated system action $\cS$ is given by the \emph{convolution sum}:
\begin{equation}\label{sys_act}
\cS:~\bu \longmapsto \by=\cS(\bu)=\bh\ast \bu,~~{\textrm{where}}~~
(\bh\ast \bu)(t) =\sum_{k=-\infty}^{\infty} h_{t-k} \bu(k),~~t \in \mathbb{Z}.
\end{equation}
Here we restrict our attention to causal systems: $\bh_k=0$, 
$k<0$; furthermore it is assumed that $\bu(t)=0$, $t<0$. Hence, one can write that
\begin{equation}\label{inp_out_form}
\by(t)=h_0\bu(t)+h_1\bu(t-1)+~\cdots~+h_k\bu(t-k)+~\cdots~,~~t\in \mathbb{Z}_+ .
\end{equation}
In the formulation above, $h_j$ denotes the $j^{\textrm th}$ Markov parameter
of the underlying system. In the time domain, the data are samples of input and output signals
\begin{equation}\label{inp_out_data}
     \bu_N = [u_0, ~\cdots,~ u_{N-1}],~~
 \by_N = [y_0, ~\cdots,~ y_{N-1}],
\end{equation}
where, for simplicity, we have used the shortened expressions $u_k := u(k)$ and $y_k := y(k)$. 
The system identification problem consists in recovering a
discrete-time linear time invariant system compatible with the
data in (\ref{inp_out_data}). We seek a minimal realization $(\bE,\bA,\bB,\bC,\bD)$:
\begin{align}\label{sys:ss}
\Hreal_D : \bE \,\bx(t+1) = \bA\bx(t) + \bB u(t),~~ y(t) = \bC\bx(t) + \bD u(t), 
\end{align}
where $\bE,\bA \in \IR^{n \times n}, \bB, \bC^T \in \IR^{n \times 1}$, 
$\bD \in \IR$, and $\bx(t) \in \IR^n$ is the state; with the transfer function 
\begin{equation}\label{sys:tf}
\bH(z) = \bC(z\bE-\bA)^{-1}\bB+\bD= 
\frac{b_mz^m+\cdots+b_1z+b_0}{z^n+\cdots+a_1z+a_0},~m\leq n.
\end{equation}
The Markov parameters in (\ref{inp_out_form}) can be explicitly written in terms of matrices from the realization in (\ref{sys:ss}), as follows:
\begin{equation}
    h_0 = \bD, \ \ h_k = \bC \bA^{k-1} \bB, \ \forall k \geq 1.
\end{equation}
Moreover, another interpretation of Markov parameters is that they encode the behavior of the transfer function in  $\bH(z)$ in (\ref{sys:tf}) at $z = \infty$. More precisely, the values $h_k$'s represent the coefficients of the following Laurent series expansion of the transfer function $\bH(z)$:
\begin{equation}
\bH(z) = h_0 + h_1 z^{-1}+ h_2 z^{-2} + \cdots +  h_k z^{-k} + \cdots
\end{equation}
The first step in formulating the data-driven identification problem is to assemble the available input-output data into matrices with special format, i.e., Hankel matrices. Consequently, we introduce $\bU_k \in \mathbb{R}^{M \times L}, \ \bY_k \in \mathbb{R}^{M \times L}$ for any $k \geq 0$, as follows
\footnotesize
\begin{equation}\label{eq:hankelUY}
	\bU_k = 
	\left[\begin{array}{cccc}
	u_k& u_{k+1}& \cdots & u_{k + L-1}\\
	u_{k+1}& u_{k+2} & \cdots & u_{k + L} \\
	\vdots & \vdots &\ddots &\vdots \\
	u_{k + M-1} & u_{k + M} & \cdots & u_{k+M+L-2}	
	\end{array}\right], \bY_k = 
	\left[\begin{array}{cccc}
	y_k& y_{k+1}& \cdots & y_{k + L-1}\\
	y_{k+1}& y_{k+2} & \cdots & y_{k + L} \\
	\vdots & \vdots &\ddots &\vdots \\
	y_{k + M-1} & y_{k + M} & \cdots & y_{k+M+L-2}	
	\end{array}\right].
\end{equation}
\normalsize

\begin{theorem}\label{th:matrixpencil}
The following results are given in \cite{ia12}. If $M \geq n + \mbox{rank}\,\bU_0$, and $z \in \IC$, then the following holds true:
\begin{itemize}
	\item[(a)] 
$ \mbox{rank} \,[z\bY_0- \bY_1,\bU_0] = n + \mbox{rank}\,\bU_0$,~~
and the rank decreases by one if ~$z$~ is a pole.
	\item[(b)] Let $\bPi_{\bU_0}$ be the orthogonal projection
onto the column space of $\bU_0$.
Then the system poles $p_j$ are the ~$n$~ finite generalized eigenvalues of the
singular pencil
\begin{align}\label{eq:thepencil}
	z\bQ_0 - \bQ_1 = (\bI-\bPi_{\bU_0})(z\bY_0-\bY_1),
\end{align}
where
$\mbox{rank}\,\bQ_0 = \mbox{rank}\,\bQ_1 = n$. It also follows that matrices $\bQ_0$, $\bQ_1$ 
have the same column and row spaces.
\end{itemize}
\end{theorem}
In order to be able to accurately extract system invariants (poles, residues, Markov parameters, etc.) from input-output data, there are certain conditions that need to be imposed to sequence of control inputs applied. For example, one of such conditions is the so-called \emph{persistence of excitation}. However, as explained in \cite{ia12}, this requirement of the input is not necessary
when the initial conditions are zero, i.e. the system is at rest before
the input is applied: $\bu(t) = 0$~ and ~$\by(t) = 0$,~ for ~$t <0$. This assumption is equivalent to $\bx(0) = 0$.

Next, as explained in \cite{ia12}, there exists matrix $\bY$, such that the matrix pencil
$(\widehat{\bQ}_0,\,\widehat{\bQ}_1)$, where 
$\widehat{\bQ}_0=\bY^*\bQ_0$, $\widehat{\bQ}_1=\bY^*\bQ_1$,
is {\emph regular} (often $\widehat{\bQ}_0$, $\widehat{\bQ}_1$
can be taken as the leading $n \times n$ 
sub-matrices of $\bQ_0$, $\bQ_1$). 
%Furthermore, let
%${\bU}_0$, ${\bY}_0$ be $M\times (n+1)$ rectangular 
%Hankel matrices according to (\ref{eq:hankelUY}).
The following result in \cite{ia12} gives a realization for a model of dimension $n$: 
\begin{theorem}\label{th:statespace}
For zero initial conditions, 
the system has a minimal realization
$$
\widetilde{\bE} = \widehat{\bQ}_0,
\quad \widetilde{\bA} = \widehat{\bQ}_1, \quad \widetilde{\bB} = \bq_0,\quad \widetilde{\bC} = [h_1, \cdots, h_n],
\quad \widetilde{\bD} = h_0,
$$
where ~$\bq_0$~ is the first column of ~$\widehat{\bQ}_0$~
and the Markov parameters $h_j$'s are obtained by solving the following linear system of equations
\begin{equation}\label{eq:find_MPs}
    \left[\begin{array}{cccc}
	u_0& &&\\
	u_1& u_0 && \\
	\vdots & \ddots & \ddots & \\
	u_n & \cdots & u_1 & u_0
\end{array}\right]
\left[\begin{array}{c}
	h_0 \\
	h_1 \\
	\vdots \\
	h_n
\end{array}\right]
=
\left[\begin{array}{c}
y_0 \\
y_1 \\
\vdots \\
y_n
\end{array}\right] .
\end{equation}
\end{theorem}
\vspace*{2mm}
\noindent
In this case, the solution of a lower triangular system of equations
is needed. It readily follows that the Markov parameters can be computed for any input $\bu$. For more details on this procedure, we refer the reader to \cite{ia12}. \\[2mm]
The  result stated in Theorem \ref{th:statespace} can indeed be further specialized for the case when 
of a very special input given by $\bu = [1,0,\cdots,0]$. Hence, when the input is an impulse, the output is a finite sequence of Markov parameters, i.e.,
$\by = [h_0,h_1,\cdots,h_{N-1}]$.
The realization in Theorem \ref{th:statespace} is hence modified appropriately, since the matrix pencil is given by two Hankel matrices. Now, let  $\tilde{\Hreal}_n$ be the new realization given by
\begin{align}\label{realiz_disc_n}
\begin{split}
\widetilde{\bE} &= \!
\left[\begin{array}{cccc}
	h_1& h_2 & \cdots & h_n\\
	h_2& h_3 & \cdots & h_{n+1} \\
	\vdots & \vdots & \ddots & \vdots \\
	h_n \hspace{-1mm}& h_{n+1}\hspace{-1mm} & \cdots & h_{2n-1}
\end{array}\right]\!,
\widetilde{\bA} = \!
\left[\begin{array}{cccc}
	h_2& h_3 & \cdots & h_{n+1}\\
	h_3& h_4 & \cdots & h_{n+2} \\
	\vdots & \vdots & \ddots & \vdots \\
	h_{n+1} \hspace{-1mm}& h_{n+2}\hspace{-1mm} & \cdots & h_{2n}
\end{array}\right]\!, \\[1mm]
\widetilde{\bC} &= \big[\,h_1,~ h_2, \cdots,~ h_n\,\big],~~ \widetilde{\bB} = \widetilde{\bC}^T,~~ \widetilde{\bD} = h_0.
\end{split}
\end{align}
As in Section \ref{ssec:loe_lti}, we could further reduce the dimension of the fitted model in (\ref{realiz_disc_n}) by means of projection (compressing the realization of order $n$ to one of order $r$ by means of orthogonal matrices computed using the  \textbf{SVD}). In this case, we talk about approximation, i.e. fitting a model which approximately explains the data. Hence, let $\bY \in \IR^{n \times  r}$ (resp. $\bX \in \IR^{ n \times  r}$) be the matrix containing the first $r$ left and respectively, right singular vectors of the Hankel matrix denoted in (\ref{realiz_disc_n}) by $\widetilde{\bE}$. The reduced-order realization $\tilde{\Hreal}_r:(\tilde{\bE}_r,\tilde{\bA}_r,\tilde{\bB}_r,\tilde{\bC}_r,\bfz)$ is computed as follows
\begin{equation}\label{realiz_disc_r}
    \tilde{\bE}_r = \bY^T \tilde{\bE} \bX \text{ , }
    \tilde{\bA}_r = \bY^T \tilde{\bA} \bX \text{ , }
    \tilde{\bB}_r = \bY^T \tilde{\bB}
    \text{ , }
    \tilde{\bC}_r = \tilde{\bC} \bX\text{, and } \tilde{\bD}_r = h_0.
\end{equation}
\begin{example}[A structural mechanics model]
{\textrm 
As a numerical test case, we consider the model of a building (the Los Angeles University Hospital) from the SLICOT MOR benchmark collection. The building has 8 floors, each
having 3 degrees of freedom, i.e, displacements in x and y directions, and rotation. The original model is hence a second-order linear system of dimension $n_0 = 24$. It can be written equivalently as a first-order linear system of dimension $n = 48$. We slightly modify the original model by scaling the vector $\bB \in \IR^{48}$ with $10^4$.

The numerical treatment goes as follows: the original continuous-time LTI model of dimension $n = 48$ is discretized using a classical Backward Euler first order scheme. The simulation time horizon is $[0,5]$s, while the time step is $\Delta t = 4 \cdot 10^{-3}$. The control input is chosen to be $u(t) = \frac{1}{10} \big{(} \cos(50t)+2\cos(20t)+3\cos(10t)\big{)}$. Hence, by means of this time-domain simulation, we collect $N = 2001$ measurements of the discretized input and output, i.e.,as in \ref{inp_out_data}. These values are depicted in the upper pane of Fig.\;\ref{fig:Build1}. The Markov parameters are extracted by following the approach in (\ref{eq:find_MPs}), and are depicted in the lower pane of Fig.\;\ref{fig:Build1} (there, the magnitude of the error between the true Markov parameters and the estimated ones is shown in orange).

\begin{figure}[H]  \vspace{-3mm}
\centering
		\includegraphics[scale=0.28]{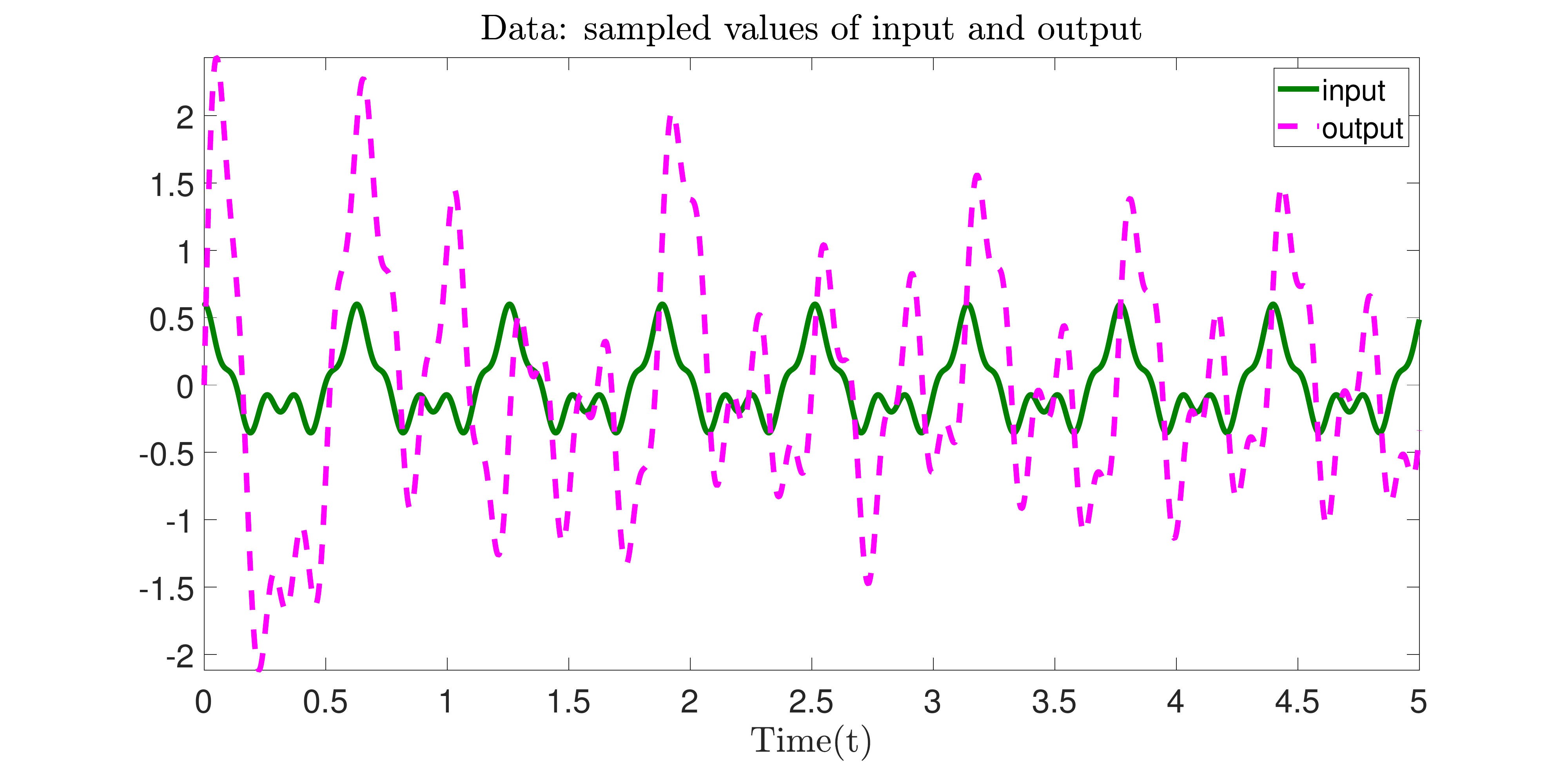}
		\hspace{-8mm}
		\includegraphics[scale=0.28]{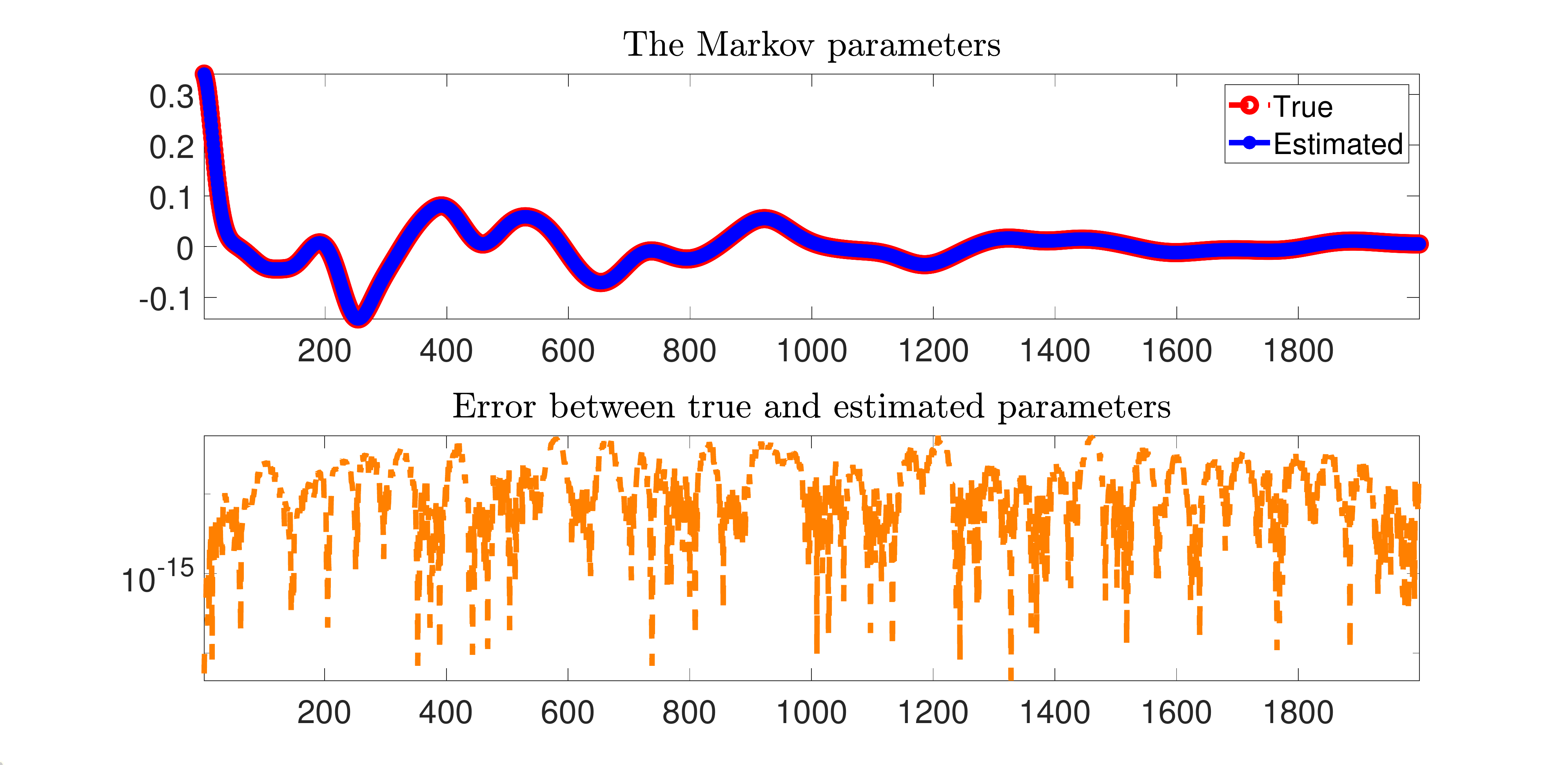}
		\vspace{-3mm}
		\caption{Samples of the input and output signals (up) and the true and recovered Markov parameters (down).}
	\label{fig:Build1}
	\vspace{-4mm}
\end{figure}

Next, form a $1000 \times 1000$ Hankel matrix as in (\ref{realiz_disc_n}). The decay of its singular values is displayed in the upper pane of Fig.\;\ref{fig:Build2}. Then, choose the truncation order $r=20$, and construct a realization of order $r$ as presented in (\ref{realiz_disc_r}). Finally, convert this discrete-time model back to the continuous time, and compare the frequency response of the original model of order $n$, with that of the reduced one of order $r$ (on a range of 500 frequency points in the interval $[10^0,10^{2}]$). The results (frequency responses and the approximation error) are presented in the lower pane of Fig.\;\ref{fig:Build2}. Indeed, the model is well approximated by means of the proposed method.

\begin{figure}[H] \vspace{-3mm}
	\centering
		\includegraphics[scale=0.28]{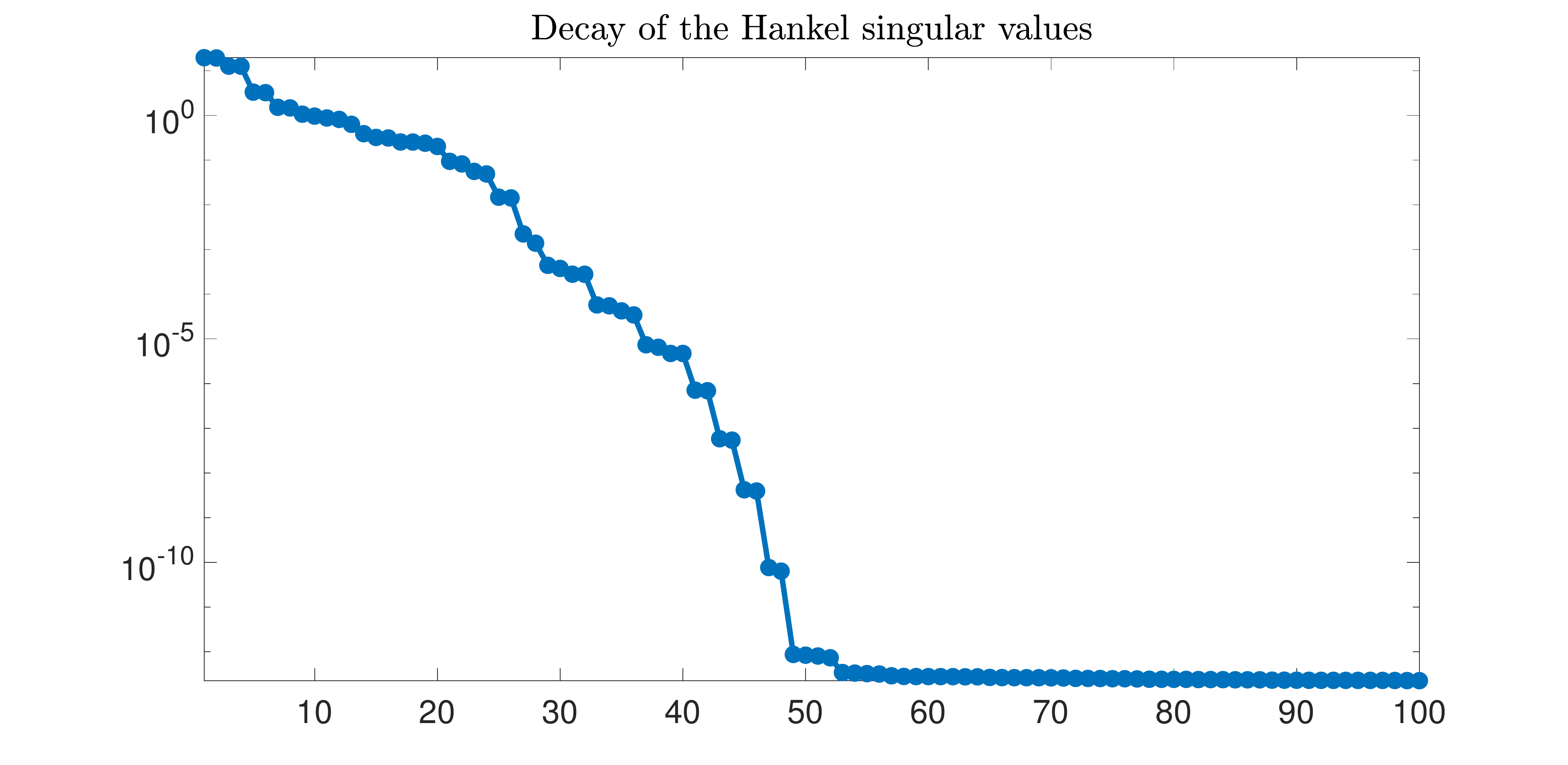}
		\hspace{-8mm}
		\includegraphics[scale=0.28]{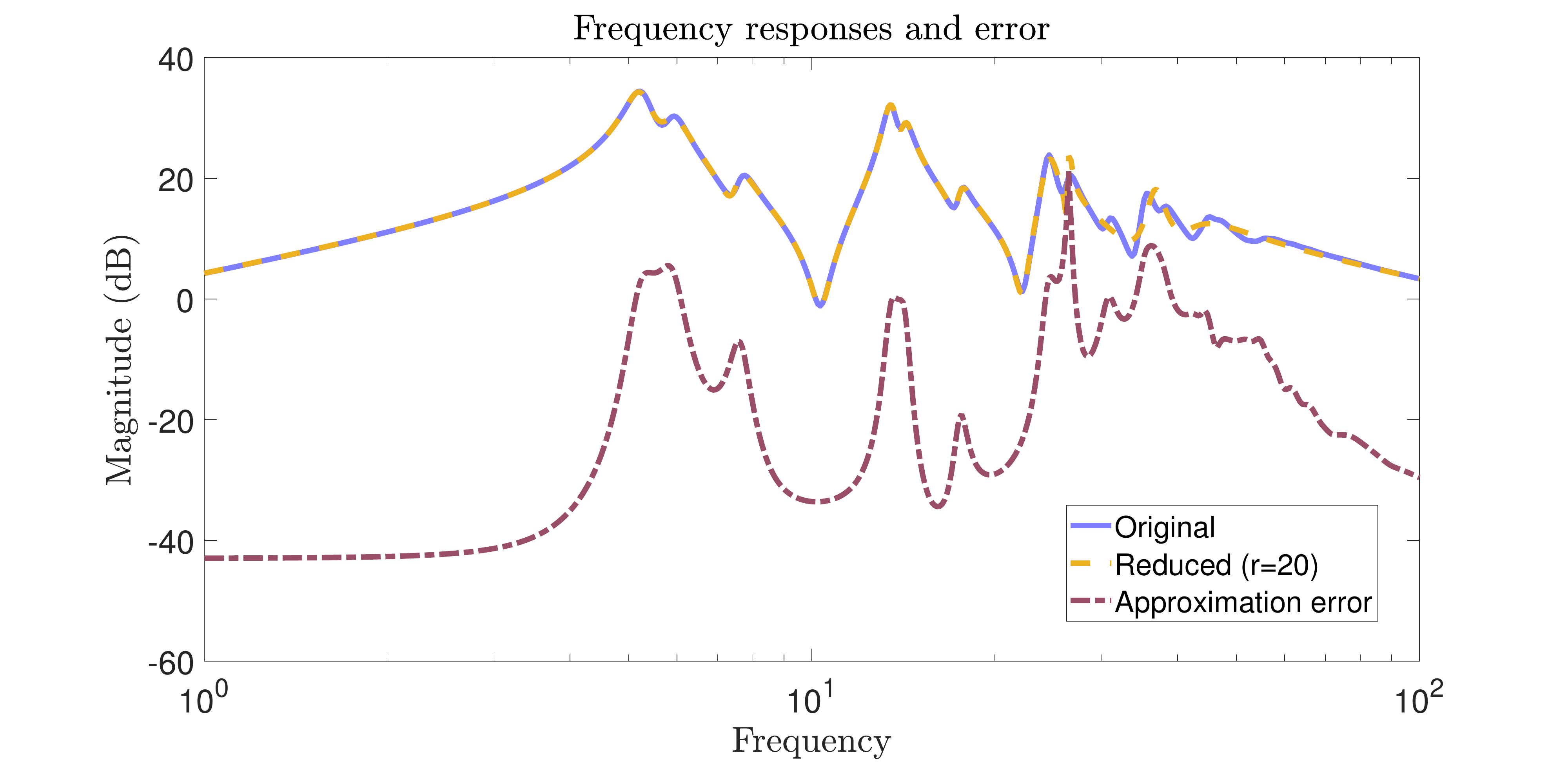}
		\vspace{-3mm}
		\caption{Decay of the Hankel singular values (up) and frequency responses computations: original, reduced and the approximation error (down).}
	\label{fig:Build2}
	\vspace{-4mm}
\end{figure}
}
\end{example}

\subsection{Extensions to nonlinear systems}
\label{ssec:loe_blti}

Consider a nonlinear system described by the following equations
\begin{equation}\label{def_nonlin_sys2_gen}
\Hreal_N : \begin{cases} \dot{\bx}(t) = \bff(\bx(t))+\bg(\bx(t)) \bu(t), \\ \by(t) = \bC \bx(t), \end{cases}
\end{equation}
where $t \geqslant 0, \ \bx(0) = \bx_0$ and the nonlinear functions $\bff,\bg :\mathbb{R}^n \rightarrow \mathbb{R}^n$ are assumed to be analytic in $\bx(t)$. We also assume that the output depends linearly on the variable $\bx(t)$, i.e., $\by(t) = \bC \bx(t)$.

In this section, we will focus on a recent extension of the Loewner framework to reducing bilinear systems. The motivation for this choice is that any smooth, nonlinear system with analytical nonlinearities can be approximated by a bilinear system. This is accomplished by means of a technique, commonly known as Carleman linearization (see \cite{car32,Ru82}). Since this is based on Taylor expansion and truncation, the resulting bilinear system will approximate the original nonlinear system depending on the number of terms kept in the expansion. In many practical applications, approximating the original system is sufficient for a large variety of tasks. We proceed by writing the truncated \emph{Taylor series} for the non-linear functions $\bff$ and $\bg$, where $N$ represents the truncation index, i.e.
\begin{equation}\label{fg_def_trunc}
\begin{cases} \bff(\bx) = \sum_{k=1}^N \bF_k \bx^{( N)} =  \bF_1 \bx+\bF_2 \bx^{(2)}+\ldots+\bF_N \bx^{(N)}, \\
\bg(\bx) = \sum_{k=0}^{N-1} \bG_k \bx^{(k)} = \bG_0+\bG_1 \bx+\ldots+\bG_{N-1} \bx^{(N-1)}. \end{cases}
\end{equation}
\noindent
where $\bG_0 \in \mathbb{R}^{n \times 1}, \bF_j, \bG_j \in \mathbb{R}^{n^j \times n^j}, \ j \geqslant 1$. Here, $\bF_1, \bG_1$ denote the Jacobian matrices of $\bff$ and $\bg$, respectively, and $\bF_k, \bG_k$ denote the matrices of higher derivatives. Moreover $\bx^{(k)}$ denotes the Kronecker product of the state variable $\bx$ with itself (k times). The next step is to introduce a new state variable $\bx^{\otimes}(t)$ as
\begin{align*}
\bx^{\otimes}(t) =  \left[\begin{array}{cccc} \bx(t) &  \bx^{(2)}(t) & \ldots & \bx^{(N)}(t)
\end{array}\right]^T \in \mathbb{R}^{n^{(N)}},
\end{align*}
where  $n^{(N)} = n + n^2 + ... +n^N = \frac{n^N-n}{n-1}$. This is obtained by concatenating all higher powers of vector $\bx$ (up to $N$). In this way, by computing derivatives of $\bx^{(k)}$, we obtain a bilinear system with the following realization
\begin{equation}
\left\{
%\begin{cases} 
\begin{array}{rcl} 
\dot{\bx}^{\otimes}(t) &=& \bA^{\otimes} \bx^{\otimes}(t) + \bN^{\otimes} \bx^{\otimes}(t) \bu(t) + \bB^{\otimes} \bu(t), \\
\by &=& \bC^{\otimes} \bx^{\otimes}(t),
\end{array}
\right.
%\end{cases}
%\vspace{-2mm}
\end{equation}
where $\bx^{\otimes}(0) = \bfz$ and the matrices  $ \bA^{\otimes}, \bN^{\otimes} \in  \mathbb{R}^{n^{(N)} \times n^{(N)}}, \bB^{\otimes} , \big{(} \bC^{\otimes} \big{)}^T \in \mathbb{R}^{n^{(N)}}$ are as in Section 2.1.1 of \cite{gosea_phd}. In what follows, we employ a more generic definition of {\emph bilinear systems} $\cS_{\mathrm B}=(\bC,\bE,\bA,\bN,\bB)$, characterized by:
\begin{equation} \label{bil11}
\Hreal_B : \ \  \bE\dot\bx(t)=\bA\bx(t)+\bN\bx(t)\bu(t)+\bB\bu(t),~~\by(t)=\bC\bx(t),
\end{equation}
where  $\bE, \ \bA, \ \bN \in \mathbb{R}^{n \times n}$, $\bB \in \mathbb{R}^{n \times m}$, $\bC \in \mathbb{R}^{p \times n}$ and $\bx\in\IR^n$, $\bu,\,\by\in\IR$. The matrix $\bE$ is assumed to be non-singular. Also, for simplicity of exposition, we will discuss only the \textbf{SISO} case. More details on bilinear system model order reduction can be found in
\cite{breiten10, breiten12,flagg15}.
Bilinear systems as in (\ref{bil11}) are equivalent an infinite collection of coupled linear time-varying systems of the form:
\begin{align} \label{bil_sys_coupled_rep}
\begin{array}{l}
\bE\dot{\bx}_1(t) =\bA\bx_1(t) + \bB\bu(t),\ \ \bE\dot{\bx}_i(t) =\bA\bx_i(t) + \bN\bx_{i-1}(t)\bu(t),\ i \geq 2.
\end{array}
\end{align}
The time-varying factor appears only in the matrices that scale the control input $\bu(t)$ at each level $i \geq 2$. Based on (\ref{bil_sys_coupled_rep}), the solution of (\ref{bil11}) is decomposed as $\bx(t)=\sum_{i=1}^{\infty}\bx_i(t)$. Furthermore, the input-output representation of the bilinear system $\cS_{\mathrm B}$ can be expressed in terms of the \emph{Volterra series representation} (\cite{Ru82,flagg15}).
 Moreover, considering $\bx_{\ell-1}(t)$ in the $\ell^{\mathrm th}$ equation as a pseudo-input for $l=1,2,\ldots$, the frequency-domain behavior is described by a series of generalized transfer functions as given also in \cite{Ru82,flagg15,AGI16}:
\begin{equation}\label{bil01}
\bH_\ell(s_1,s_2,\ldots,s_\ell)=\bC\,\bPhi(s_1)\,\bN\,\bPhi(s_2)\,\bN\,~
\cdots~\,\bN\,\bPhi(s_\ell)\,\bB,
\end{equation}
where the {\emph resolvent} of the pencil $(\bA,\bE)$ is denoted by $\bPhi(\xi)=\left(\xi\bE-\bA\right)^{-1}$. The characterization of bilinear systems by means of the rational functions in (\ref{bil01}) suggests that reduction of such systems can be performed by means of the Loewner framework. In what follows, we will review some highlights of the procedure originally presented in \cite{AGI16}. We use the concept of multi-tuples, composed of multiple interpolation points corresponding to evaluations of the transfer functions in (\ref{bil01}). For simplicity, we will assume that one set of right multi-tuples $\blambda$, and one set of left multi-tuples $\bmu$ with the same number of interpolation points (denoted with $k$), are given as
\begin{align}\label{tuples}
	\begin{array}{l}
	\blambda=\left\{\{\lambda_1\},\{\lambda_2,\lambda_1\},~\ldots,~\{\lambda_k,\ldots,\lambda_2,\lambda_1\}
	\right\},\\[1mm]
	\bmu=\left\{\{\mu_1\},\{\mu_1,\mu_2\},~\ldots,~\{\mu_1,\mu_2,\ldots,\mu_k\}
	\right\}.
	\end{array}
\end{align}
For the tuples in (\ref{tuples}), we introduce the associated generalized controllability and observability matrices, denoted with $\cR \in \IC^{n\times k}$, and respectively with $\cO  \in\IC^{k\times n}$, as in \cite{AGI16}, i.e.:
	\begin{align}
	\cR&=\left[~\bPhi(\lambda_1)\bB,~\bPhi(\lambda_2)\bN\bPhi(\lambda_1)\bB,~\cdots,~
	\bPhi(\lambda_k)\bN\bPhi(\lambda_{k-1})\bN\,\cdots\,\bN\bPhi(\lambda_1)\bB
	\right],\nonumber \\[2mm]
	\cO&=\left[\begin{array}{l}
	\bC\bPhi(\mu_1)\\
	\bC\bPhi(\mu_1)\bN\bPhi(\mu_2)\\
	\qquad\vdots\\
	\bC\bPhi(\mu_1)\bN \bPhi(\mu_2)\bN \ \cdots \ \bN\bPhi(\mu_k)\\
	\end{array}\right].
 \label{reach_obs_bil}
\end{align}
As shown in \cite{AGI16}, the matrices $\cR$ and $\cO$ defined in (\ref{reach_obs_bil}), satisfy the following generalized Sylvester equations:
	\begin{align}
	\begin{split}
	   	&\bA \,\cR + \bN\, \cR\, \bS_\bR + \bB\, \bR = \bE \,\cR\, \bLambda \\
	&\cO \,\bA + \bS_\bL \,\cO\, \bN + \bL\, \bC = \bM\, \cO\, \bE .
		\end{split}
	\end{align}
 
\subsubsection{The generalized Loewner pencil} \label{subsec:4.3.2}

Given the above notations, we introduce the following matrices, i.e., the generalized { Loewner matrix} $\IL$, and the generalized { shifted Loewner} matrix $\sIL$
	\begin{equation} \label{loewner}
	\IL= - \cO\,\bE\,\cR \in \IC^{k \times k},~~\sIL= - \cO\,\bA\,\cR \in \IC^{k \times k}.
	\end{equation}
	In addition we define the quantities
	\begin{equation}\label{loewner2}
	\IT=\cO\,\bN\,\cR \in \IC^{k \times k},~~\IV=\cO\,\bB \in \IC^{k}~~\mbox{and}~~\IW=\bC\,\cR \in \IC^{1 \times k}.
	\vspace{-1mm}
	\end{equation}
Note that $\IL$ and $\sIL$ as defined above are indeed Loewner matrices, that is, they can be expressed as divided differences of appropriate transfer function values of the underlying bilinear system; the following equalities hold:
\small
	\begin{align}
%	\hspace*{-8mm}
\begin{split}
    	\IL(j,i)&=
	\frac{\displaystyle \bH_{j+i-1}(\mu_1,\ldots,{ \mu_j},\lambda_{i-1},\ldots,\lambda_1)-
		\bH_{j+i-1}(\mu_1,\ldots,\mu_{j-1},{ \lambda_i},\ldots,\lambda_1)}{\displaystyle{  \mu_j}-{ \lambda_i}}\\[2mm]
	\sIL(j,i)&=
	\frac{\displaystyle { \mu_j}
		\bH_{j+i-1}(\mu_1,\ldots,{ \mu_j},\lambda_{i-1},\ldots,\lambda_1)-
		{ \lambda_i}\bH_{j+i-1}(\mu_1,\ldots,\mu_{j-1},{ \lambda_i},\ldots,\lambda_1)}{\displaystyle { \mu_j}-{ \lambda_i}},
		\end{split}
	\end{align}
	\normalsize
	while $\bV(j,1)$ $=$ $\bH_j(\mu_1,\ldots,\mu_{j-1},{ \mu_j})$,
	$\bW(1,i)$ $=$ $ \bH_i({ \lambda_i},\lambda_{i-1},\ldots,\lambda_1)$, and\\
	$\IT(j,i)$ $=$ $\bH_{j+i}(\mu_1,\ldots,\mu_{j-1},{ \mu_j},
	{ \lambda_i},\lambda_{i-1},\ldots,\lambda_1)$. This result shows that all quantities of the bilinear Loewner surrogate model  can be indeed computed using only data, and the realization is written concisely as
	\begin{equation}\label{eq:real_bil}
	   	\hat\bE=-\IL,~~\hat\bA=-\sIL,~~\hat\bN=\IT,~~\hat\bB=\IV,~~\hat\bC=\IW.
		\end{equation}
It was shown in \cite{AGI16}, that the bilinear model of dimension $k$ in (\ref{eq:real_bil})  matches a total of $2k+k^2$ transfer function values of the original bilinear system of dimension $n$.

If necessary, the model given is (\ref{eq:real_bil}) is further reduced similarly to the classical linear case, e.g., as in (\ref{eq:proj1}). This is done by projecting with special matrices using the singular value decay of the Loewner pencil involved. This provides a useful indicator for choosing the truncation order (\cite{AGI16}).

\begin{example} [An illustrative example] {\textrm Given a \textbf{SISO} bilinear system as in (\ref{bil11}), given by
		$(\bC,\bE,\bA,\bN,\bB)$ of order $n$, consider the {tuples} of left and right interpolation points:
		$\left[\begin{array}{cc}
		\{\mu_{1}\} &
		\{\mu_{1},\mu_{2}\}
		\end{array}\right]$, 
		$\left[\begin{array}{cc}\{\lambda_{1}\},&\{\lambda_{2},\lambda_{1}\}\end{array}\right]$. The  generalized observability and controllability matrices are
		\small
		$$
		\begin{array}{l}
		\cO=\left[
		\begin{array}{c}
		\bC(\mu_{1}\bE-\bA)^{-1}\\
		\bC(\mu_{1}\bE-\bA)^{-1}\bN(\mu_{2}\bE-\bA)^{-1}
		\end{array}\right],\\
		\cR=\left[\begin{array}{cc}
		(\lambda_{1}\bE-\bA)^{-1}\bB,&(\lambda_{2}\bE-\bA)^{-1}\bN(\lambda_{1}\bE-\bA)^{-1}\bB
		\end{array}\right].
		\end{array}
		$$
		\normalsize
		The Loewner model matrices can be written in terms of data as:
		\begin{align*}
			&& \IL = \left[ \begin{array}{cc}
				\frac{\bH_1(\mu_1)-\bH_1(\lambda_1)}{\mu_1-\lambda_1} & \frac{\bH_2(\mu_1,\lambda_1)-\bH_2(\lambda_2,\lambda_1)}{\mu_1-\lambda_2}  \\ \frac{\bH_2(\mu_1,\mu_2)-\bH_2(\mu_1,\lambda_1)}{\mu_2-\lambda_1} & \frac{\bH_3(\mu_1,\mu_2,\lambda_1)-\bH_3(\mu_1,\lambda_2,\lambda_1)}{\mu_2-\lambda_2}
			\end{array}  \right] = -\mathcal{O} \bE \mathcal{R} ,\\[1mm]
			&& \sIL = \left[ \begin{array}{cc}
				\frac{\mu_1 \bH_1(\mu_1)-\lambda_1 \bH_1(\lambda_1)}{\mu_1-\lambda_1} & \frac{\mu_1 \bH_2(\mu_1,\lambda_1)- \lambda_2 \bH_2(\lambda_2,\lambda_1)}{\mu_1-\lambda_2}  \\ \frac{\mu_2 \bH_2(\mu_1,\mu_2)-\lambda_1 \bH_2(\mu_1,\lambda_1)}{\mu_2-\lambda_1} & \frac{\mu_2 \bH_3(\mu_1,\mu_2,\lambda_1)- \lambda_2 \bH_3(\mu_1,\lambda_2,\lambda_1)}{\mu_2-\lambda_2}
			\end{array}  \right] = -\mathcal{O} \bA \mathcal{R} ,\\[1mm]
			&& \IT = \left[ \begin{array}{cc}
				\bH_2(\mu_1,\mu_2) & \bH_3(\mu_1,\lambda_2,\lambda_1) \\ \bH_3(\mu_1,\mu_2,\lambda_1) & \bH_4(\mu_1,\mu_2,\lambda_2,\lambda_1)
			\end{array} \right] = \mathcal{O} \bN \mathcal{R} ,\\[1mm]
			&& \IV = \left[ \begin{array}{c}
				\bH_1(\mu_1)  \\ \bH_2(\mu_1,\mu_2)
			\end{array} \right] = \mathcal{O} \bB  ,\ \IW = \left[ \begin{array}{cc}
				\bH_1(\lambda_1)  & \bH_2(\lambda_2,\lambda_1)
			\end{array} \right] = \bC \mathcal{R}  .
		\end{align*}
		The surrogate bilinear system constructed as in (\ref{eq:real_bil}) matches eight transfer function values (\ref{bil01}) of the original system, namely: 
		$$
		\begin{array}{rl}
		\text{two of} \ ~\bH_1:& \bH_1(\mu_1),~ \bH_1(\lambda_1), \\
		\text{three of} \ ~\bH_2:& \bH_2(\mu_1,\mu_2), ~\bH_2(\mu_1,\lambda_1), ~\bH_2(\lambda_2,\lambda_1), \\
		\text{two of} \ ~\bH_3:& \bH_3(\mu_1,\mu_2,\lambda_1), ~\bH_3(\mu_1,\lambda_2,\lambda_1), ~\text{and}\\
		\text{one of} \ ~\bH_4:& \bH_4(\mu_1,\mu_2,\lambda_2,\lambda_1).
		\end{array}
		$$ }
		\end{example}

\begin{example}[Viscous (bi)linearized Burgers' equation model]
{\textrm 
We choose as a numerical test-case example, a discretized model of the viscous Burgers' equation (previously presented also in \cite{AGI16}). The original partial differential equation is given by
\vspace{-1mm}
\begin{equation} \label{burg_bil}
\frac{\partial v(x,t)}{\partial t}+v(x,t) \frac{\partial v(x,t)}{\partial x} =\frac{\partial}{\partial x} \Big{(} \nu 
\frac{v(x,t)}{\partial x} \Big{)},\  \ \ \ (x,t) \in (0,1) \times (0,T)\;, 
\normalsize
\end{equation}
\vspace{-1mm}
subject to the initial and boundary conditions given by
\vspace{-1mm}
\begin{equation*}
v(x,0) = f(x), \ x \in [0,1],\ v(0,t) = u(t),\  v(1,t) = 0, \ t \geqslant 0\;.
\vspace{-1mm}
\end{equation*}
The above system occurs in the area of fluid dynamics where it can be used for modeling gas dynamics and traffic flow. The solution $v(x, t)$ can be interpreted as a function describing the velocity at $(x, t)$. In general, the viscosity coefficient $\nu(x, t)$ might depend on space and time as well. 

Some simplifications are performed, and the viscosity coefficient $\nu(x, t) = \nu$ is assumed to be constant. Furthermore, a zero initial condition on the system, i.e., $f(x) = 0$, is considered. Finally, we assume that the left boundary is subject to a control.
 
Start with a spatial discretization of equation (\ref{burg_bil}), using an equidistant step size $h = \frac{1}{n+1}$ where n denotes the number of interior points of the interval $(0,1)$. By using first-order derivative approximations schemes,
a nonlinear model is obtained (with quadratic-bilinear nonlinearities).
Next, use the Carleman bilinearization technique to approximate this $n^{\text{th}}$order nonlinear system with a bilinear system of order $\mathcal{N} = n^2+n$. 
 
Denote with $\Si_B$ the  $4970^{\mathrm th}$ order initial bilinear system obtained by means of the Carleman bilinearization. The first step is to collect samples from generalized bilinear transfer functions up to order two; the 400 interpolations points are chosen logarithmically spaced in the interval $[10^{-3},10^3]\imath$. Next, we construct the bilinear Loewner matrices as presented in this section, and display the singular value decay in the upper pane of Fig.\;\ref{fig:Burgers1}. We construct a reduced-order model of order $r = 32$; the poles are depicted 
in the lower pane of Fig.\;\ref{fig:Burgers1}.

\begin{figure}[H]  
	\centering
		\includegraphics[scale=0.28]{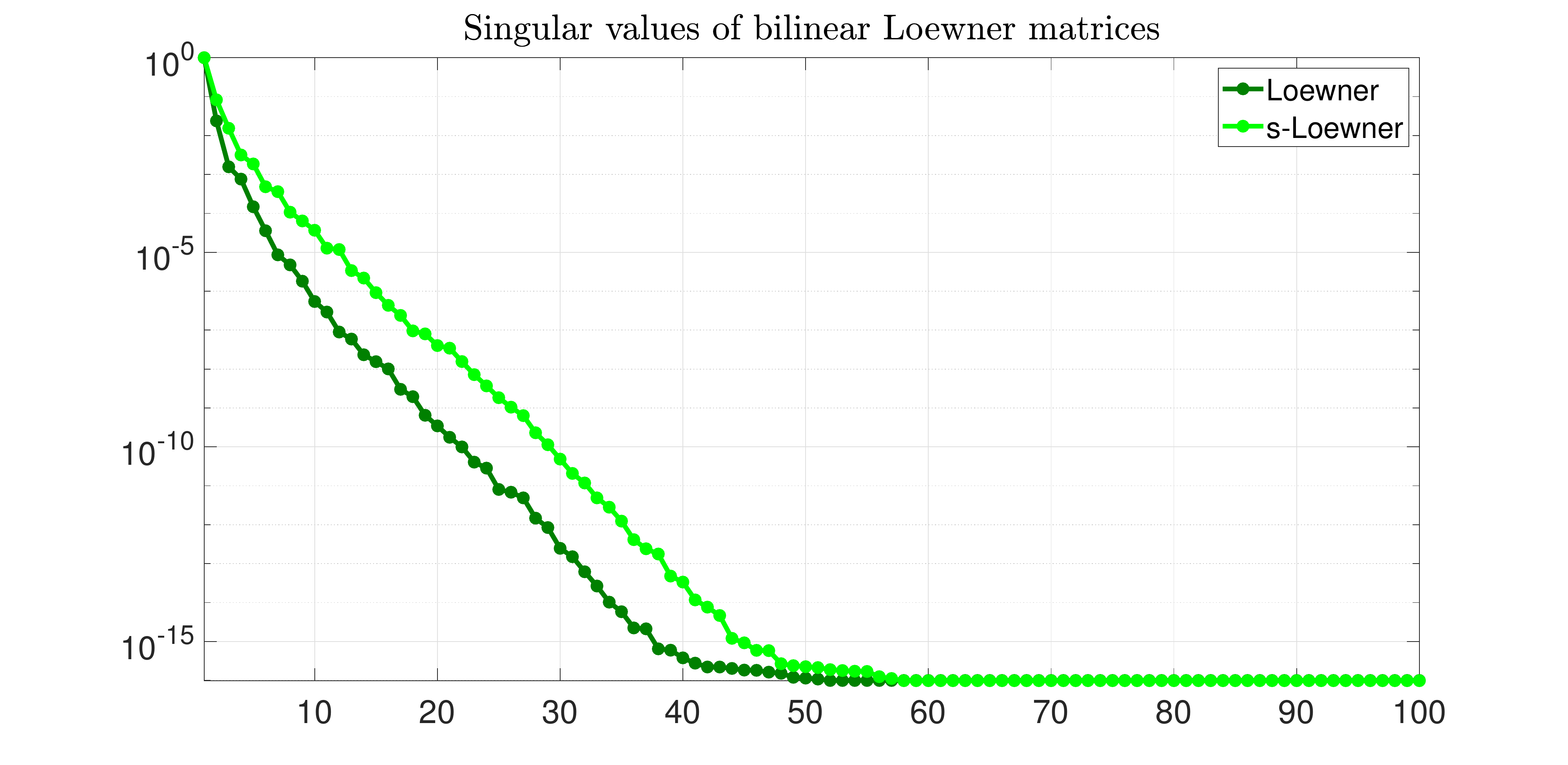}
		\hspace{-8mm}
		\includegraphics[scale=0.28]{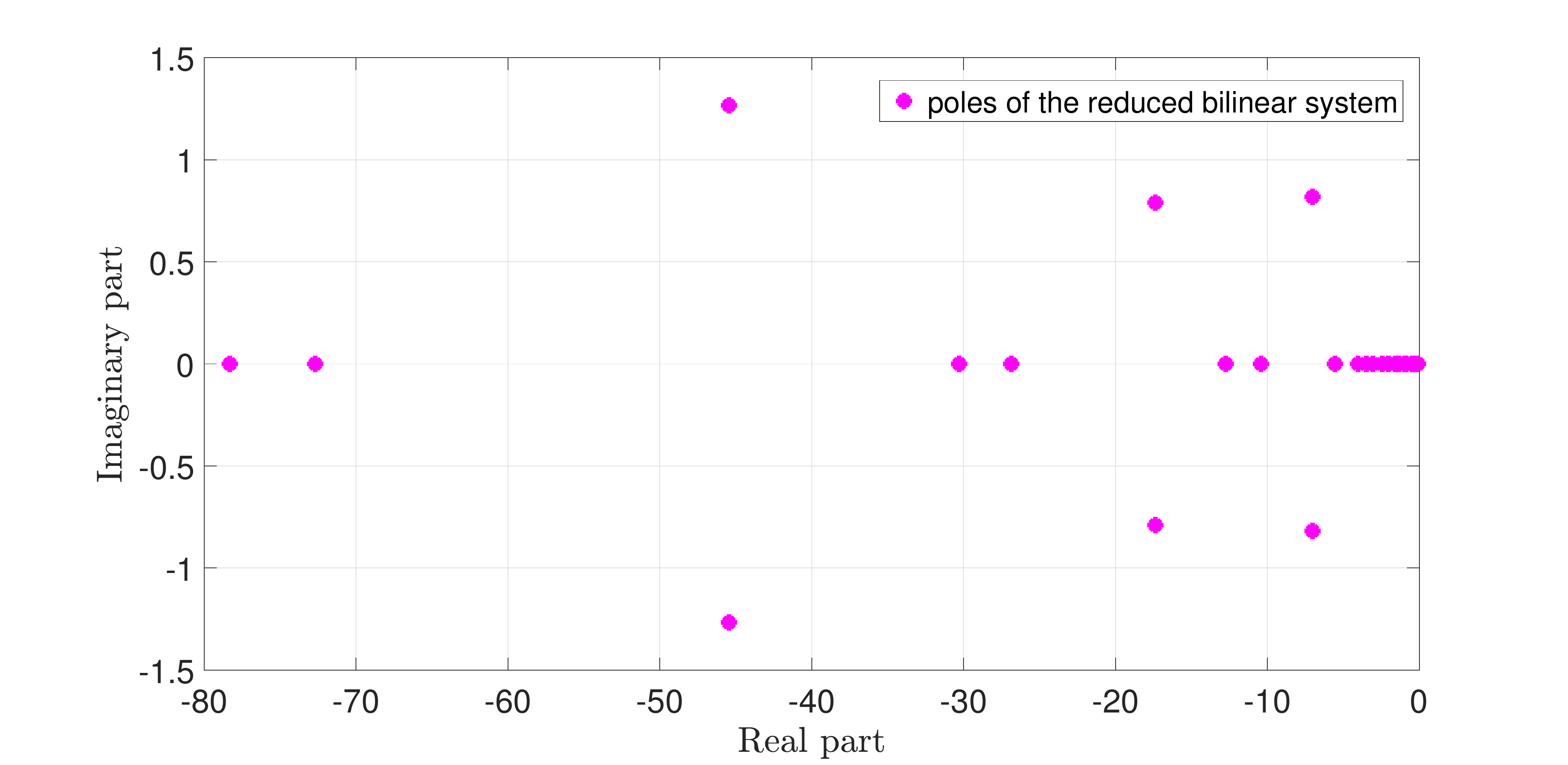}
		\vspace{-3mm}
		\caption{The first 100 singular values of the Loewner matrices (up) and the poles of the reduced-order model (down).}
		\label{fig:Burgers1}
 \vspace{-4mm}
\end{figure}

Finally, perform a time-domain simulation for a control input given by $u(t) = \frac{1}{5}(\cos(2\pi t)+\sin(20 \pi t) e^{-t/5})$, and on a chosen time span of $[0,10]$s. The observed outputs for both the original and of the reduced-order bilinear systems are displayed in the upper pane of Fig.\;\ref{fig:Burgers2}, while the approximation error is depicted in the lower pane of Fig.\;\ref{fig:Burgers2}.

\begin{figure}[H] 
	\centering
		\includegraphics[scale=0.28]{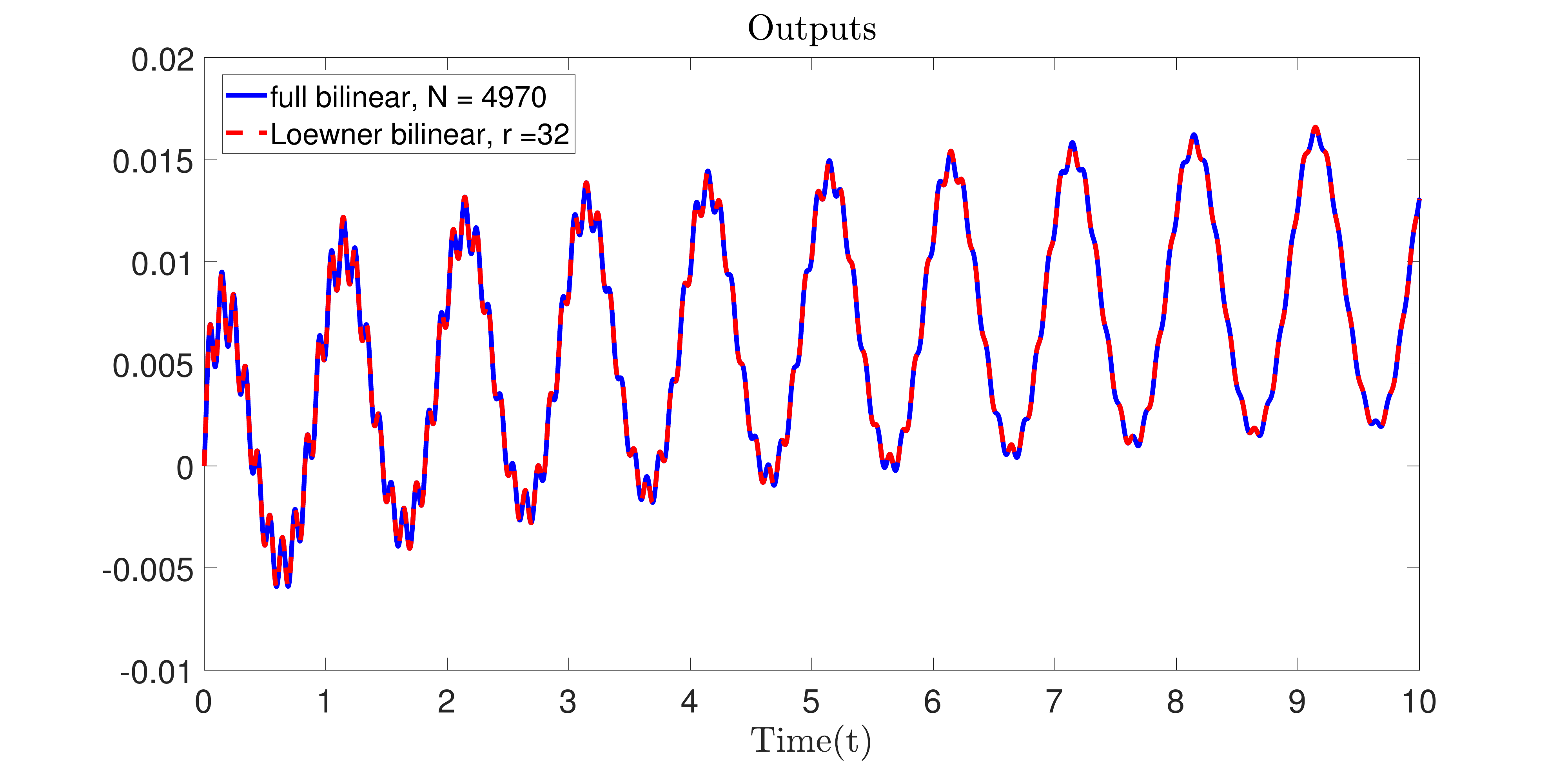}
		\hspace{-8mm}
		\includegraphics[scale=0.28]{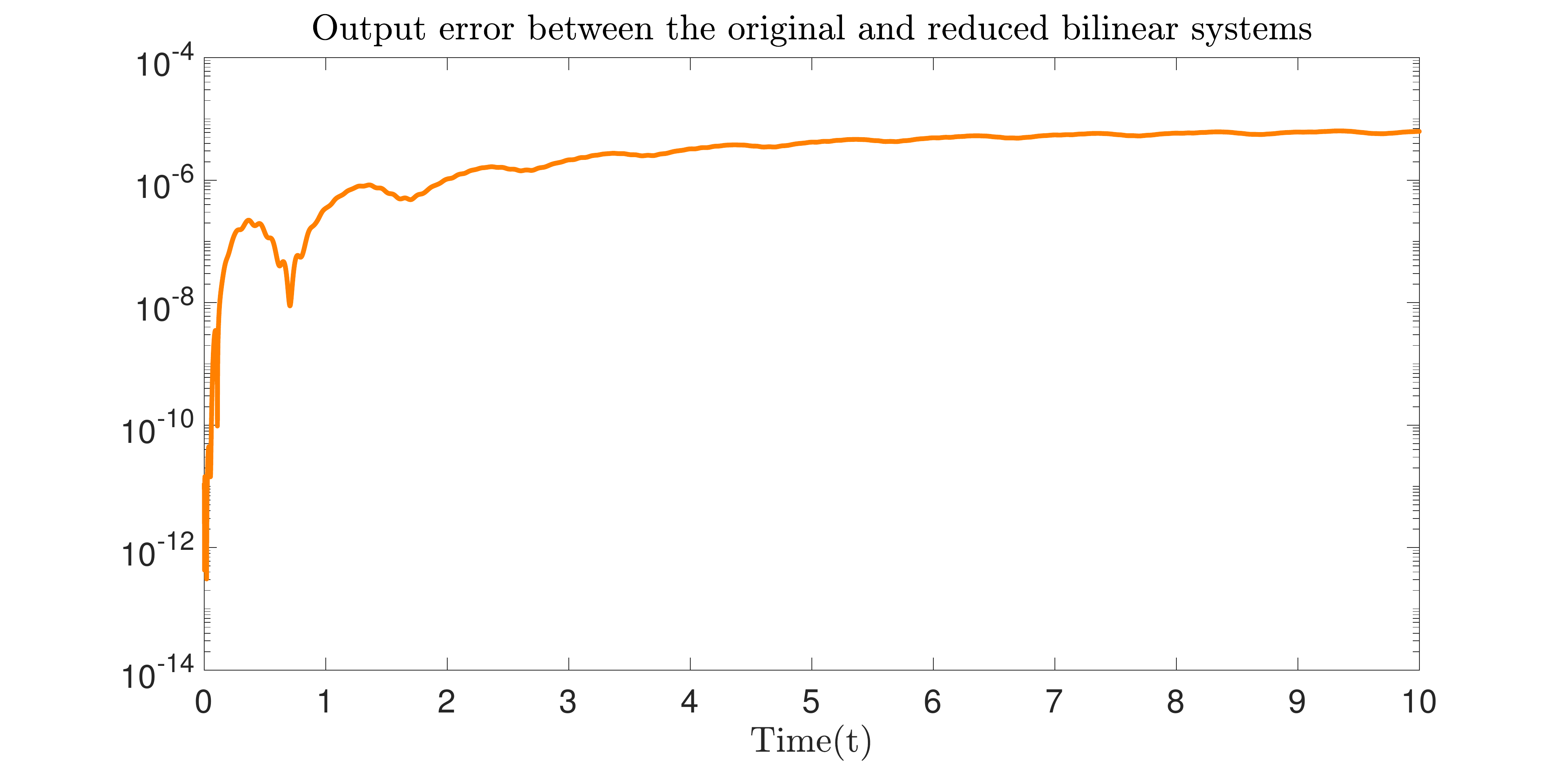}
		\vspace{-3mm}
		\caption{Time-domain simulations: the observed outputs (up) and the approximation error in the time domain (down).}
		\label{fig:Burgers2}
	 \vspace{-4mm}
\end{figure}

}
		\end{example}

\section{Examples of model reduction of large-scale systems}
\label{sec:examples}
In this section, we will demonstrate how Loewner-based rational approximation and reduction features have been successfully applied on real-life industrial problems. First, two benchmarks sequentially involving a generic business jet aircraft model and measurements data obtained by Dassault-Aviation, a French aircraft supplier, are considered (see \cite{PoussotALCOSP:2013,MeyerMOVIC:2016,MeyerIFASD:2017,PoussotSIADS:2021}). Second, a benchmark involving a simplified open channel model constructed by Electricit\'e De France, the French electricity supplier is involved (see \cite{DalmasECC:2016}). More specifically, a gust oriented model described by an non-rational transfer function is considered (section \ref{ssec-bizjetGust}), then ground vibration experimental data (in section \ref{ssec-bizjetGVT}) and finally, linear partial differential equations (in section \ref{ssec-EDF}).

%%%%%%%%%%%%%%%%%%%%%%
\subsection{Gust load oriented generic business jet aircraft model}
\label{ssec-bizjetGust}

%\subsubsection{Problem and model overview}

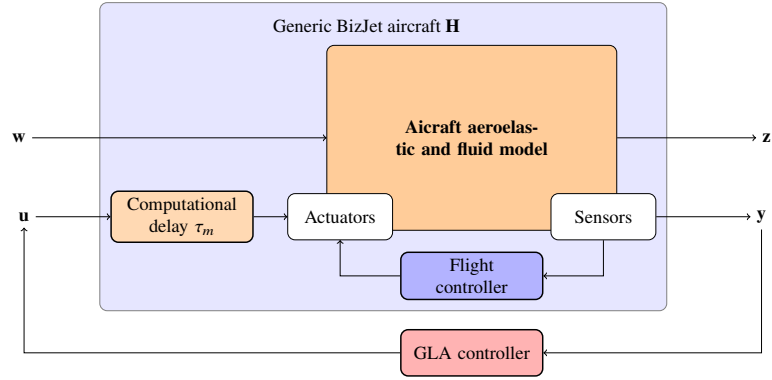
\begin{figure}[H]
  \centering
  \scalebox{.7}{\pgfdeclarelayer{background}
\pgfdeclarelayer{foreground}
\pgfsetlayers{background,main,foreground}

% Define a few styles and constants
%\tikzstyle{sysProp}  = [draw=green, fill=white,text width=6em,text centered,minimum height=5em,rounded corners]
\tikzstyle{system}  = [draw=black, fill=orange!40,text width=15em,text centered,minimum height=10em,rounded corners]
\tikzstyle{actProp}  = [draw=black, fill=white,text width=5em,text centered,minimum height=2.5em,rounded corners]
\tikzstyle{contPropFLIGHT} = [draw=black,thick, fill=blue!30,text width=7em,text centered,minimum height=2.5em,rounded corners]
\tikzstyle{contPropLOAD} = [draw=black,thick, fill=red!30,text width=7em,text centered,minimum height=2.5em,rounded corners]
\tikzstyle{delayProp} = [draw=black,thick, fill=orange!30,text width=7em,text centered,minimum height=2.5em,rounded corners]
%\tikzstyle{contPropVIBRATION} = [draw=black,thick, fill=orange!30,text width=7em,text centered,minimum height=2.5em,rounded corners]
\tikzstyle{ann}      = [above, text width=5em]
\def\blockdist{2.5cm}

\begin{tikzpicture}
    %% SYSTEM
    \node (sys) {};
    %\path (sys) node (image) []  {{\includegraphics[width=.35\columnwidth]{figures/GLA/7x}}};
    \path (sys) node (image) [system]  {\textbf{Aicraft aeroelastic and fluid model}};
    \path (sys)+(-\blockdist,-1.5cm) node (actuator) [actProp] {Actuators};
    \path (sys)+(\blockdist,-1.5cm) node (sensor) [actProp] {Sensors};
    \path (actuator.west)+(-2cm,0)  node (controllerDelay) [delayProp] {Computational delay $\tau_m$};
   
    %% INPUT OUTPUT ARROWS
    \path (sensor.0)+(2cm,0) node (outputY) [] {$\y$};
    \draw [->] (sensor.0) -- (outputY);
    \path (sys)+(5.6cm,0) node (outputZ) [] {$\mathbf z$};
    %\draw [->] (sys)+(2cm,0) -- (outputZ);
    \draw [->] (image.east) -- (outputZ);
    \path (controllerDelay)+(-3cm,0cm)  node (inputU) [] {$\u$}; 
    \draw [->] (inputU) -- (controllerDelay.west);
    \draw [->] (controllerDelay.east) -- (actuator.west);
    \path (sys)+(-8.6cm,0)  node (inputW) [] {$\mathbf w$};
    %\draw [->] (inputW) -- (sys);
    \draw [->] (inputW) -- (image.west);

    %% Controller Flight
    \path (sys.south)+(0,-2.5cm)  node (controllerFlight) [contPropFLIGHT] {Flight \\ controller};
    \draw [->] (sensor.-90) |- (controllerFlight.east);
    \draw [->] (controllerFlight.west) -| (actuator.-90);

   %% Controller Load
    \path (controllerFlight.south)+(0cm,-1cm)  node (controllerLoad) [contPropLOAD] {GLA controller};
    \draw [->] (outputY.-90) |- (controllerLoad.east);
    \draw [->] (controllerLoad.west) -| (inputU.-90);
    %\path (controllerLoad.north)+(0cm,1cm)  node (h) [] {$h$};
    %\draw [->] (h.south) -| (controllerLoad.north);
    
     %% BLOCK ENVELOPPE
    \path (sys.north)+(-2cm,2) node (globalModel) {Generic BizJet aircraft $\Htran$};
    \begin{pgfonlayer}{background}
        \path (controllerDelay.west |- globalModel.north)+(-0.2,0.2) node (a) {};
        \path (controllerFlight.south -| sensor.east)+(0.2,-0.2) node (b) {};
        \path [fill=blue!10,rounded corners, draw=black!50] (a) rectangle (b);
    \end{pgfonlayer}
\end{tikzpicture}}
  \caption{Closed-loop architecture of the \textbf{GLA} problem. The complete aeroservoelastic dynamical aircraft model $\Htran$ includes the "Flight controller", "Actuators", "Sensors" and "Computational delay $\tau_m$". The "GLA controller" is \textbf{GLA} function to be computed. Signals $\mathbf w$, $\u$, $\mathbf z$ and $\y$ denote the exogenous inputs, control inputs, performance outputs and measurements, respectively. Then $h$ denotes the sampling time for the \textbf{GLA}.}
  \label{fig:gla_global}
\end{figure}

An important aircraft design criterion concerns the so-called \emph{gust load envelope} monitoring. Prior to any test or exploitation, aircraft structural integrity should be guaranteed. One important certificate is to preserve and limit the worst case loads along the wings in response to vertical gust episodes. To this aim, it is standard to consider vertical gust disturbances $\mathbf w$, modelled through the so-called "1-cosine" profiles \cite{PoussotSIADS:2021}. The \emph{gust load envelope} is simply the worst case load responses along the wing span in reaction to the set of many differently chosen time-domain vertical wind gust profiles affecting the aircraft structure. In the preliminary conception step, the aircraft is designed by experts so that the wings support a given nominal load envelope, dictated by physical considerations such as desired aircraft manoeuvrability, gust, and many other manufacturing constraints. The larger the supported loads are, the larger the structural stiffeners and mass reinforcements should be. The aircraft mass is consequently bigger and its consumption during flight increased. In this context, gust load alleviation (\textbf{GLA}) control function plays an important role in the aircraft conception: it is aimed at lowering the loads envelope and thus at reducing the aircraft overall mass, consumption and emissions (see \cite{PoussotSIADS:2021} for details). To achieve this \textbf{GLA} function, as illustrated in Figure \ref{fig:gla_global}, model-based control design approaches are usually preferred. In this section, following \cite{PoussotSIADS:2021}, we illustrate through a generic business jet aircraft model constructed by Dassault-Aviation, how the Loewner framework is a pivotal tool used in the industry to simplify the complexity of these dynamical models, prior control design and analysis.

At each flight and mass configuration, a gust load oriented linear dynamical model considering aerodynamical, structural and actuator dynamics is constructed. Generic aircraft models have the following continuous-time realization
\begin{equation}
\Hreal:\left \lbrace
\begin{array}{l}
\bE\dx(t) = \bA_0\x(t)+\bA_1\x(t-\tau_1)+\bA_2\x(t-\tau_2) + \bB_u\u(t)+\bB_w\mathbf w(t)  \text{ , } \\
\y(t) = \bC_0\x(t) +\bC_1\x(t-\tau_m)  \text{ where , }\\
\bE,\bA_0,\bA_1,\bA_2 \in \IR^{n\times n}, \bB_u \in \IR^{n\times n_u}, \bB_w \in \IR^{n\times n_w}, \bC_0,\bC_1 \in \IR^{p\times n}.
\end{array}
\right. 
\label{eq:gla_model}
\end{equation}
where $\x(t)\in\Real^{n}$, $\u(t)\in\Real^{n_u}$, $\mathbf w(t)\in\Real^{n_w}$ ($m=n_u+n_w)$ and $\y(t)\in\Real^{p}$ are the internal variables, control input, exogenous gust input and output signals, respectively. In the considered case, $n_u=3$, $n_w=1$ ($m=4$), $p=5$ and $n\approx500$. The presence of internal delays is caused by the physical restitution of the gust impact over the fuselage at three different locations which are function of the aircraft velocity. Moreover, due to the model construction method (see \eg  \cite{QueroJFS:2021} or \cite{PoussotSIADS:2021}), the $\bE$ matrix may also be rank deficient. Here, due to the additional double derivative and delay structure added to accurately describe the gust disturbance effect along the fuselage, $\rank \,\bE=n-6$. Following \eqref{eq:gla_model}, the gust load model transfer associated function $\Htran$, from $[\u^T,\mathbf w^T]^T$ to $\y$ thus reads, 
\begin{equation}
\Htran(s) =\big(\bC_0 + \bC_1e^{-\tau_m s}\big) \big(s\bE - \bA_0-\bA_1e^{\tau_1s}-\bA_2e^{\tau_2s} \big)^{-1}\bB \in \IC^{p\times m} 
\label{eq:gla_modelTF}
\end{equation}

We seek a simplified rational model description to be used in place of \eqref{eq:gla_modelTF} for fast simulation, control design and (modal) analysis while avoiding dealing with an  infinite number of eigenvalues and transcendental equations related to the resolvant $\bPhi(s)=\big(s\bE - \bA_0-\bA_1e^{\tau_1s}-\bA_2e^{\tau_2s} \big)^{-1}$. The first step in the process consists in gridding the interpolation (support points) along the imaginary axis and collecting the associated response as follows (with $\overline n=\underline n=n=500$, $2n=N$ and $\omega_i\neq\omega_j$): 
\begin{equation}
\begin{array}{rcl}
\{z_k\}_{k=1}^{N} &=& \{\imath \omega_i,-\imath \omega_i\}_{i=1}^{n/2} \cup \{\imath \omega_j,-\imath \omega_j\}_{j=1}^{n/2} \text{ and }\\
\{\Phi_k\}_{k=1}^{N} &=& \{\Phi_i,-\overline{\Phi_i}\}_{i=1}^{n/2} \cup \{\Phi_j,-\overline{\Phi_j}\}_{j=1}^{n}.
\end{array}
\end{equation}
where $\omega_i,\omega_j \in \Real_+$ are the frequencies at which one evaluates each transfer $\Htran$. In our application $\omega_i$ and $\omega_j$ are selected to be logarithmically spaced. %This choice allows to focus on the frequency range of interest. Indeed, in the case of irrational models, the method is efficient for interpolation but not necessary for extrapolation. 

%In addition, as the stability of the obtained rational model $\{\Htran_i^{n_i}\}_{i=1}^{n_s}$ is not guaranteed by the Loewner interpolatory procedure, a \emph{post stabilisation} is performed using the procedure presented in \cite{Kohler:2014}. This latter consists in projecting the rational models $\{\Htran_i^{n_i}\}_{i=1}^{n_s}$ onto their closest stable model, here using the $\Hinf$-norm, leading to a set of stable models of the same dimensions. Mathematically, given a realization $\Hreal$ associated to $\Htran\in \Linf$, one aims at finding $P_\infty(\Htran) \in \Hinf$ such that,
%\begin{equation}
%$P_\infty(\Htran) = \arg \inf_{\mathbf G \in \Hinf}\norm{\Htran-\mathbf G}_{\Linf}$.
%\label{eq:stableApprox}
%\end{equation}
%Technical details and assumption can be found in \cite{Kohler:2014}. 

\begin{figure}[H]
\centering
\includegraphics[width=.74\columnwidth]{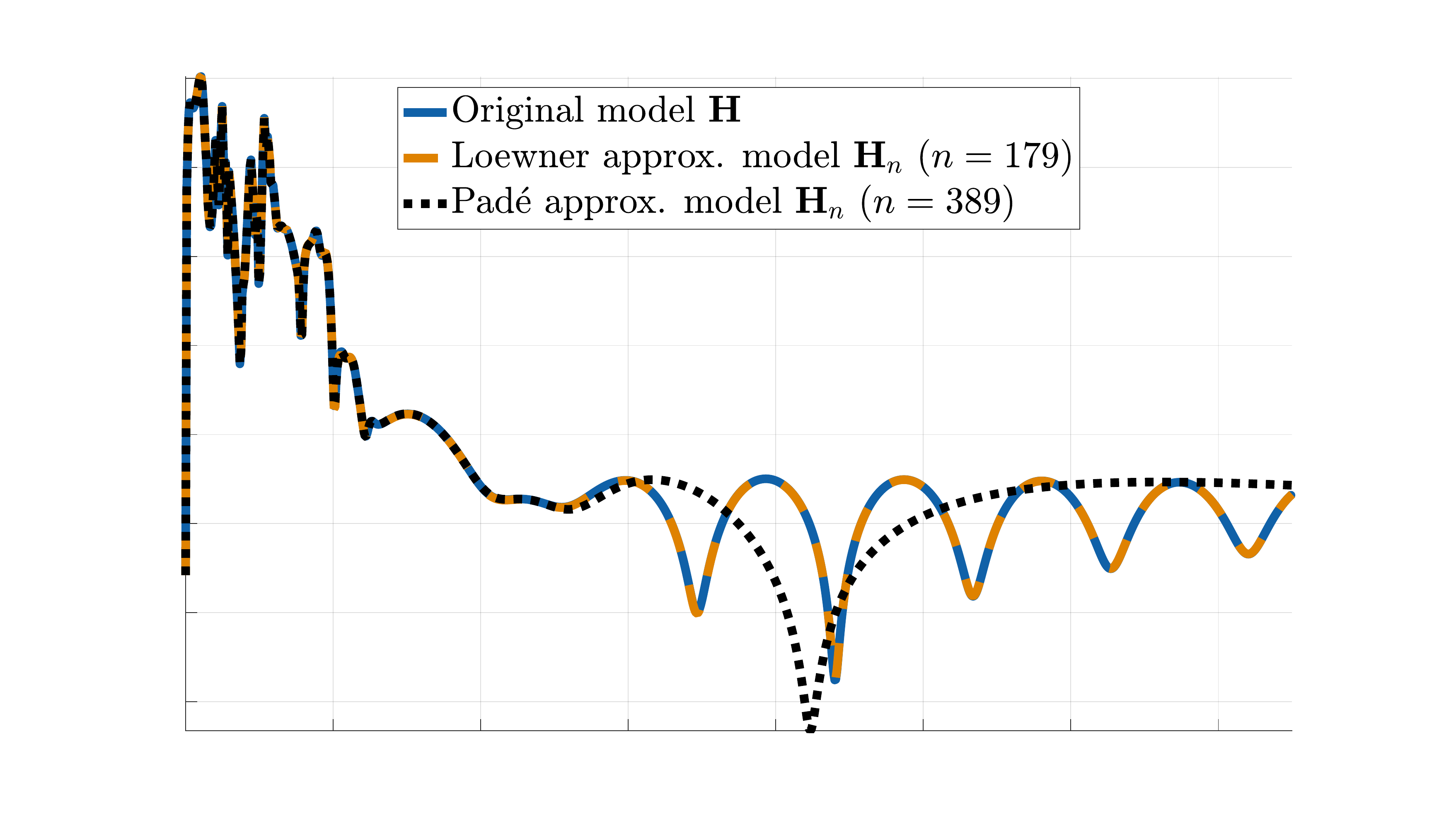}
\vspace{-2mm}
\includegraphics[width=.74\columnwidth]{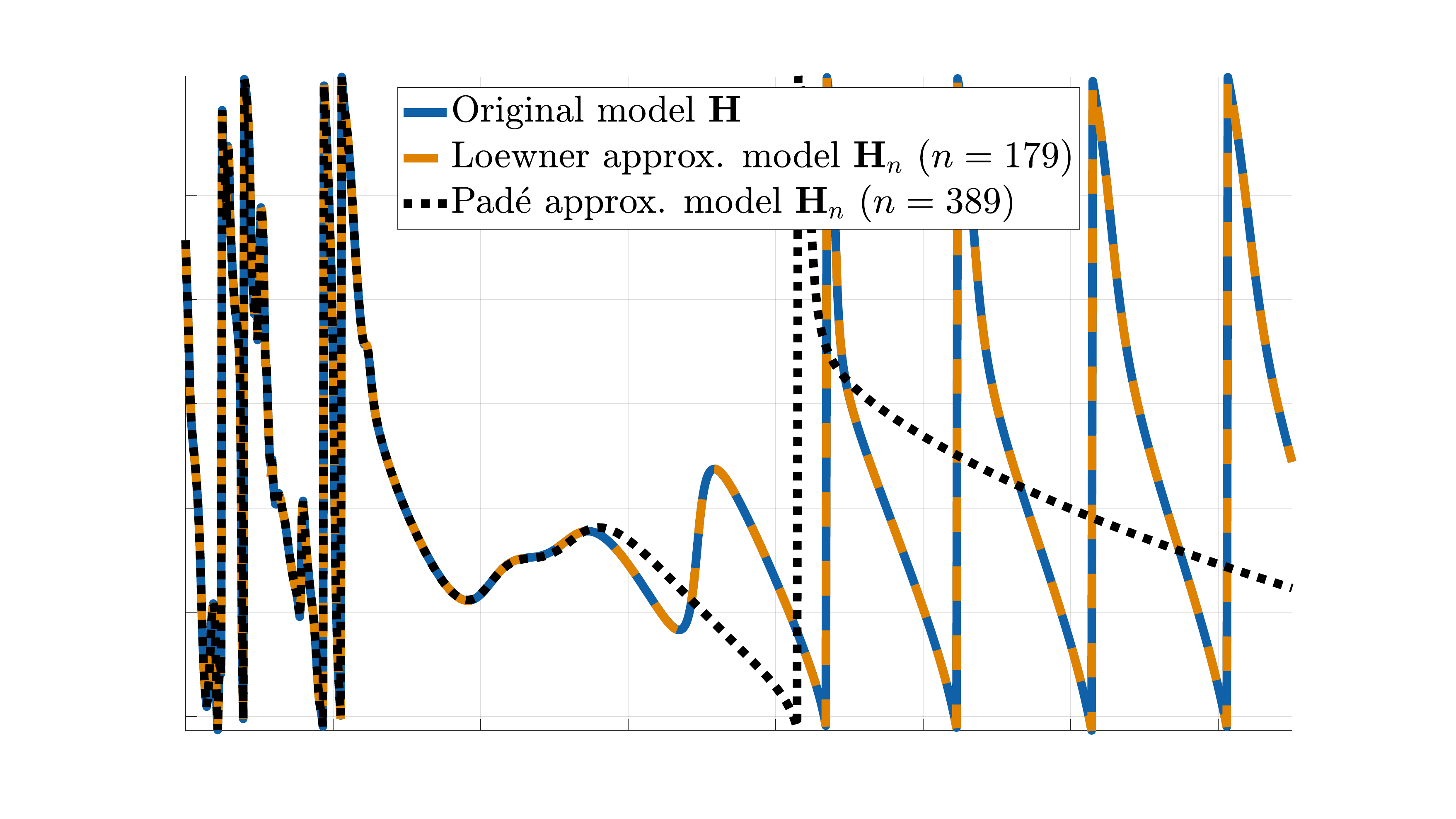}
\vspace{-4mm}
\caption{Top: frequency response gain (left) and impulse response (right). Bottom: frequency phase response. Comparison of the original model with the rational approximation obtained by Loewner interpolation and Pad\'e.}
\label{fig:gla}
\end{figure}

\begin{remark}[About a Pad\'e delay approximation]
One option is to replace the delays with a Pad\'e approximation, which preserves the gain but modifies the phase. While this is classically used in many applications, it is, to the authors experience, not the most accurate way to deal with internal and external delays. Indeed, Pad\'e often results in significant error in the phase, which can be inappropriate for flexible structures. In addition, the use of Pad\'e will drastically increase the model internal dimension which in turn is not appropriate for model reduction. Therefore, the accuracy / complexity ratio is not in favour of Pad\'e approximation (see also Figure \ref{fig:gla}).
\end{remark}

Figure \ref{fig:gla} illustrates the transfer function from the gust disturbance to a wing bending moment output, used to monitor the gust envelope. It compares the responses of the original irrational model $\Htran$ with its rational approximate $\Htran_{n}$ constructed with Loewner and its rational approximation $\Htran_{\text{Pad\'e}}$ obtained with Pad\'e. 

Figure \ref{fig:gla} emphasises the good performance of the rational model obtained by Loewner after reducing the complexity of the model (internal variable reduced). Most interestingly, the phase is much well captured by the Loewner approach than with Pad\'e, using  even less internal variables. In \cite{PoussotSIADS:2021}, this rational model is then used for frequency-limited reduction and \textbf{GLA} controller synthesis, leading to an impressive load envelope reduction which is not achievable without the use of a Loewner interpolatory approach. This result emphasizes the importance of the Loewner framework for aircraft consumption reduction objective.

%%%%%%%%%%%%%%%%%%%%%%
\subsection{Ground vibration tests on Business jet aircraft}
\label{ssec-bizjetGVT}

We continue on the business aircraft benchmark provided by Dassault-Aviation. Now we move from the gust load problem to the vibration one. While the former is more related to (the low frequency) structure and consumption issues, the latter is related to (the medium frequency) fatigue and comfort issues. Anti-vibration controllers are usually designed using model-based approaches in order to reduce the undesirable amplifications of the aerodynamical effects on the fuselage around some specified frequencies (see \cite{PoussotALCOSP:2013} for details). After such a model-based design and validation step, Ground Vibration Tests (\textbf{GVT}) are performed to validate the control performance, but also to validate the model.

\begin{figure}[H]
\centering
\includegraphics[width=.74\columnwidth]{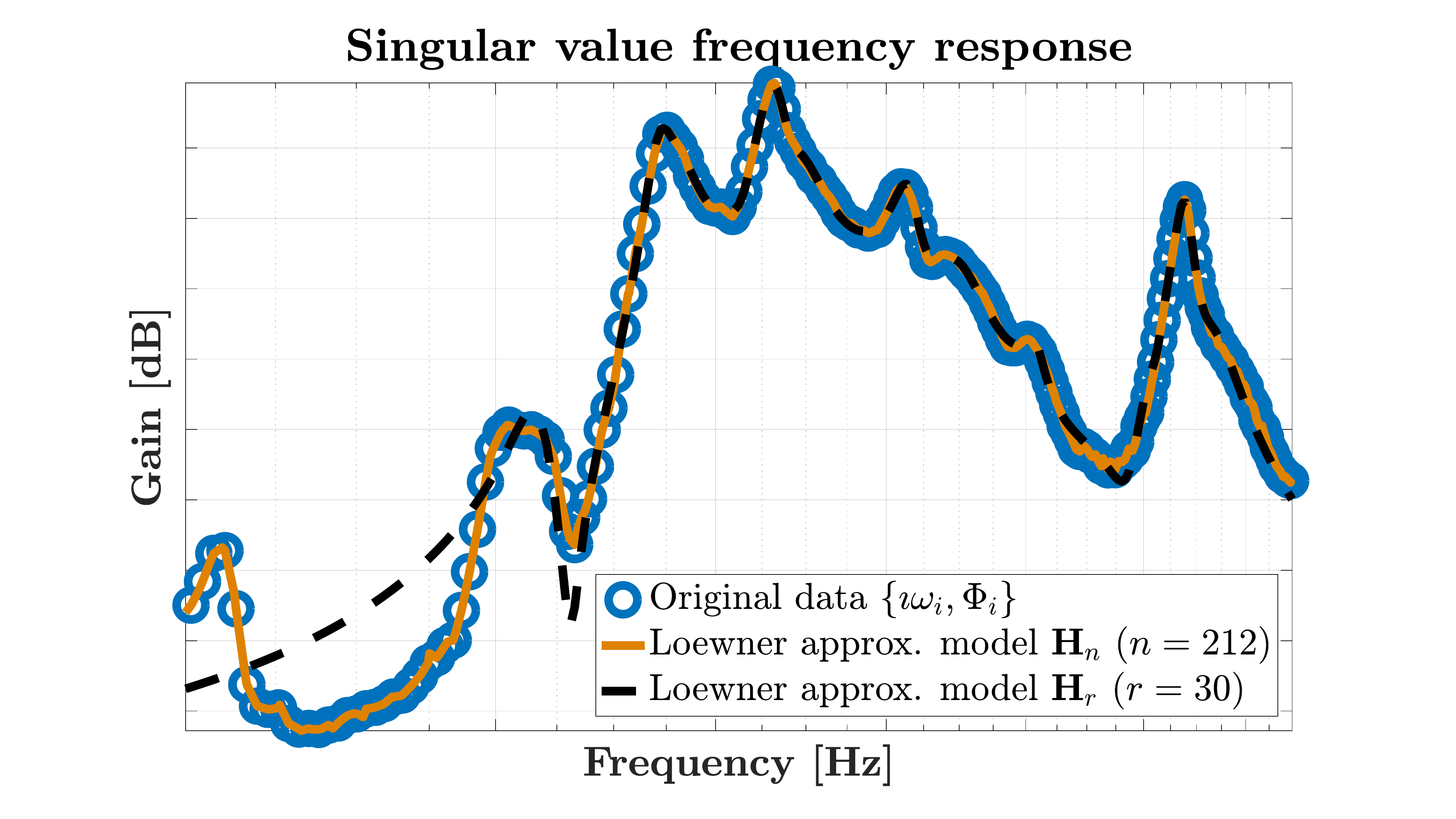} \vspace{-2mm}
\includegraphics[width=.74\columnwidth]{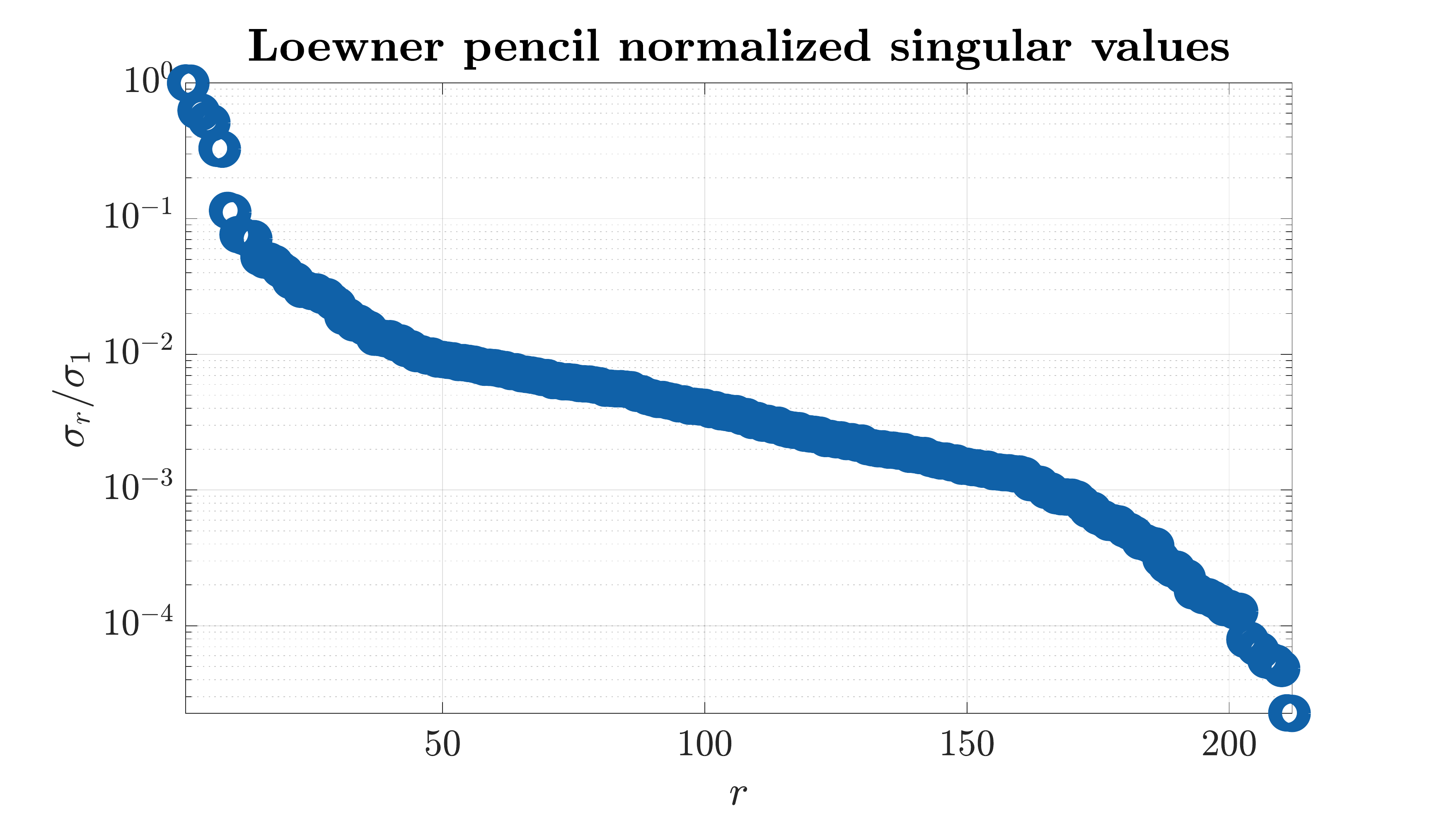}
\vspace{-2mm}
\caption{Top: singular value frequency response of the data (blue circles) the minimal Mc Millan degree rational function (solid orange) and reduced 30-th order rational model (dashed black). Bottom: singular values drop of the Loewner pencil.}
\label{fig:gvt}
\end{figure}

The benchmark considered here illustrates the generic business jet \textbf{GVT}, performed on a Falcon 7X at Istres, France, in 2015 \cite{MeyerMOVIC:2016,MeyerIFASD:2017}\footnote{Flight test have been performed in 2017, validating the results.}. The first step consists in designing an anti-vibration controller aimed at attenuating the vibrations at the passenger cabin and specified fatigue locations in response to aerodynamics turbulence occurring at specified frequencies. At the next step  Dassault-Aviation engineers implemented  the control law on the real business jet aircraft. Then, using shakers applied at some  aircraft locations, the structure was excited, thus simulating aerodynamic disturbances. Hundreds of sensors were positioned on the aircraft and used for analysis\footnote{\url{https://drive.google.com/file/d/1H2GqlYkiny_PZND2ekB6swSetoGcmFTK/view} shows a video that illustrates the kinematic effect of the control law acting on the tail surface to reduce the vibrations.}. Figure \ref{fig:gvt} (top) shows the frequency response of the data collected from a single-input and 100-outputs; this is compared with the frequency response of a rational model of minimal complexity constructed, in open loop \eg  without anti-vibration devices. The singular values drop is also illustrated in Figure \ref{fig:gvt} (bottom). In both cases, the truncation and rank computation are performed via \textbf{SVD}\footnote{Notice that other methods can be considered such as \textbf{CUR}, \textbf{EV}, see \eg \cite{Karachalios:2020}.}. 

Additional information may be found in \cite{MeyerMOVIC:2016,MeyerIFASD:2017} or in \S 2.4.7 of \cite{PoussotHDR:2019}. In this industrial challenging case, one important feature of the Loewner framework illustrated here is to be able to recover the transfer function from raw data, and perform modal (residue) analysis. In the considered industrial application, such a feature allows engineers to re-adjust the theoretical models accordingly to the collected real data, detect some new phenomena and re-adjust the control law. This step contributes to the quest for a so-called \emph{digital twin}.

%%%%%%%%%%%%%%%%%%%%%%%
\subsection{Hydroeletricity open-channel benchmark}
\label{ssec-EDF}

In this third example, we consider a model representing the level $h$ of an open-channel as a function of the inflow $q_i$ and outflow $q_o$ inputs. Such a model is used by hydro-electricity engineers from Electricit\'e De France to monitor the level of a river in order to control the available energy (note that in real applications, these model come in a network). One important feature of open-channels is that they can be viewed as easily available energy tanks. Indeed, unlike windmills or nuclear factories, energy is available on demand, and unlike solar panels, energy (water) can be stored. In France, in May 2021, the hydraulic energy represented about 10\% of the total produced  energy\footnote{\url{https://www.rte-france.com/eco2mix/la-production-delectricite-par-filiere}.}. Well understanding the underlying dynamics in view of energy management is therefore crucial in the global warming frame.

Mathematically such models for such benchmarks belong to
the class of linear partial differential equations (\textbf{PDE}). Such a models come from the so-called Saint-Venant equations, used to model the dynamics of open channel flow (see \cite{DalmasECC:2016} for a detailed description). They consist of two nonlinear hyperbolic \textbf{PDE}s. For a channel of length $L$ and bottom slope $I$, we have
\begin{equation}
\begin{array}{rcll}
\dfrac{\partial S}{\partial t} + \dfrac{\partial Q}{\partial x} &=& 0 & \text{(mass conservation)} \\
\dfrac{\partial Q}{\partial t} + \dfrac{\partial (Q^2/S)}{\partial x}+ gS\dfrac{\partial H}{\partial x} &=& gS(I-J)  & \text{(momentum conservation)},
\end{array}
\label{eq:exHydroStVenant}
\end{equation}
where $x \in [0,L]$  is the spatial variable, $t$ the time variable, $H(x,t)$ the water depth, $S(x,t)$ the wetted area, $Q(x,t)$ the discharge, $g$ the gravity acceleration and $J$ the Manning-Strickler friction\footnote{Numerical values of this model are provided at \url{https://morwiki.mpi-magdeburg.mpg.de/morwiki/index.php/Hydro-Electric_Open_Channel}}.

These equations are quite complex to simulate and analyse. Under mild assumptions a linearization around an equilibrium point $(Q_0,H_0)$, detailed in \cite{DalmasECC:2016}, expresses the variation relations $(q,h)$, between inflow ($q_e$, being $q$ at $x=0$), outflow ($q_s$, being $q$ at $x=L$) and the water depth ($h$, at a given measurement point $x$) as follows,
\begin{equation}
h(x,s) = \mathbf G_e(x,s)q_e(s) - \mathbf G_s(x,s)q_s(s),
\label{eq:exHydroH}
\end{equation}
where
\begin{equation}
\begin{array}{rcl}
\mathbf G_i(x,s)&=& \dfrac{\lambda_1(s)e^{\lambda_2(s)L+\lambda_1(s)x}-\lambda_2(s)e^{\lambda_1(s)L+\lambda_2(s)x}}{B_0s(e^{\lambda_1(s)L}-e^{\lambda_2(s)L})} \text{ and }\\
\mathbf G_o(x,s)&=& \dfrac{\lambda_1(s)e^{\lambda_1(s)x}-\lambda_2(s)e^{\lambda_2(s)x}}{B_0s(e^{\lambda_1(s)L}-e^{\lambda_2(s)L})}.
\end{array}
\label{eq:exHydroGeGs}
\end{equation}

\begin{figure}[H]
\centering
\includegraphics[width=.9\columnwidth]{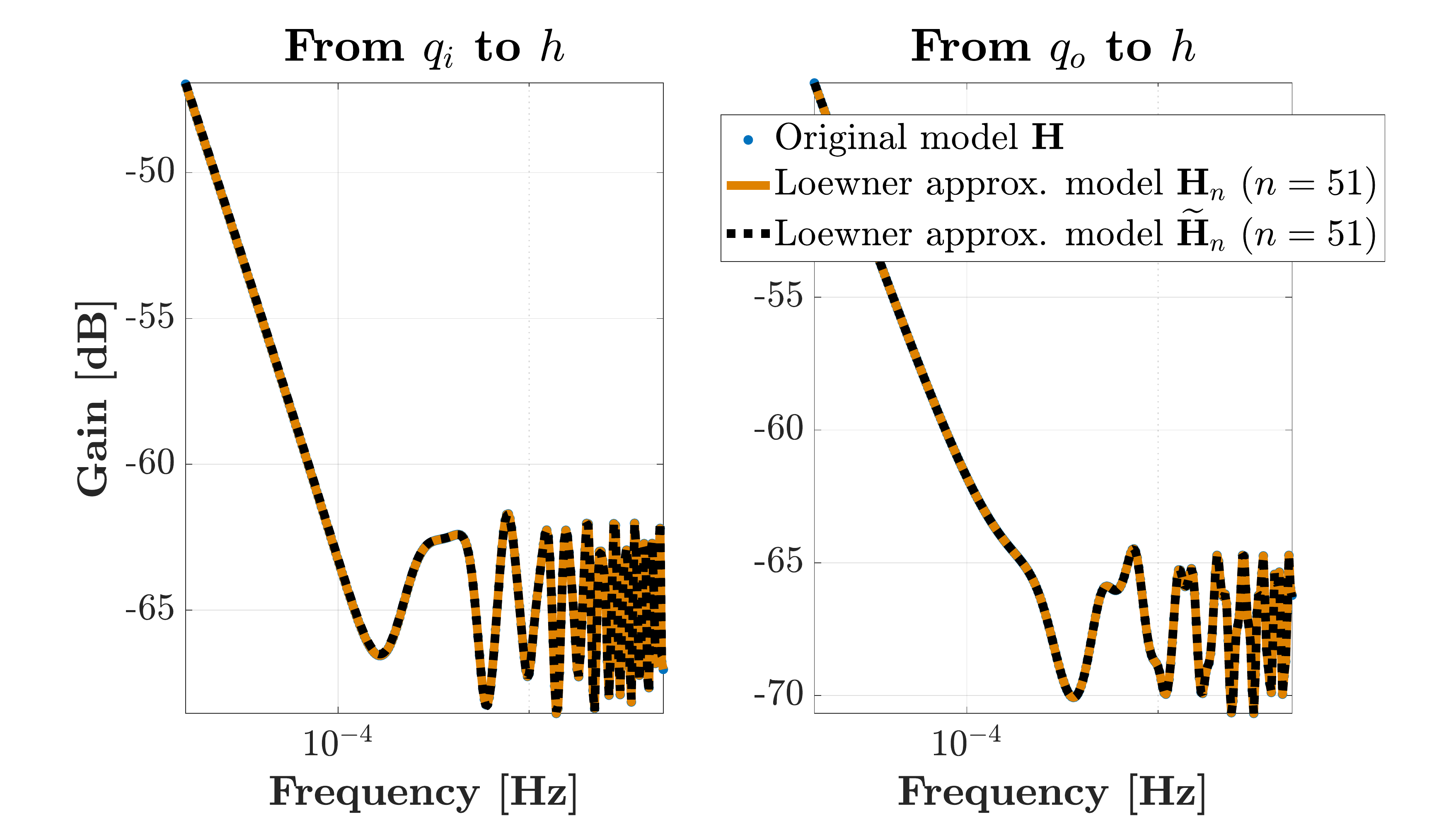}\\
\includegraphics[width=.6\columnwidth,align=c]{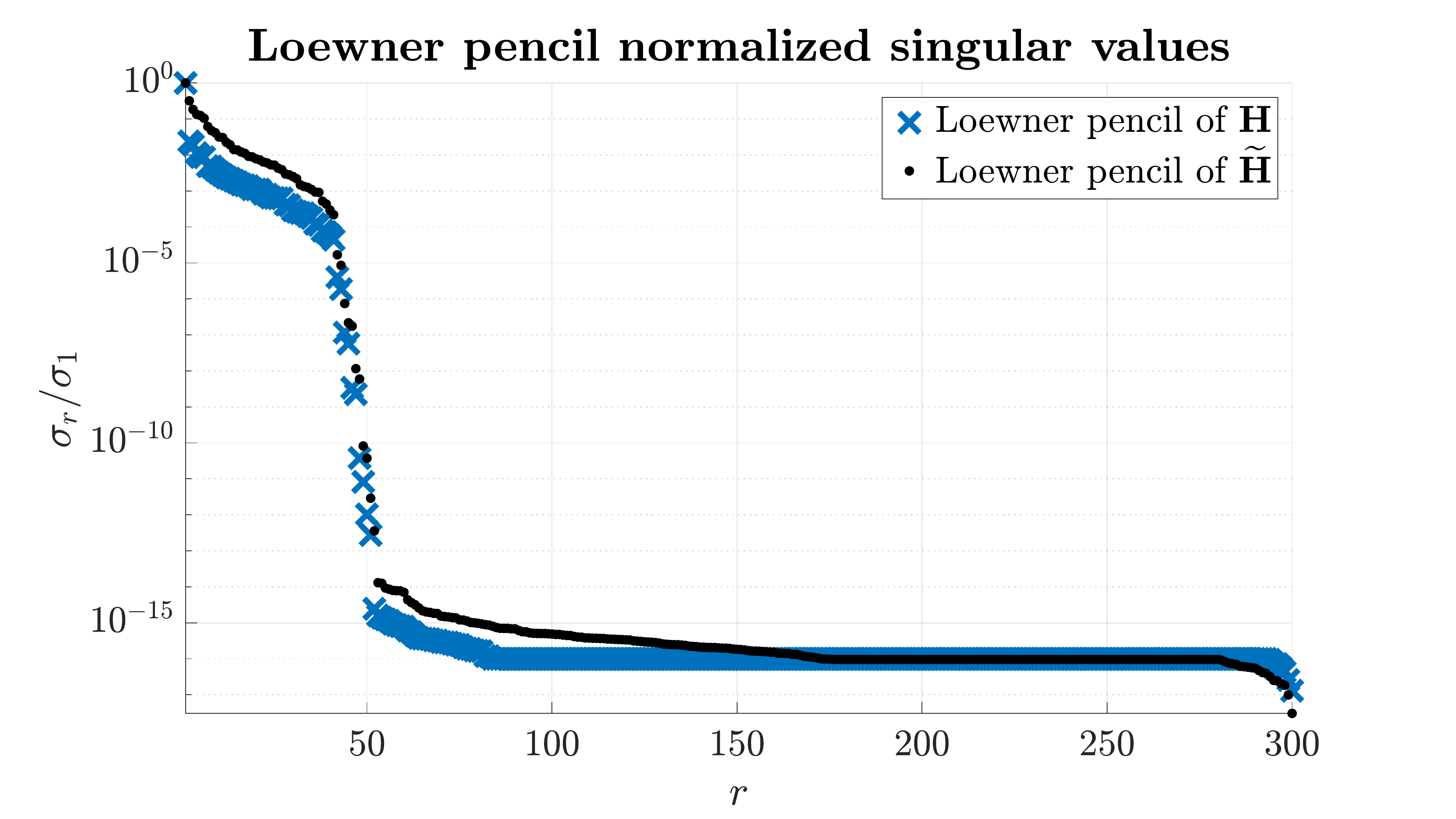}
\includegraphics[width=.3\columnwidth,align=c]{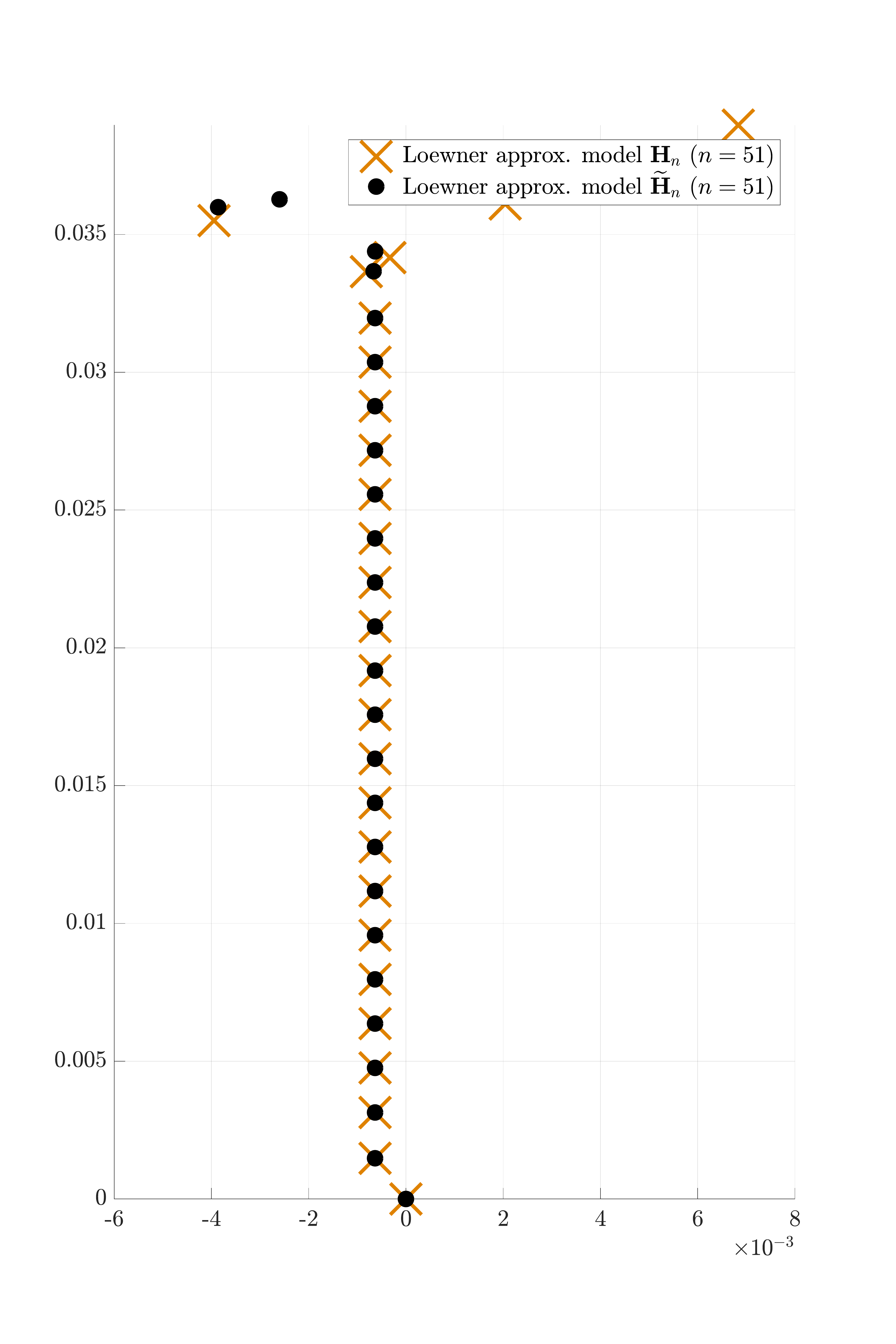}
\caption{Top: frequency response comparison between the original irrational model an two approximated Loewner models. Bottom left: Singular values drop of the Loewner pencil for the two models. Bottom right: eigenvalues of the resulting minimal order rational approximation.}
\label{fig:hydro}
\end{figure}

Clearly $\mathbf G_i$ and $\mathbf G_o$ yield a non-rational infinite dimensional model. For a frozen measurement point $x=x_m$, then one has 
\begin{equation}
h_{x_m}(s) =\Htran(s) \bu(s) = \mathbf G_i(x_m,s)q_i(s) + \mathbf G_o(x_m,s)q_o(s).
\label{eq:exHydroFreq_xm}
\end{equation}
where $\bu(s)$ contains the two inputs $q_i(s)$ and $q_o(s)$ and where $\Htran$ is now a one output two inputs complex-valued transfer function. Figure \ref{fig:hydro} illustrates the approximation features and accurate reconstruction of the open-channel phenomenon. To obtain this result, we consider complex conjugated points $\{z_k\}_{k=1}^N=\{\imath \omega_k,-\imath \omega_k\}_{k=1}^{N/2}$ (where $\overline n=\underline n=300=N/2$)  sampled between $10^{-4}$ and $10^{1.5}$ in logarithmic space. Then, the responses  
\begin{equation}
    \Htran(s) \text{ and } \widetilde \Htran(s)=\Htran(s) \dfrac{s}{(s+10^{-2})(s+10^{-3})},
\end{equation}
are computed. Dealing with $\Htran$ remains standard with the framework presented so far. By approximating $\widetilde \Htran$ removes the integral action and enforces roll-off in high frequency, and thus allows to deal with limited energy functions ($\widetilde \Htran\in\mathcal H_2$). Therefore, the resulting interpolated model should be post processed as $\widetilde \Htran_n\leftarrow \Htran_n\frac{(s+10^{-2})(s+10^{-3})}{s}$ to recover the original one.

As illustrated on Figure \ref{fig:hydro}, both approaches lead to a perfect matching of the irrational transfer. Interestingly, working with $\widetilde \Htran$ instead of $\Htran$ leads to a model with all singularities on the left hand side plus the 0 one. Working with the shifted function $\widetilde \Htran$ illustrates how one can perform grey box identification by simply shifting the original data. Here, the integral action (physically known from open-channel models) is removed and added afterward. The trick of working with $\mathcal H_2$ functions instead of $\mathcal H_\infty$ ones (as $\Htran$ is) is more numerical than theoretical as is avoids bad conditioning of exact 0 singularities and focusing on low dynamics first.  Moreover, in the similar flavour, one may also remove the delay part of such a transfer by pre-multiplying by $e^{s\tau}$, where $\tau\in\mathbb R_+$ is the estimated delay of the function, and thus dealing with $\widetilde \Htran(s)=\Htran(s) \dfrac{se^{s\tau}}{(s+10^{-2})(s+10^{-3})}$ instead. This feature is relevant for real-life applications.

\section{Control in the Loewner framework}
\label{sec:control}
Let us now deviate from the original purpose of the Loewner framework, initially introduced to provide solutions to the identification, approximation and reduction problems through the lens of rational function construction. Here instead, such a framework is used for feedback controller design. More specifically it is used as in some traditional loop shaping methods, to fit a reference controller \cite{ZhouBook:1997,McFarlane:1992,Francis:1987,Doyle:1981a}. However, in the proposed setup, the reference controller is not computed by means of a model but rather involving input-output data of the system. 

\subsection{Data-driven control, virtual reference model and Loewner framework}

In this section, the Loewner framework will be used for synthesizing a controller directly from measured data. Hence, this a data-driven control (\textbf{DDC}) framework\footnote{The reader may notice that \textbf{DDC} methods have a long history dating to the proportional, integral, derivative (\textbf{PID}) tuning method by Ziegler-Nichols in early 40's or the self tuning regulator by \AA str\"om in the 90's (see \eg \S 3 of \cite{KergusPhD:2019} for more details and references).}. Data-driven control consists in recasting the control design problem as an identification one. Major advantages of this strategy are: \emph{(i)} it provides a controller tailored to the actual system and \emph{(ii)} that is not dependent of the underlying mathematical model description. This change of paradigm shifts the model identification / simplification process to the controller directly. 

The considered technique belongs to the so-called \emph{reference model approaches} and more specifically relies on the definition of a so-called \emph{ideal controller}, derived from a reference model. %Such an approach is known as the virtual reference feedback tuning (\textbf{VRFT}).  It was first developed in \cite{Campi:2002} and deployed in a series of papers using time-domain measured input-output data of the process (see \eg \cite{Formentin:2013,Formentin:2014} and references therein). 
Recently \cite{KergusWC:2017,VuilleminWC:2020} moved the formulation in the frequency-domain, with the use of the Loewner framework as the identification tool, allowing to skip the controller complexity selection thanks to its rank properties (see section \ref{sec:loewner}). The Loewner data-driven control (\textbf{L-DDC}) is thus a combination of determining the ideal controller from frequency-domain data via a reference model and the use of the Loewner framework \cite{MA07} to construct a reduced order controller. Such an interpolatory-based data-driven control design solves problems faced by practitioners: \emph{(i)} the controller design is directly obtained using open-loop raw data collected on the experimental setup, \emph{(ii)} without any prior controller structure or order specification. This approach has proven to be effective on infinite dimensional systems \cite{GoseaECC:2021}, for digital control \cite{VuilleminWC:2020}, experimental application \cite{PoussotCST:2021} and relates to data-driven stability analysis \cite{PoussotGTA:2021}.

\subsection{The \textbf{L-DDC} rationale at a glance}
%boils down

The \textbf{L-DDC} procedure boils down to two steps: first deriving the \emph{ideal controller} definition and second the \emph{controller identification} via interpolation in the Loewner framework (in \cite{GoseaECC:2021} the use of Loewner in this context is compared with \textbf{AAA} and \textbf{VF}).  We recall the mains steps in the \textbf{SISO} case. Following Figure \ref{fig:ddc_problem}, the objective is to find a controller $\mathbf{K} \in \IC$ that minimizes the difference between the resulting closed-loop and a given user-defined reference model $\mathbf{M}\in \IC$. This is made possible through the definition of the ideal controller $\mathbf{K}^\star$, being the \textbf{LTI} controller that would have given the desired reference model behaviour if inserted in the closed-loop. The latter is defined as $\mathbf{K}^\star=\Htran^{-1}\mathbf{M}(I-\mathbf{M})^{-1}$. In the data-driven case, this definition may be recast as a discrete set of equations (where $\{z_k\}_{k=1}^N\in \IC$, $k=1,\dots,N$)
\begin{equation}
    \mathbf{K}^\star(z_k)=\Htran(z_k)^{-1}\mathbf{M}(z_k)(I-\mathbf{M}(z_k))^{-1},
    \label{eq:Kideal}
\end{equation}
where $\Htran(z_k)$ is the evaluation of the considered model, if available. In an experimental context, one usually considers sampling $\Htran$ at $z_k=\imath \omega_k$ ($\omega_k\in\IR_+$). In this case input-output measurements are given as $\Htran(\imath\omega_k)=\overline \by(\imath\omega_k)/\overline \bu(\imath\omega_k)$, where $\overline \bu$ and $\overline \by$ are the Fourier transform of $\bu$ and $\by$, respectively. Finding a controller $\mathbf K$ that fits $\mathbf{K}^\star(z_k)$ can be considered to be an identification problem. Thus, in the Loewner framework, the control design boils down to finding a rational function $\mathbf K$ interpolating \eqref{eq:Kideal}.

\begin{figure}[H]
\centering
\scalebox{.78}{\tikzstyle{block} = [draw, thick,fill=blue!20, rectangle, minimum height=4em, minimum width=6em,rounded corners]
\tikzstyle{sum} = [draw, thick,fill=blue!20, circle, node distance=1cm]
\tikzstyle{input} = [coordinate]
\tikzstyle{output} = [coordinate]
\tikzstyle{pinstyle} = [pin edge={to-,thick,black}]
\tikzstyle{connector} = [->,thick]

% The block diagram code is probably more verbose than necessary
\begin{tikzpicture}[auto, node distance=3cm,>=latex']
    % We start by placing the blocks
    \node [input, name=input] {};
    \node [sum, right of=input] (sum) {};
    \node [block, right of=sum] (controller) {$\mathbf K(\xi)$};
    \node [block, right of=controller, node distance=4cm] (system) {$\mathbf H(\xi)$};
    \node [sum, right of=system, node distance=2cm] (sum2) {};
    \node [block, above right of=controller, node distance=3cm and 2cm] (obj) {$\mathbf M(\xi)$};
    % We draw an edge between the controller and system block to 
    % calculate the coordinate u. We need it to place the measurement block. 
    \draw [connector] (controller) -- node[name=u] {${\u}$} (system);
    \node [output, right of=sum2, node distance=1cm] (output) {};

    % Once the nodes are placed, connecting them is easy. 
    \draw [connector] (input) -- node {${\mathbf r}$} (sum);
    \draw [connector] (sum) -- node {$\mathbf e$}(controller);
    \draw [connector] (system) -- node [name=y] {${\mathbf y}$}(sum2);
    %\draw [->] (y) |-| (measurements);
    %\draw [->] (measurements) -| node[pos=0.99] {$-$} node [near end] {$y_m$} (sum);
    \draw [connector] (sum2)+(-0.3cm,0) -- ++(-0.3cm,-1.5cm) -| node [near start] {} (sum.south);
    \draw [connector] (sum2) -- node {$\boldsymbol \varepsilon$}(output);
    \draw [connector] (input)+(0.3cm,0) |- (obj) -| (sum2);
    \draw [connector] (input)+(0.3cm,0) |- (obj);
    
\end{tikzpicture}}
\caption{Data-driven control problem formulation: $\mathbf{M}$ is the reference model (objective) and $\mathbf{K}$ the controller to be designed.}
\label{fig:ddc_problem}
\end{figure}
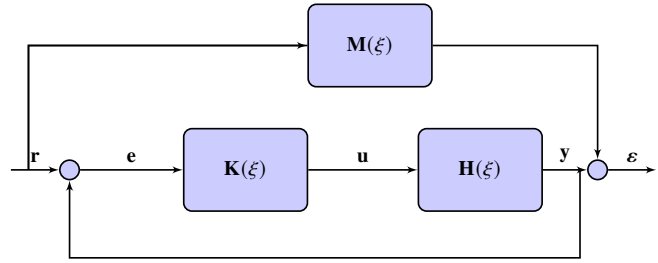

In what follows, two \textbf{L-DDC} applications are illustrated. The first one involves experimental data and considers the design of a reference tracking controller applied on a pulsed fluidic actuator (in short, \textbf{PFA}), see section \ref{ssec-ex_pfa}, \cite{PoussotCST:2021}). The second case considers a numerical benchmark representing the boundary control a wave equation, described by an infinite dimensional equation. For this latter case, equivalence with a model-based approach is also illustrated (see also \cite{PoussotGTA:2021}).

%%%%%%%%%%%%%%%%%%%%%%
\subsection{Pulsed fluidic actuator}
\label{ssec-ex_pfa}

\begin{figure}[H]
\centering
\scalebox{.5}{\input{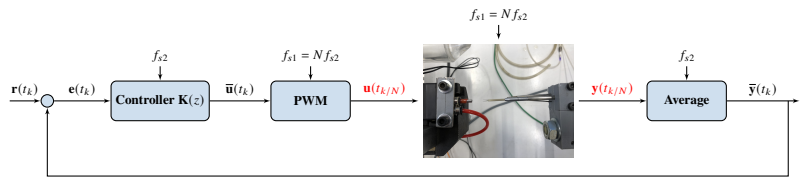}}
\caption{\textbf{PFA} control setup. \textbf{Controller} $\mathbf K(z)$ is the sampled-time control law to be computed (sampled at $f_{s2}$), Pulsed Width Modulation (\textbf{PWM}) block transforms the continuous signal into on/off values (sampled at frequency $f_{s1}$) and \textbf{Average} block is a down-sampling  function  providing the mean value of the input signal. The system is illustrated by its top view photo, where the left side represents the \textbf{PFA} and the right side, the Pressure Sensor (\textbf{PS}).}
    \label{fig:pfa_setup}
\end{figure}

The design of active closed-loop flow controllers constitutes an important field of research in fluid mechanics (see \eg \cite{Sipp:2016,Willcox:2002}). The possible objectives are to maintain laminarity or delay transition to turbulence, decrease turbulence level, reduce noise, increase lift and decrease drag, enhance mixing and heat release, etc. Without detailing the methodology employed in each case, in most cases, both the sensor(s) and the actuator(s) are supposed to be lumped and ideal (\ie sensors deliver instantaneous accurate measurements and actuators deliver the exact control signals with no delay, no noise, continuous control signal and unbounded intervals). These developments are relevant for academic and methodological purposes. However, to move towards experimental applications and real-life validations, it is essential to consider realistic set-ups. Considering the actuator-sensor combination is necessary and is the core contribution of \cite{PoussotCST:2021}, where the \textbf{L-DDC} is applied on a  \textbf{PFA}. \textbf{PFA} are on/off actuators that blow air to modify the pressure in a flow setup. They are typically used to control fluidic phenomena. The control setup considered is schematized on Figure \ref{fig:pfa_setup}

\begin{figure}[H]
    \centering
    \includegraphics[width=.73\columnwidth,align=c]{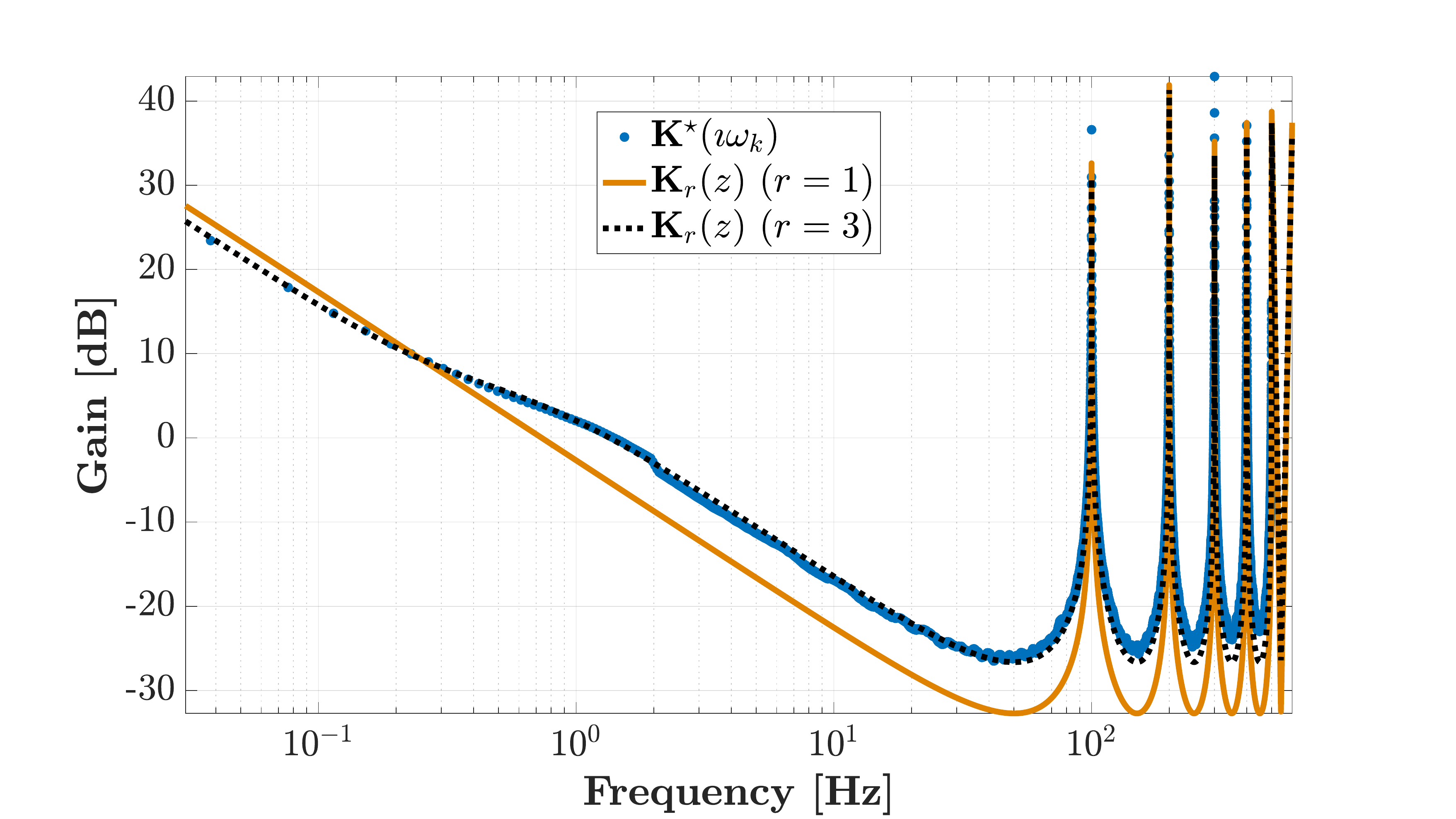}
    \includegraphics[width=.73\columnwidth,align=c]{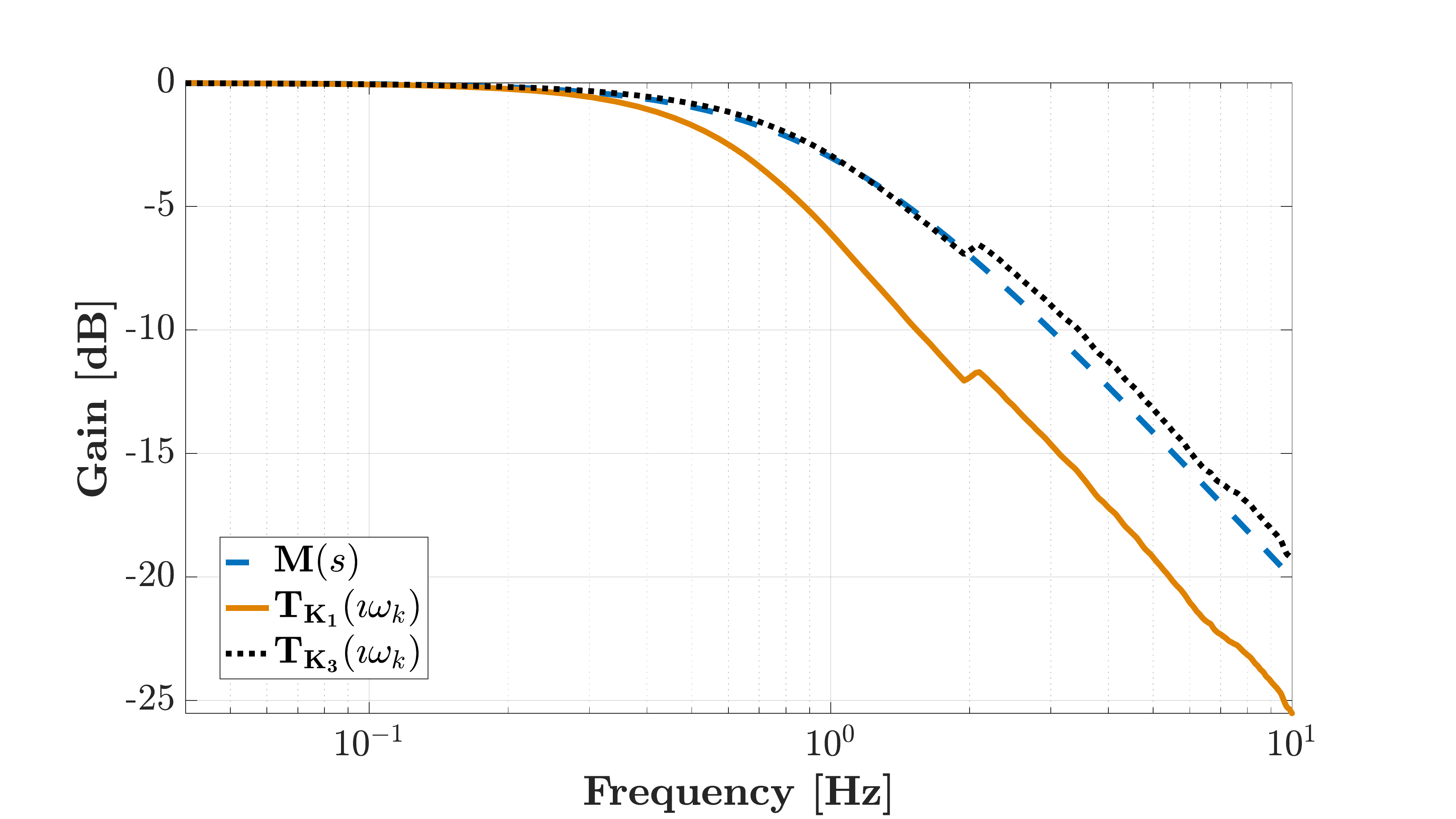}
    \includegraphics[width=.73\columnwidth,align=c]{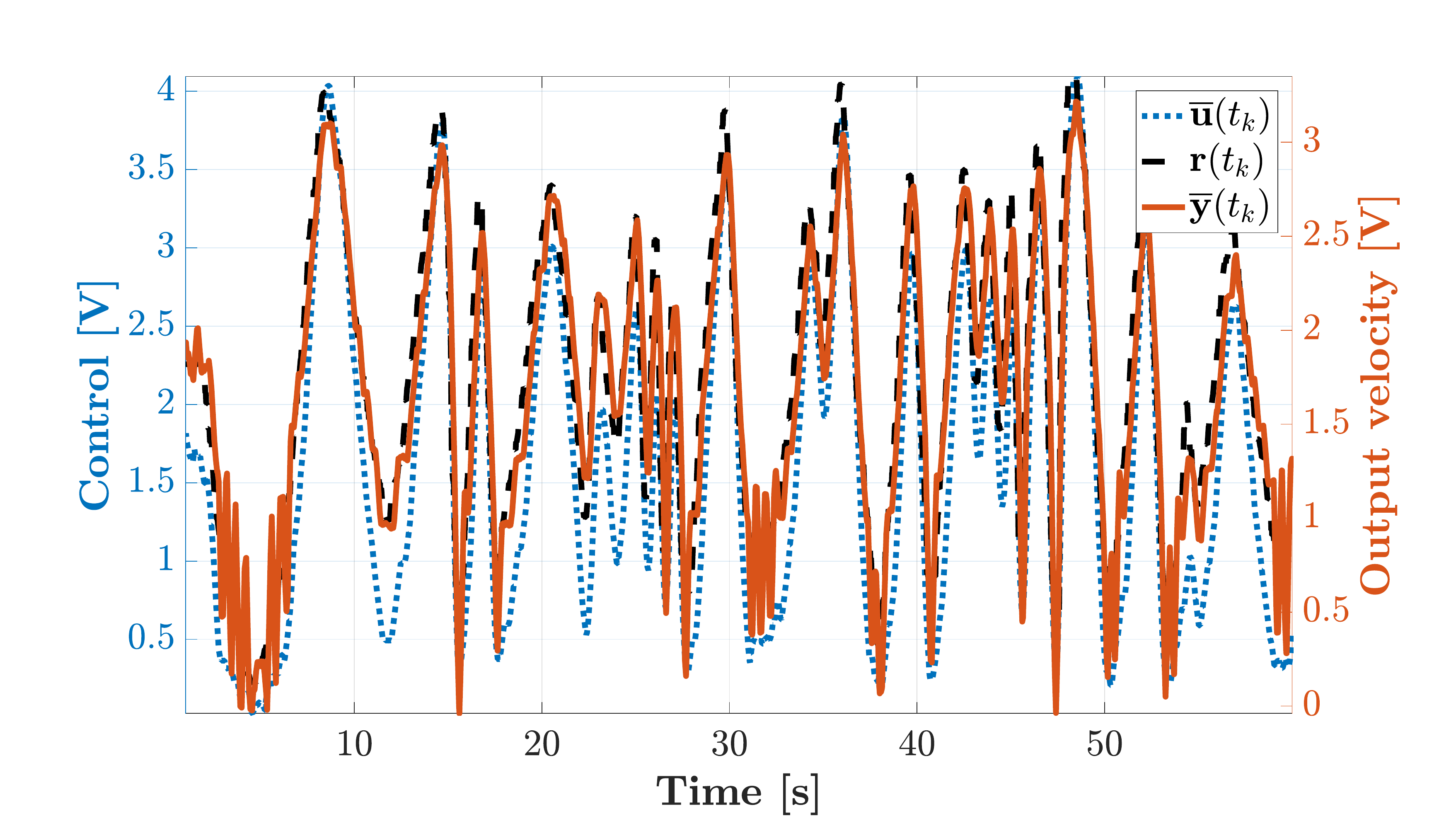}
    \caption{Top: gain of the frequency responses of the ideal controller $\mathbf K^\star$ evaluated at the available frequencies (blue dots) an of the estimated controller $\mathbf{\tilde K_{r}}(s)$ of order $r=1$ (solid orange) and $r=3$ (dotted black). Middle: closed-loop response estimation using controller $\mathbf{K_{1}}$ and $\mathbf{K_{3}}$ of the averaged output $\overline{\y}(t_k)$ (solid orange an dotted black) based on the measured data. Time-domain response to a variable reference trajectory $\mathbf r(t_k)$ (dashed black), averaged control signal $\overline{\u}(t_k)$ (dotted blue) and averaged output (solid orange).}
    \label{fig:pfa_K}
\end{figure}

After exciting the \textbf{PFA} using a pseudo random binary sequence $\bu(t_{k/n})$, output data $\by(t_{k/n})$ are collected. The corresponding frequency responses $\overline\bu$ and $\overline\by$ are computed and transfer function values $\Htran(\imath\omega_k)$ are thus obtained. Applying \eqref{eq:Kideal} with $z_k=\imath\omega_k$ and the Loewner approach, it leads to a singular value decay indicating that a first or third order model is sufficient to recover the main dynamics (see Figure \ref{fig:pfa_K}).

One important result is the ability of the \textbf{L-DDC} to construct, directly from raw open-loop data, a control law performing well on an experimental setup. Relevant in this context is that the  \textbf{L-DDC} structure and complexity is almost automatically chosen by the Loewner framework, and no pole pre-assignment is required.

%%%%%%%%%%%%%%%%%%%%%%
\subsection{Transport phenomena benchmark}
\label{ssec-ex_transport}

Finally, let us consider the case of a one dimensional transport equation controlled at its left boundary through a second order actuator. This model is used in \cite{PoussotGTA:2021} or \cite{GoseaECC:2021} and detailed in \S 2, Example 7 of \cite{PoussotHDR:2019}. This phenomenon is represented by a linear \textbf{PDE} with constant coefficients interconnected with a second order linear \textbf{ODE} actuator, as described in \eqref{eq:exTransportTime}. 
\begin{equation}
\begin{array}{rcll}
\dfrac{\partial \tilde y(x,t)}{\partial x} + 2x\dfrac{\partial \tilde y(x,t)}{\partial t} &=& 0 & \text{~~(transport equation)}\\ 
\tilde y(x,0)&=&0& \text{~~(initial condition)} \\
\tilde y(0,t)&=&\dfrac{1}{\sqrt{t}} \tilde u_f(0,t)& \text{~~(boundary control input)}\\ 
\dfrac{\omega_0^2}{s^2+m\omega_0s+\omega_0^2}u(0,s)&=&u_f(0,s) &\text{~~(actuator model)},
\end{array}
\label{eq:exTransportTime}
\end{equation}
where $x\in[0~L]$ ($L=3$) is the spatial variable. Then, $\omega_0=3$ and $m=0.5$ are the input actuator parameters. The scalar input of the model is the vertical force applied at the left boundary, \emph{i.e.} at $x=0$. We denote the input $\tilde \bu(0,t)$ in the time-domain or $\bu(0,s)$ in the complex one. Similarly, the output at location $x$ is given as $\tilde \by(x,t)$ for the time-domain and $\by(x,s)$ in the complex one. Such a transport equation set may be used to represent a simplified one dimensional wave equation used in telecommunications, traffic jams prediction, \emph{etc}. 

By applying the Laplace transform, one obtains the transfer function from the input $\bu(0,s)$ to the output $\by(x,s)$:
\begin{equation}
\by(x,s) = \dfrac{\sqrt{\pi}}{\sqrt{s}}e^{-x^2s} \dfrac{\omega_0^2}{s^2+m\omega_0s+\omega_0^2}\bu(0,s) = \mathbf G(x,s) u(0,s).
\label{eq:exTransportFreq}
\end{equation}

Relation \eqref{eq:exTransportFreq} links the (left boundary) input to the output through an irrational transfer function $\mathbf G(x,s)$ for any value $x$\footnote{Interestingly, the exact time-domain solution of \eqref{eq:exTransportTime}, along $x$, is given by $\tilde \by(x,t)=\tilde u_f^{t-x^2}/\sqrt{t}$, where $\tilde \bu_f$ is the output of the second order actuator transfer function, in response to $u$.}. Let us now consider that one single sensor is available and is located at $x_m=1.9592$ along the $x$-axis\footnote{In the rest of the chapter, $x$ will be discretized with 50 points from 0 to $L=3$, and $x_m$ has been chosen to be located at $x(\lfloor{ 50 \times 2/3}\rfloor )$.}. The transfer from the same input $\bu(0,s)$, denoted by $\bu(s)$ to $\by(x_m,s)$ denoted by $\by(s)$ then reads $\by(s)=\by(x_m,s)=\mathbf G(s,x_m) \bu(0,s)=\Htran(s) \bu(s)$, where $\Htran(s)$ is now a \textbf{SISO} complex-valued irrational transfer function. 

%%%%%%%%%%%%%%%%%%%%%%%%%%%%%%%%%%%%%%%%%%%%%%%%%%%%%%%%%%%%%% GTA

\subsubsection{A model-driven approximation and control}
\label{sec-model-based}

By Loewner interpolation, the transfer function $\Htran$ can be approximated by a rational function $\Htran_{r}$ ($r=33$). Then, standard feedback synthesis methods can be applied. In this example, the \textsc{hinfstruct} function (embedded in the \textsc{MATLAB} Robust Control Toolbox) has been used \cite{Apkarian:2006}. It allows designing fixed structure controllers while minimising some $\Hinf$-norm oriented performance criterion. Starting from $\Htran_{r}$, let us first define the following generalised plant $\mathbf{T}=\Htran_{r} \mathbf W_o$, where $\mathbf W_o$ is the weighting filter defining the output signals on which the $\mathcal H_\infty$-norm optimisation will be performed. $\mathbf W_o$ is constructed to define the desired closed-loop performances attenuation and its bandwidth which share a similar architecture as the one on Figure \ref{fig:ddc_problem}. Using the same notation, the performance transfer from $\mathbf r$ to $\mathbf e$, is defined as $\mathbf T_{\mathbf r\mathbf e}=\Htran_{r} \mathbf W_o$. In the case considered, one aims at tracking the reference signal $\mathbf r$ and limiting the control action $\mathbf u$. One can then construct $W_o=\textbf{blkdiag}\big(W_e,W_u\big)=\textbf{blkdiag}\big(10\frac{s+1}{s},\frac{s+10}{s+1000}\big)$ describing performance  output $\mathbf z=\textbf{blkdiag}\big(W_e\mathbf e,W_u\u\big)$. The $W_e$ weighting filter has been chosen to weight the sensitivity function and guarantee no steady-state error (\eg roll-off in low frequencies) and a bandwidth around $10^{-1}$ rad/s. $W_u$ is used to weigh the actuator action in high frequencies (here the actuator will roll-off above $10$rad/s). Notice that this is also a fairly standard way of weight selection. The $\Hinf$ control design consists in finding the controller $\mathbf K$, mapping $\mathbf e$ to $\u$, such that, $\mathbf K := \arg \min_{\tilde{\mathbf{K}}\in{\mathcal K}} \norm{\mathcal F_l\big(\mathbf{T}_{\mathbf r\mathbf z},\tilde{\mathbf{K}}\big)}_{\mathcal H_\infty}$, where $\mathcal F_l(\cdot,\cdot)$ is the lower fractional operator defined as (for appropriate partitions of $M$ and $K$) by ${\cal F}_l(M,K) =  M_{11}+M_{12}K(I-M_{22}K)^{-1}M_{21}$ \cite{MagniLFR:2006}. Moreover, it is possible to define the class $\mathcal K$ of $\mathbf K$ to be restricted to the filtered proportional integral (\textbf{PI}), meaning that one is seeking $\mathbf K$ with the following form, $\mathbf K(s) = (k_p + k_i\frac{1}{s})\frac{1}{s/a+1}$, where $k_p,k_i,a \in \Real$. After optimisation, one obtains $k_p=0.1914$, $k_i=0.0251$ and $a=5667.2$ (note also that in this case, the optimal attenuation reached is $\gamma_\infty=66.9558$) \footnote{The optimisation is done using the \textsc{hinfstruct} routine, allowing minimising the closed-loop interconnection of $\mathbf{T_{rz}}$ with $\mathbf{\tilde K}$. In general, we seek for $\norm{\mathcal F_l\big(\mathbf{T}_{\mathbf r\mathbf z},{\mathbf{K}}\big)}_{\mathcal H_\infty}=\gamma_\infty\leq 1$. Here, we simply aim to reaching stability and tracking performances.}.

\subsubsection{Data-driven control}

Let us now apply the \textbf{L-DDC} rationale, instead of a model based control design. As explained in \S 6.1-6.2 of \cite{KergusPhD:2019}, the reference model choice is a key factor for the \textbf{L-DDC}  success, as for any other model reference control procedure. Indeed, the latter should not only represent a desirable closed-loop behaviour, but also achievable dynamics of the considered system (\ie the ideal controller should not internally destabilise the plant). A reference model is said to be achievable by the plant if the corresponding ideal controller internally stabilises the plant. Here let us skip this point and focus on the equivalence of model vs. data-based design. Let the reference model $\mathbf M$ be the closed-loop rational function obtained by the previous approach interconnecting $\Htran_{r}$ with the obtained filtered \textbf{PI} control law obtained in the above section. 

\begin{figure}[H]
\centering
\includegraphics[width=.74\textwidth]{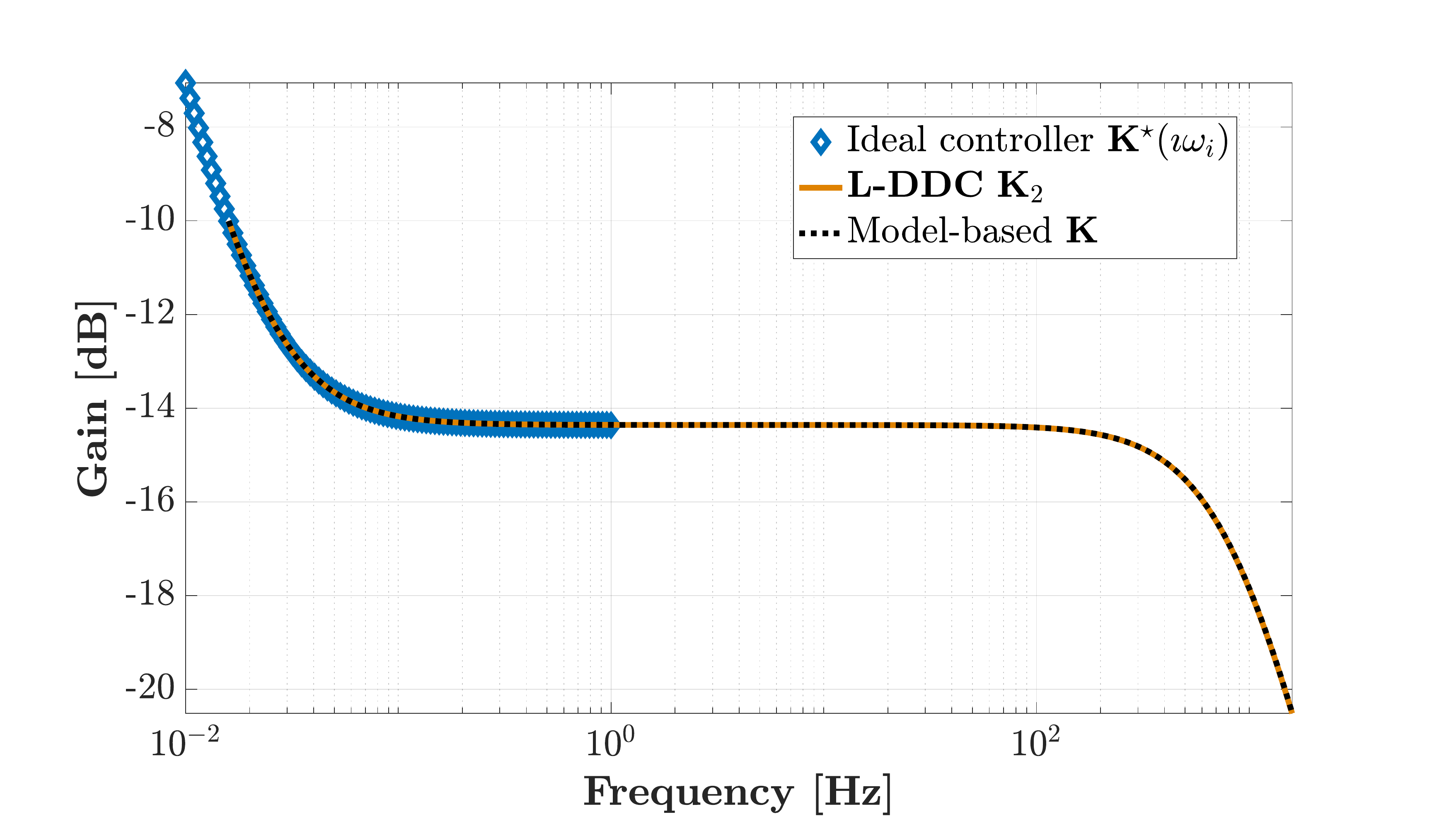}
\includegraphics[width=.74\textwidth]{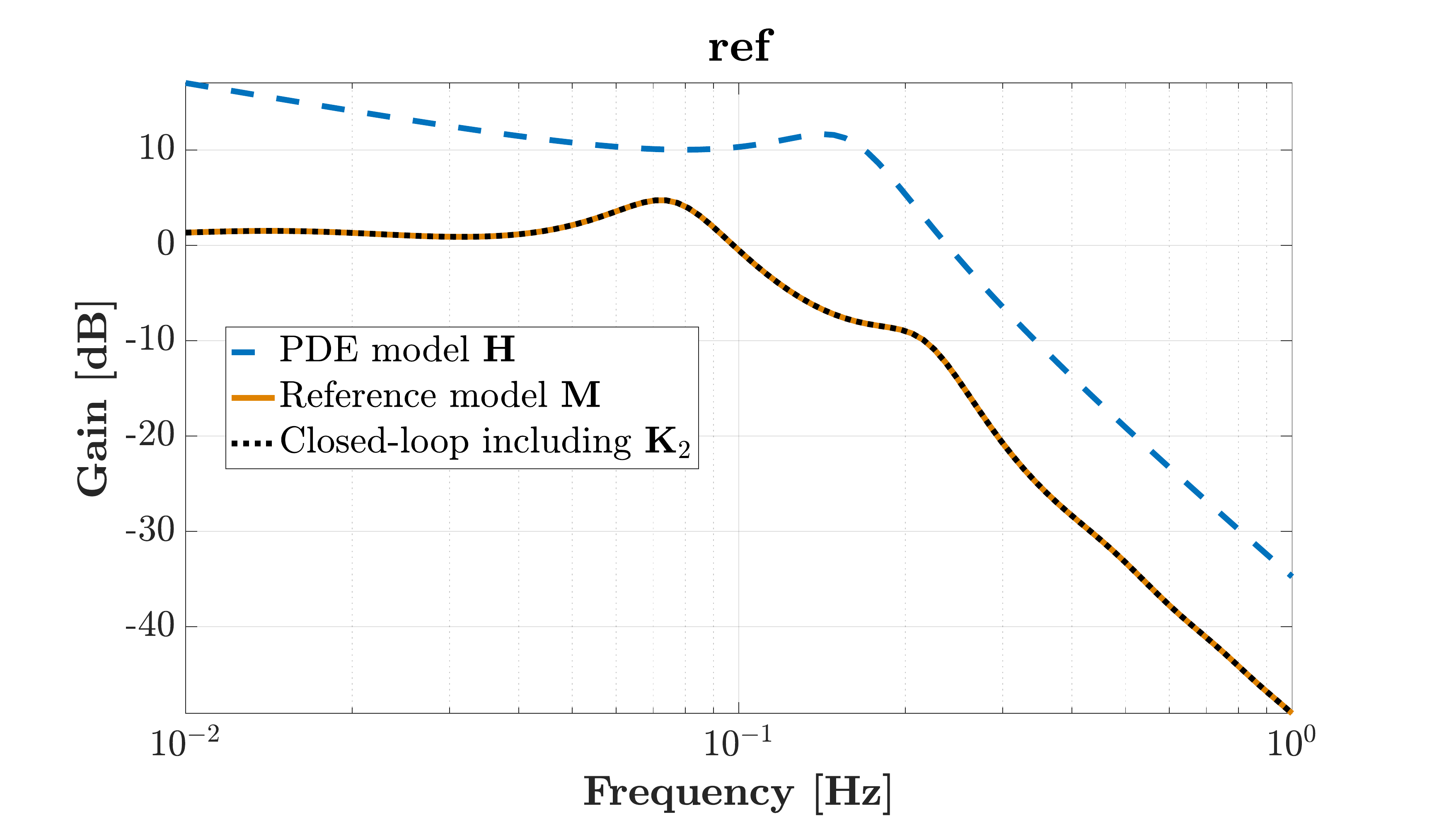}
\caption{Top: frequency response of the controller (ideal, model-based and data-driven). Bottom: open-loop vs. closed-loop frequency responses.}
\label{fig:lddc_K}
\end{figure}

By computing the ideal controller through \eqref{eq:Kideal}, we again compute the Loewner pencil, leading to a minimal realization with $n=42$. Obviously, such a control order is prohibitive for classical control applications. The singular values decay indicates that an order $r=2$ is enough to catch the main dynamics of the underlying controller. One obtains  $\mathbf{K}_r$ ($r=2$) with transfer function 
\begin{equation}
    \mathbf{K}_2(s)=\frac{1082.7(s+0.1313)}{s(s+5656)},
    \label{LDDC_controllers}
\end{equation}
being very close to the numbers obtained by the model-based approach\footnote{The model based approach yield to $\frac{1084.9(s+0.1313)}{s(s+5667)}$}. The controller and resulting close-loop frequency response gains are illustrated on Figure \ref{fig:lddc_K}.

Interestingly, with reference to Figure \ref{fig:lddc_K}, $\mathbf K_r$ perfectly recovers the model-based requested performances of $\mathbf M$ with a controller of rational order two (indeed, we expected to observe this result since we knew from the model-based approach presented in Section \ref{sec-model-based} that a rational control of order leading to this performance is achievable). 

This example demonstrates how the Loewner framework can be effectively used, either for model-based, or for data-driven control. Interestingly, by choosing the closed-loop performances $\mathbf{M}$ obtained with the model-based approach, the controller $\mathbf K_r$  exactly recovers the original properties, while skipping the model construction step and the order selection. This property reduces the model construction step and allows a quick design of the controller. However, this main advantage is balanced by the fact that in the model-based approach, the stability assessment is usually carried out using the approximate model, here $\Htran_{r}$. The latter being very accurate, the eigenvalues computation is traditionally enough for concluding stability, robustness. On the contrary, in the second data-driven approach, stability cannot be analysed as easily. However, \cite{PoussotGTA:2021} suggests an approach based on the combination of Loewner with optimal $\mathcal H_\infty$ projections.

%Both approaches can be viewed as equivalent since they lead to the same  controller. Moreover, in both cases, the interpolatory framework offered by the Loewner matrices is the major ingredient for the success of the design. One may consider these approaches as complementary: the model-based approach may be privileged for critical systems where model understanding is of major  importance and for which engineering time can be spent, while the data-driven one should be the best solution for fast computation, preliminary design, for which neither safety nor critical issues are in the scope.

\section{Summary and Conclusions}
\label{sec:conc}
\label{sec:conc}

In this work, we have provided an inventory of selected extensions and applications of the Loewner framework. The main philosophy of this approach is as follows: use the available data to construct a model or a controller; if needed, apply compression techniques to reduce the complexity of the model or of the controller. The Loewner framework was shown to be applicable for reducing large-scale dynamical systems from computational fluid dynamics (such as the linearized Navier-Stokes model with more than half a million degrees of freedom), to data-driven modeling in aeronautics applications, and to various benchmarks described by complicated dynamics (characterized by irrational transfer functions, having multiple delays, with many input or output ports, with nonlinear terms etc.).  The key observation here is that one can accomplish all of these successful endeavours by having access only to compressed data (transfer function measurements, Markov parameters, etc.), and nothing else. Moreover, the Loewner data-driven control approach was shown to faithfully recover the performance attained by other classical model-based control approaches. Thus, one advantage is the data-driven characteristic, and another is the robustness of the approach. The Loewner framework is hence a valid alternative to intrusive methodologies, and can be successfully used when data are available.

%\appendix 
%\section{Appendix}
%\label{sec:appendix}
%\input{sec-appendix}
\footnotesize

\bibliographystyle{plain}
\bibliography{_biblioPoussot.bib,_biblioCPV.bib,IVG_ref.bib}

\end{document}